\documentclass[aps,prx,showpacs,twocolumn,superscriptaddress]{revtex4}%
\usepackage{graphics}
\usepackage{amsmath}
\usepackage{graphicx}
\usepackage{amsfonts}
\usepackage{amssymb}
\providecommand{\U}[1]{\protect\rule{.1in}{.1in}}

\usepackage{hyperref}
\begin{document}
\title{Environment-assisted bosonic quantum communications}

\begin{abstract}
We consider a quantum relay which is used by two parties to perform several
continuous-variable protocols of quantum communication, from entanglement
distribution (swapping and distillation), to quantum teleportation, and
quantum key distribution. The theory of these protocols is suitably extended
to a non-Markovian model of decoherence characterized by correlated Gaussian
noise in the bosonic environment. In the worst case scenario where bipartite
entanglement is completely lost at the relay, we show that the various
protocols can be reactivated by the assistance of classical (separable)
correlations in the environment. In fact, above a critical amount, these
correlations are able to guarantee the distribution of a weaker form of
entanglement (quadripartite), which can be localized by the relay into a
stronger form (bipartite) that is exploitable by the parties. Our findings are
confirmed by a proof-of-principle experiment where we show, for the first
time, that memory effects in the environment can drastically enhance the
performance of a quantum relay, well beyond the single-repeater bound for
quantum and private communications.
\end{abstract}

\author{Stefano Pirandola}
\email{stefano.pirandola@york.ac.uk}
\affiliation{Department of Computer Science, University of York, York YO10 5GH, United Kingdom}
\author{Carlo Ottaviani}
\affiliation{Department of Computer Science, University of York, York YO10 5GH, United Kingdom}
\author{Christian S. Jacobsen}
\affiliation{Department of Physics, Technical University of Denmark, Fysikvej, 2800 Kongens
Lyngby, Denmark}
\author{Gaetana Spedalieri}
\affiliation{Department of Computer Science, University of York, York YO10 5GH, United Kingdom}
\author{Samuel L. Braunstein}
\affiliation{Department of Computer Science, University of York, York YO10 5GH, United Kingdom}
\author{Tobias Gehring}
\affiliation{Department of Physics, Technical University of Denmark, Fysikvej, 2800 Kongens
Lyngby, Denmark}
\author{Ulrik L. Andersen}
\affiliation{Department of Physics, Technical University of Denmark, Fysikvej, 2800 Kongens
Lyngby, Denmark}
\maketitle


The concept of a relay is at the basis of network information
theory~\cite{CoverThomas}. Indeed the simplest network topology is composed by
three nodes: two end-users, Alice and Bob, plus a third party, the relay,
which assists their communication. This scenario is inherited by quantum
information
theory~\cite{book1,book2,book3,book4,book5,book6,book7,book8,RMP,RMP2,hybrid1,hybrid2}%
, where the mediation of a quantum relay can be found in a series of
fundamental protocols. By sending quantum systems to a middle relay, Alice and
Bob may perform entanglement
swapping~\cite{Zukowski,EntSwap,EntSwap2,GaussSWAP}, entanglement
distillation~\cite{Briegel}, quantum
teleportation~\cite{Tele,Tele2,telereview} and quantum key distribution
(QKD)~\cite{mdiQKD,Lo,Untrusted,TFQKD,SNSwang,QKDrev}.

Quantum relays are crucial elements for quantum network architectures at any
scale, from short-range implementations on quantum chips to long-distance
quantum communication. In all cases, their working mechanism has been studied
assuming Markovian decoherence models, where the errors are independent and
identically distributed (iid). Removing this iid approximation is one of the
goals of modern quantum information theory.

In a quantum chip (e.g., photonic~\cite{photonic1,photonic2} or
superconducting~\cite{chip3}), quantum relays can distribute entanglement
among registers and teleport quantum gates. Miniaturizing this architecture,
correlated errors may come from unwanted interactions between quantum systems.
A common bath may be introduced by a variety of imperfections, e.g., due to
diffraction, slow electronics etc. It is important to realize that
non-Markovian dynamics~\cite{Petruccione} will become increasingly important
as the size of quantum chips further shrinks.

At long distances (in free-space or fibre), quantum relays intervene to assist
quantum communication, entanglement and key distribution. Here,
noise-correlations and memory effects may naturally arise when optical modes
are employed in high-speed communications~\cite{UlrikCORR}, or propagate
through atmospheric turbulence~\cite{Tyler09,Semenov09,Boyd11} and
diffraction-limited linear systems.
Most importantly, correlated errors must be considered in relay-based QKD,
where an eavesdropper (Eve) may jointly attack the two links with the relay
(random permutations and de Finetti arguments~\cite{Renner1,Renner2} cannot
remove these residual correlations). Eve can manipulate the relay itself as
assumed in measurement-device independent QKD~\cite{mdiQKD,Lo,Untrusted}.
Furthermore, Alice's and Bob's setups may also be subject to correlated
side-channel attacks.

For all these reasons, we generalize the study of quantum relays to
non-Markovian conditions, developing the theory for continuous variable (CV)
systems~\cite{RMP} (qubits are discussed in the Supplemental Material). We
consider an environment whose Gaussian noise may be correlated between the two
links. Our model is formulated as a spatial non-Markovian model, where
spatially-separated bosonic modes are subject to correlated errors, but could
also be connected to a time-like model where the parties use the same channel
at different times. In this scenario, while the relay always performs the same
measurement, the parties may implement different protocols (swapping,
distillation, teleportation, or QKD) all based, directly or indirectly, on the
exploitation of bipartite entanglement.

We find a surprising behavior in conditions of extreme decoherence. We
consider entanglement-breaking links~\cite{EBchannels,HolevoEB}, so that no
protocol can work under Markovian conditions. We then induce non-Markovian
effects by progressively increasing the noise correlations in the environment
while keeping their nature separable (so that there is no external reservoir
of entanglement). While these correlations are not able to re-establish
bipartite entanglement (or tripartite entanglement) we find that a critical
amount reactivates quadripartite entanglement, between the setups and the
modes transmitted. In other words, by increasing the separable correlations
above a `reactivation threshold' we can retrieve the otherwise lost
quadripartite entanglement (it is in this sense that we talk of `reactivated'
entanglement below). The measurement of the relay can then localize this
multipartite entanglement into a bipartite form, shared by the two remote
parties and exploitable for the various protocols.

As a matter of fact, we find that all the quantum protocols can be
reactivated. In particular, their reactivation occurs in a progressive
fashion, so that increasing the environmental correlations first reactivates
entanglement swapping and teleportation, then entanglement distillation and
finally QKD. Our theory is confirmed by a proof-of-principle experiment which
shows the reactivation of the most nested protocol, i.e., the QKD\ protocol.
In particular, we show that the key rate of this environmental-assisted
protocol outperforms the single-repeater upper-bound for private
communication~\cite{RepBound}, i.e., the maximum secret key rate that is
achievable in the presence of memory-less links.

\section*{Results}

\begin{figure}[ptbh]
\vspace{-1.5cm}
\par
\begin{center}
\includegraphics[width=0.50\textwidth] {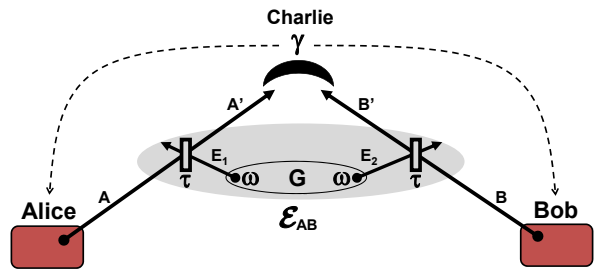}
\end{center}
\par
\vspace{-1.5cm}\caption{\textbf{Quantum relay.} Alice and Bob connect their
devices (red boxes) to a quantum relay, Charlie, for implementing a CV
protocol. On the received modes, Charlie always performs a CV Bell detection
whose outcome $\gamma$ is broadcast. \textbf{Separable Gaussian environment.}
The travelling modes are subject to a joint Gaussian channel $\mathcal{E}%
_{AB}$. This is realized by two beam splitters with tranmissivity $\tau$ which
mix $A$ and $B$ with two ancillary modes, $E_{1}$ and $E_{2}$, respectively.
These ancillas inject thermal noise with variance $\omega$ and belong to a
correlated (but separable) Gaussian state $\rho_{E_{1}E_{2}}$.
\textbf{Entanglement breaking.} For $\omega>\omega_{\text{EB}}(\tau)$,
bipartite (and tripartite) entanglement cannot survive at the relay. In
particular, $A^{\prime}$ is disentangled from Alice's device, and $B^{\prime}$
is disentangled from Bob's, no matter if the environment is correlated or not.
\textbf{Non-Markovian reactivation.} Above a critical amount of separable
correlations, quadripartite entanglement is reactivated between Alice's and
Bob's devices and the transmitted modes, $A^{\prime}$ and $B^{\prime}$. Bell
detection can localize this multipartite resource into a bipartite form and
reactivate all the protocols.}%
\label{generalFIG}%
\end{figure}

\textit{General scenario}.--~As depicted in Fig.~\ref{generalFIG}, we consider
two parties, Alice and Bob, whose devices are connected to a quantum relay,
Charlie, with the aim of implementing a CV\ protocol (swapping, distillation,
teleportation, or QKD). The connection is established by sending two modes,
$A$ and $B$, through a joint quantum channel $\mathcal{E}_{AB}$, whose outputs
$A^{\prime}$ and $B^{\prime}$ are subject to a CV Bell
detection~\cite{BellFORMULA}. This means that modes $A^{\prime}$ and
$B^{\prime}$ are mixed at a balanced beam splitter and then homodyned, one in
the position quadrature $\hat{q}_{-}=(\hat{q}_{A^{\prime}}-\hat{q}_{B^{\prime
}})/\sqrt{2}$ and the other in the momentum quadrature $\hat{p}_{+}=(\hat
{p}_{A^{\prime}}+\hat{p}_{B^{\prime}})/\sqrt{2}$. The classical outcomes
$q_{-}$ and $p_{+}$ can be combined into a complex variable $\gamma
:=q_{-}+ip_{+}$, which is broadcast to Alice and Bob through a classical
public channel. 

The joint quantum channel $\mathcal{E}_{AB}$ corresponds to an environment
with correlated Gaussian noise. This is modelled by two beam splitters (with
transmissivity $0<\tau<1$) mixing modes $A$ and $B$ with two ancillary modes,
$E_{1}$ and $E_{2}$, respectively (see Fig.~\ref{generalFIG}). These ancillas
are taken in a zero-mean Gaussian state~\cite{RMP} $\rho_{E_{1}E_{2}}$ with
covariance matrix (CM) in the symmetric normal form
\[
\mathbf{V}_{E_{1}E_{2}}(\omega,g,g^{\prime})=\left(
\begin{array}
[c]{cc}%
\omega\mathbf{I} & \mathbf{G}\\
\mathbf{G} & \omega\mathbf{I}%
\end{array}
\right)  ,~%
\begin{array}
[c]{c}%
\mathbf{I}:=\mathrm{diag}(1,1),\\
\mathbf{G}:=\mathrm{diag}(g,g^{\prime}).
\end{array}
\]
Here $\omega\geq1$ is the variance of local thermal noise, while the block
$\mathbf{G}$ accounts for noise-correlations.

For $\mathbf{G}=\mathbf{0}$ we retrieve the standard Markovian case, based on
two independent lossy channels~\cite{EntSwap,EntSwap2,GaussSWAP}. For
$\mathbf{G}\neq\mathbf{0}$, the lossy channels become correlated, and the
local dynamics cannot reproduce the global non-Markovian evolution of the
system. Such a separation becomes more evident by increasing the correlation
parameters, $g$ and $g^{\prime}$, whose values are bounded by the bona-fide
conditions $|g|<\omega$, $|g^{\prime}|<\omega$, and $\omega\left\vert
g+g^{\prime}\right\vert \leq\omega^{2}+gg^{\prime}-1$%
~\cite{TwomodePRA,NJPpirs}. In particular, we consider the realistic case of
separable environments ($\rho_{E_{1}E_{2}}$ separable), identified by the
additional constraint $\omega\left\vert g-g^{\prime}\right\vert \leq\omega
^{2}-gg^{\prime}-1$~\cite{NJPpirs}. The amount of separable correlations can
be quantified by the quantum mutual information $I(g,g^{\prime})$.

To analyse entanglement breaking, assume the asymptotic infinite-energy
scenario where Alice's (Bob's) device has a remote mode $a$ ($b$) which is
maximally entangled with $A$ ($B$). We then study the separability properties
of the global system composed by $a$, $b$, $A^{\prime}$ and $B^{\prime}$. In
the Markovian case ($\mathbf{G}=\mathbf{0}$), all forms of entanglement
(bipartite, tripartite~\cite{tripartite}, and
quadripartite~\cite{quadripartite}) are absent for $\omega>\omega_{\text{EB}%
}(\tau):=(1+\tau)/(1-\tau)$, so that no protocol can work. In the
non-Markovian case ($\mathbf{G}\neq\mathbf{0}$) the presence of separable
correlations does not restore bipartite or tripartite entanglement when
$\omega>\omega_{\text{EB}}(\tau)$. However, a sufficient amount of these
correlations is able to reactivate $1\times3$ quadripartite
entanglement~\cite{quadripartite}, in particular, between mode $a$ and the set
of modes $bA^{\prime}B^{\prime}$. See Fig.~\ref{ENCcorr}. \begin{figure}[ptbh]
\vspace{-0.64cm}
\par
\begin{center}
\includegraphics[width=0.35\textwidth]{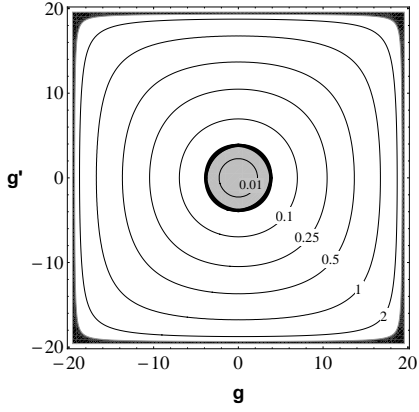}
\end{center}
\par
\vspace{-0.7cm}\caption{Non-Markovian reactivation of $1\times3$ quadripartite
entanglement. Assuming maximally-entangled states for the parties, and
entanglement-breaking conditions (here $\tau=0.9$ and $\omega=1.02\times
\omega_{\text{EB}}=19.38$), we show how quadripartite entanglement is
reactivated by increasing the separable correlations of the environment (bits
of quantum mutual information, which are constant over the concentric contour
lines). Inside the gray region there is no quadripartite entanglement with
respect to any $1\times3$ grouping of the four modes $abA^{\prime}B^{\prime}$.
Outside the gray region all the possible $1\times3$ groupings are entangled.
The external black region is excluded, as it corresponds to entangled or
unphysical environments.}%
\label{ENCcorr}%
\end{figure}

Once quadripartite entanglement is available, the Bell detection on modes
$A^{\prime}$ and $B^{\prime}$ can localize it into a bipartite form for modes
$a$ and $b$. For this reason, entanglement swapping and the other protocols
can be reactivated by sufficiently-strong separable correlations. In the
following, we discuss these results in detail for each specific protocol,
starting from the basic scheme of entanglement swapping. For each protocol, we
first generalize the theory to non-Markovian decoherence, showing how the
various performances are connected. Then, we analyze the protocols under
entanglement breaking conditions.

\textit{Entanglement swapping.}--~The standard source of Gaussian entanglement
is the two-mode squeezed vacuum (TMSV) state, which is a realistic
finite-energy version of the ideal EPR state~\cite{RMP}. More precisely, this
is a two-mode Gaussian state with zero mean-value and CM%
\[
\mathbf{V}(\mu)=\left(
\begin{array}
[c]{cc}%
\mu\mathbf{I} & \sqrt{\mu^{2}-1}\mathbf{Z}\\
\sqrt{\mu^{2}-1}\mathbf{Z} & \mu\mathbf{I}%
\end{array}
\right)  ,~\mathbf{Z}:=\mathrm{diag}(1,-1),
\]
where the variance $\mu\geq1$ quantifies its entanglement. Indeed the
log-negativity~\cite{logNEG,logNEG1,logNEG2} is strictly increasing in $\mu$:
It is zero for $\mu=1$ and tends to infinity for large $\mu$.

\begin{figure}[t]
\vspace{-0.6cm}
\par
\begin{center}
\includegraphics[width=0.53\textwidth] {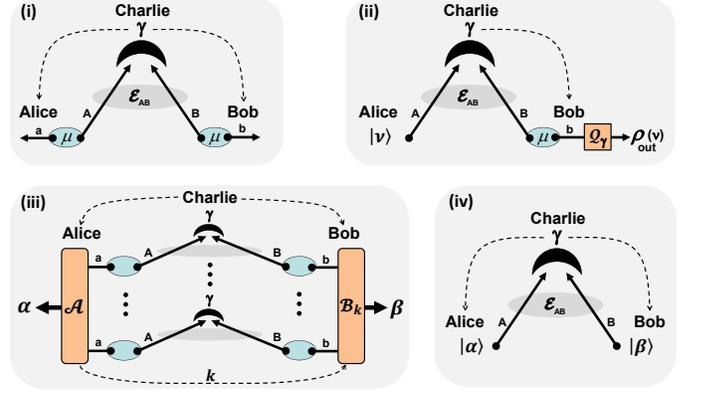}
\end{center}
\par
\vspace{-1.0cm}\caption{Relay-based quantum protocols in a correlated Gaussian
environment.\textbf{ (i)~Entanglement swapping.} Alice and Bob possess two
TMSV states with variance $\mu$. Modes $A$ and $B$ are sent through the joint
channel $\mathcal{E}_{AB}$\ and received by Charlie. After the outcome
$\gamma$ is broadcast, the remote modes, $a$ and $b$, are projected into a
conditional state $\rho_{ab|\gamma}$. \textbf{(ii) Quantum teleportation}.
Alice's coherent state $\left\vert \nu\right\rangle $ is teleported into Bob's
state $\rho_{\text{out}}(\nu)$, after the communication of $\gamma$ and the
action of a conditional quantum operation $\mathcal{Q}_{\gamma}$.
\textbf{(iii) Entanglement/key distillation}. In the limit of many uses of the
relay, Alice performs a quantum instrument on her modes $a$, communicating a
classical variable $k$ to Bob, who performs a conditional quantum operation on
his modes $b$. This is a non-Gaussian quantum repeater where entanglement
swapping is followed by optimal one-way distillation.\textbf{ (iv) Practical
QKD}. Alice and Bob prepare Gaussian-modulated coherent states to be sent to
Charlie. The communication of the outcome $\gamma$ creates remote classical
correlations which are used to extract a secret key. Here the role of Charlie
could be played by Eve, so that the relay becomes an MDI-QKD\ node.}%
\label{swap}%
\end{figure}

Suppose that Alice and Bob have two identical TMSV states, $\rho_{aA}(\mu)$
describing Alice's modes $a$ and $A$, and $\rho_{bB}(\mu)$ describing Bob's
modes $b$ and $B$, as in Fig.~\ref{swap}(i). They keep $a$ and $b$, while
sending $A$ and $B$ to Charlie through the joint channel $\mathcal{E}_{AB}$ of
the Gaussian environment. After the broadcast of the outcome $\gamma$, the
remote modes $a$ and $b$ are projected into a conditional Gaussian state
$\rho_{ab|\gamma}$, with mean-value $\mathbf{x}=\mathbf{x}(\gamma)$ and
conditional CM $\mathbf{V}_{ab|\gamma}$. In the Supplemental Material, we
compute
\begin{equation}
\mathbf{V}_{ab|\gamma}=\left(
\begin{array}
[c]{cc}%
\mathbf{A} & \mathbf{C}\\
\mathbf{C}^{T} & \mathbf{B}%
\end{array}
\right)  , \label{CMmain}%
\end{equation}
where the $2\times2$ blocks are given by%
\begin{align}
\mathbf{A}  &  =\mathbf{B}=\mathrm{diag}\left[  \mu-\frac{\mu^{2}-1}%
{2(\mu+\kappa)},\mu-\frac{\mu^{2}-1}{2(\mu+\kappa^{\prime})}\right]
,\label{blockBmain}\\
\mathbf{C}  &  =\mathrm{diag}\left[  \frac{\mu^{2}-1}{2(\mu+\kappa)}%
,-\frac{\mu^{2}-1}{2(\mu+\kappa^{\prime})}\right]  , \label{blockCmain}%
\end{align}
and the $\kappa$'s contain all the environmental parameters%
\begin{equation}
\kappa:=(\tau^{-1}-1)(\omega-g),~\kappa^{\prime}:=(\tau^{-1}-1)(\omega
+g^{\prime}).
\end{equation}

From $\mathbf{V}_{ab|\gamma}$ we compute the log-negativity $\mathcal{N}%
=\max\{0,-\log_{2}\varepsilon\}$ of the swapped state, in terms of the
smallest partially-transposed symplectic\ eigenvalue $\varepsilon$~\cite{RMP}.
In the Supplemental Material, we derive%
\begin{equation}
\varepsilon=\left[  \frac{(1+\mu\kappa)(1+\mu\kappa^{\prime})}{(\mu
+\kappa)(\mu+\kappa^{\prime})}\right]  ^{1/2}. \label{epsFINITI}%
\end{equation}
For any input entanglement ($\mu>1$), swapping is successful ($\varepsilon<1$)
whenever the environment has enough correlations to satisfy the condition
$\kappa\kappa^{\prime}<1$. The actual amount of swapped entanglement
$\mathcal{N}$ increases in $\mu$, reaching its asymptotic optimum for large
$\mu$, where
\[
\varepsilon\simeq\varepsilon_{\text{opt}}:=\sqrt{\kappa\kappa^{\prime}}.
\]

\textit{Quantum teleportation}.--~As depicted in Fig.~\ref{swap}(ii), we
consider Charlie acting as a teleporter of a coherent state $\left\vert
\nu\right\rangle $ from Alice to Bob. Alice's state and part of Bob's TMSV
state are transmitted to Charlie through the joint channel $\mathcal{E}_{AB}$.
After detection, the outcome $\gamma$ is communicated to Bob, who performs a
conditional quantum operation~\cite{book1} $\mathcal{Q}_{\gamma}$ on mode $b$
to retrieve the teleported state $\rho_{\text{out}}(\nu)\simeq\left\vert
\nu\right\rangle \left\langle \nu\right\vert $. In the Supplemental Material,
we find a formula for the teleportation fidelity $F=F(\mu,\kappa
,\kappa^{\prime})$, which becomes asymptotically optimal for large $\mu$,
where%
\begin{equation}
F\simeq F_{\text{opt}}:=\left[  (1+\kappa)(1+\kappa^{\prime})\right]
^{-1/2}\leq(1+\varepsilon_{\text{opt}})^{-1}. \label{fidEPS}%
\end{equation}
Thus, there is a direct connection between the asymptotic protocols of
teleportation and swapping: If swapping fails ($\varepsilon_{\text{opt}}\geq
1$), teleportation is classical ($F_{\text{opt}}\leq1/2$~\cite{RMP}). We
retrieve the relation $F_{\text{opt}}=(1+\varepsilon_{\text{opt}})^{-1}$ in
environments with antisymmetric correlations $g+g^{\prime}=0$.

\textit{Entanglement distillation}.--~Entanglement distillation can be
operated on top of entanglement swapping as depicted in Fig.~\ref{swap}(iii).
After the parties have run the swapping protocol many times and stored their
remote modes in quantum memories, they can perform a one-way entanglement
distillation protocol on the whole set of swapped states $\rho_{ab|\gamma}$.
This consists of Alice locally applying an optimal quantum
instrument~\cite{Qinstrument} $\mathcal{A}$ on her modes $a$, whose quantum
outcome $\boldsymbol{\alpha}$\ is a distilled system while the classical
outcome $k$ is communicated. Upon receipt of $k$, Bob performs a conditional
quantum operation $\mathcal{B}_{k}$\ transforming his modes $b$ into a
distilled system $\boldsymbol{\beta}$.

The process can be designed in such a way that the distilled systems are
collapsed into entanglement bits (ebits), i.e., Bell state pairs~\cite{book1}.
The optimal distillation rate (ebits per relay use) is
lower-bounded~\cite{Qinstrument} by the coherent information $I_{\mathcal{C}}%
$~\cite{CohINFO,CohINFO2} computed on the single copy state $\rho_{ab|\gamma}%
$. In the Supplemental Material, we find a closed expression $I_{\mathcal{C}%
}=I_{\mathcal{C}}(\mu,\kappa,\kappa^{\prime})$ which is maximized for large
$\mu$, where $I_{\mathcal{C}}\simeq-\log_{2}(e\varepsilon_{\text{opt}})$.
Asymptotically, entanglement can be distilled for $\varepsilon_{\text{opt}%
}<e^{-1}\simeq0.367$.

\textit{Secret key distillation}.--~The scheme of Fig.~\ref{swap}(iii) can be
modified into a key distillation protocol, where Charlie (or Eve~\cite{mdiQKD}%
) distributes secret correlations to Alice and Bob, while the environment is
the effect of a Gaussian attack. Alice's quantum instrument is here a
measurement with classical outputs $\boldsymbol{\alpha}$ (the secret key) and
$k$ (data for Bob). Bob's operation is a measurement conditioned on $k$, which
provides the classical output $\boldsymbol{\beta}$ (key estimate). This is an
ideal key-distribution protocol~\cite{KeyCAP} whose rate is lower-bounded by
the coherent information, i.e., $K\geq I_{\mathcal{C}}$~(see Supplemental Material).

\textit{Practical QKD}.--~The previous key-distribution\ protocol can be
simplified by removing quantum memories and using single-mode measurements, in
particular, heterodyne detections. This is equivalent to a run-by-run
preparation of coherent states, $\left\vert \alpha\right\rangle $ on Alice's
mode $A$, and $\left\vert \beta\right\rangle $ on Bob's mode $B$, whose
amplitudes are Gaussianly modulated with variance $\mu-1$. As shown in
Fig.~\ref{swap}(iv), these states are transmitted to Charlie (or
Eve~\cite{mdiQKD}) who measures and broadcasts $\gamma\simeq\alpha-\beta
^{\ast}$.

Assuming ideal reconciliation~\cite{RMP}, the secret key rate $R=R(\mu
,\kappa,\kappa^{\prime})$ increases in $\mu$. Modulation variances $\mu
\gtrsim50$ are experimentally achievable and well approximate the asymptotic
limit for $\mu\gg1$, where the key rate is optimal and satisfies~(see
Supplemental Material)%
\begin{equation}
R_{\text{opt}}\gtrsim\log_{2}\left(  \frac{F_{\text{opt}}}{e^{2}%
\varepsilon_{\text{opt}}}\right)  +h(1+2\varepsilon_{\text{opt}}),
\label{rateTEXT}%
\end{equation}
with $h(x):=\frac{x+1}{2}\log_{2}\frac{x+1}{2}-\frac{x-1}{2}\log_{2}\frac
{x-1}{2}$. Using Eq.~(\ref{fidEPS}), we see that the right hand side of
Eq.~(\ref{rateTEXT}) can be positive only for $\varepsilon_{\text{opt}%
}\lesssim0.192$. Thus the practical QKD protocol is the most difficult to
reactivate: Its reactivation implies that of entanglement/key distillation and
that of entanglement swapping. This is true not only asymptotically but also
at finite $\mu$ as we show below.

\begin{figure}[ptbh]
\vspace{0.15cm}
\par
\begin{center}
\includegraphics[width=0.48\textwidth]{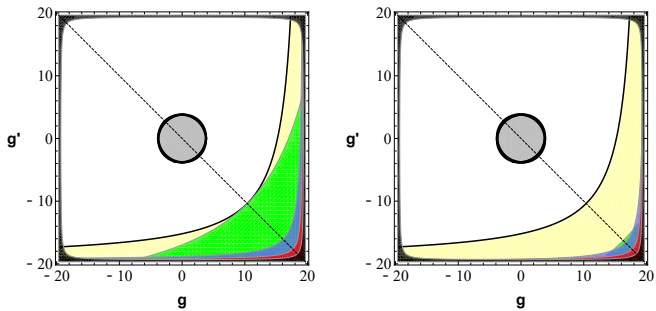}
\end{center}
\par
\vspace{-0.6cm}\caption{Non-Markovian reactivation of quantum protocols from
entanglement-breaking (here $\tau=0.9$ and $\omega=1.02\times\omega
_{\text{EB}}=19.38$). Each point of the correlation plane corresponds to a
Gaussian environment with separable correlations. In panel \textbf{a} we
consider the optimal scenario of large $\mu$ (asymptotic protocols). Once
quadripartite $1\times3$ entanglement has been reactivated (outside the gray
ring), we have the progressive reactivation of entanglement swapping
($\mathcal{N}>0$, yellow region), quantum teleportation of coherent states
($F>1/2$, green region), entanglement/key distillation ($I_{\mathcal{C}}$,
$K>0$, blue region) and practical QKD ($R>0$, red region). Panel \textbf{b} as
in \textbf{a} but refers to a realistic scenario with experimentally
achievable values of $\mu$. We consider $\mu\simeq6.5$%
~\cite{TopENT,TobiasSqueezing} as input entanglement for the
entanglement-based protocols, and $\mu\simeq50$ as modulation for the
practical QKD\ protocol. The reactivation phenomenon persists and can be
explored with current technology. Apart from teleportation, the other
thresholds undergo small modifications.}%
\label{total}%
\end{figure}

\textit{Reactivation from entanglement breaking}.--~Once the theory of the
previous protocols has been extended to non-Markovian decoherence, we can
study their reactivation from entanglement breaking conditions. Consider an
environment with transmissivity $\tau$ and entanglement-breaking thermal noise
$\omega>\omega_{\text{EB}}(\tau)$, so that no protocol can work for
$\mathbf{G}=\mathbf{0}$. By increasing the separable correlations in the
environment, not only can quadripartite entanglement be reactivated but, above
a certain threshold, it can also be localized into a bipartite form by the
relay's Bell detection. Once entanglement swapping is reactivated, all other
protocols can progressively be reactivated. As shown in Fig.~\ref{total},
there are regions of the correlation plane where entanglement can be swapped
($\mathcal{N}>0$), teleportation is quantum ($F>1/2$), entanglement and keys
can be distilled ($I_{\mathcal{C}}$, $K>0$), and practical QKD can be
performed ($R>0$). This occurs both for large and experimentally-achievable
values of $\mu$. 

Note that the reactivation is asymmetric in the plane only because of the
specific Bell detection adopted, which generates correlations of the type
$g>0$ and $g^{\prime}<0$. Using another Bell detection (projecting onto
$\hat{q}_{+}$ and $\hat{p}_{-}$), the performances would be inverted with
respect to the origin of the plane. Furthermore, the entanglement localization
(i.e., the reactivation of entanglement swapping) is triggered for
correlations higher than those required for restoring quadripartite
entanglement, suggesting that there might exist a better quantum measurement
for this task. The performances of the various protocols improve by increasing
the separable correlations of the environment, with the fastest reactivation
being achieved along the diagonal $g+g^{\prime}=0$, where swapping and
teleportation are first recovered, then entanglement/key distillation and
practical QKD, which is the most nested region.

\textit{Correlated additive noise}.--~The phenomenon can also be found in
other types of non-Markovian Gaussian environments. Consider the limit for
$\tau\rightarrow1$ and $\omega\rightarrow+\infty$, while keeping constant
$n:=(1-\tau)\omega$, $c:=g(\omega-1)^{-1}$ and $c^{\prime}:=g^{\prime}%
(\omega-1)^{-1}$. This is an asymptotic environment which adds correlated
classical noise to modes $A$ and $B$, so that their quadratures undergo the
transformations%
\[
\left(  \hat{q}_{A},\hat{p}_{A},\hat{q}_{B},\hat{p}_{B}\right)  \rightarrow
\left(  \hat{q}_{A},\hat{p}_{A},\hat{q}_{B},\hat{p}_{B}\right)  +(\xi_{1}%
,\xi_{2},\xi_{3},\xi_{4}).
\]
Here the $\xi_{i}$'s are zero-mean Gaussian variables\ whose covariances
$\left\langle \xi_{i}\xi_{j}\right\rangle $ are specified by the classical CM%
\begin{equation}
\mathbf{V}\left(  n,c,c^{\prime}\right)  =n\left(
\begin{array}
[c]{cc}%
\mathbf{I} & \mathrm{diag}(c,c^{\prime})\\
\mathrm{diag}(c,c^{\prime}) & \mathbf{I}%
\end{array}
\right)  , \label{CMadditive}%
\end{equation}
where $n\geq0$ is the variance of the additive noise, and $-1\leq c,c^{\prime
}\leq1$ quantify the classical correlations. The entanglement-breaking
condition becomes $n>2$.

\begin{figure}[ptbh]
\vspace{+0.13cm}
\par
\begin{center}
\includegraphics[width=0.46\textwidth]{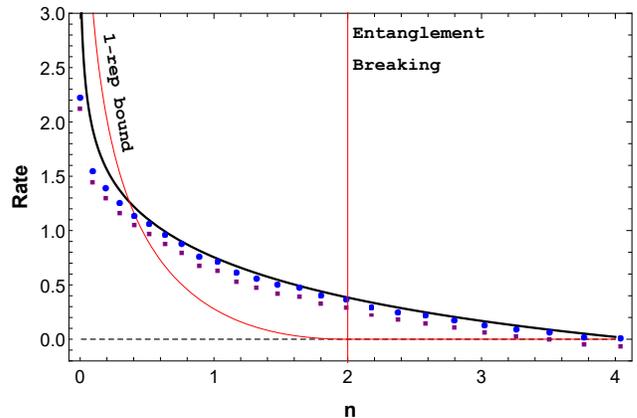}
\end{center}
\par
\vspace{-0.4cm}\caption{Plot the secret-key rate $R$ (bits per relay use) as a
function of the additive noise $n$. The solid black curve is the theoretical
rate computed for a correlated-additive environment ($c=c^{\prime}=1$)
and\ realistic signal modulation ($\mu\simeq52$). This rate is positive after
entanglement breaking ($n>2$) and beats the single-repeater
bound~\cite{RepBound} (based on memoryless links). Points are
experimental data: Blue circles refer to ideal reconciliation, and purple
squares to achievable reconcilation efficiency ($\simeq0.97$).
Error bars on the $x$-axis are smaller than the point size.
Due to loss at the untrusted relay, the experimental key rate is slightly below the
theoretical curve (associated with the correlated side-channel attack). }%
\label{ExpOUT}%
\end{figure}

To show non-Markovian effects, we consider the protocol which is the most
difficult to reactivate, the practical QKD protocol. We can specify its key
rate $R(\mu,n,c,c^{\prime})$ for $c=c^{\prime}=1$ and assume a realistic
modulation $\mu\simeq52$. We then plot $R$ as a function of the additive noise
$n$ in Fig.~\ref{ExpOUT}. As we can see, the rate decreases in $n$ but remains
positive in the region $2<n\leq4$ where the links with the relay become
entanglement-breaking. As we show below, this behaviour persists in the
presence of loss, as typically introduced by experimental imperfections.

Recall that, for an additive Gaussian channel with added noise $n$, the secret
key capacity (and any other two-way assisted quantum capacity) is
upper-bounded by
\begin{equation}
\Phi(n):=\frac{(n/2)-1}{\ln2}-\log_{2}(n/2),\label{Kadd}%
\end{equation}
for $n\leq2$ and zero otherwise. The bound $\Phi(n)$ in Eq.~(\ref{Kadd}) has
been proven in Ref.~\cite[Eq.~(29)]{PLOB} and here reported in our different
vacuum units. In the presence of a relay/repeater, where each link is
described by an independent bosonic Gaussian channel, Ref.~\cite{RepBound}
established that the secret key capacity assisted by the repeater
$K_{\text{1-rep}}$ is upper-bounded by the minimum secret key capacity of the
links. In the present setting, we therefore have the single-repeater bound
$K_{\text{1-rep}}\leq\Phi(n)$. As we show in Fig.~\ref{ExpOUT}, the presence
of classical (separable) correlations in the Gaussian environment lead to the
violation of the bound $\Phi(n)$\ when $n\gtrsim0.369$ (for the theoretical
curve) and $n\gtrsim0.4$ (for the experimental results).

\begin{figure}[t]
\vspace{+0.15cm}
\par
\begin{center}
\includegraphics[width=0.45\textwidth]{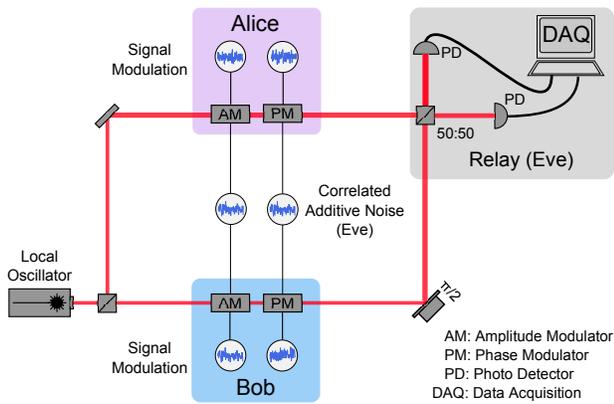}
\end{center}
\par
\vspace{-0.1cm}\caption{\textbf{Experimental setup}. Alice and Bob receive
1064 nm light from the same laser source (local oscillator). At both stations,
the incoming beams are Gaussianly modulated in phase and amplitude using
electro-optical modulators driven by uncorrelated signal generators. In
addition, the phase and amplitude modulators for Alice and Bob have correlated
inputs respectively, such that a noisy modulation identical for both Alice and
Bob is added to the phase and amplitude signals (side-channel attack). The
magnitudes of the correlated noise modulations are progressively increased
(from $n=0$ to $4$), and kept symmetrical between the quadratures, while the
signal modulations are kept constant at the same level in both quadratures for
Alice and Bob ($\mu\simeq52$). At the untrusted relay, the modes are mixed at
a balanced beam splitter and the output ports photo-detected, with an overall
efficiency of $\simeq98\%$. Photocurrents are then processed to realize a CV
Bell measurement. See Supplemental Material for details.}%
\label{SetupFIGmain}%
\end{figure}

\textit{Experimental results}.--~Our theoretical results are confirmed by a
proof-of-principle experiment, whose setup is schematically depicted in
Fig.~\ref{SetupFIGmain}. We consider Alice and Bob generating Gaussianly modulated
coherent states by means of independent electro-optical modulators, applied to
a common local oscillator. Simultaneously, the modulators are subject to a
side-channel attack: Additional electrical inputs are introduced by Eve, whose
effect is to generate additional and unknown phase-space displacements. In
particular, Eve's electrical inputs are correlated so that the resulting
optical displacements introduce a correlated-additive Gaussian environment
described by Eq.~(\ref{CMadditive}) with $c\simeq1$ and $c^{\prime}\simeq1$.
The optical modes then reach the midway relay, where they are mixed at a
balanced beam splitter and the output ports photo-detected. Although the
measurement is highly efficient, it introduces a small loss ($\simeq2\%$)
which is assumed to be exploited by Eve in the worst-case
scenario.

From the point of view of Alice and Bob, the side-channel attack and the
additional (small)\ loss at the relay are jointly perceived as a global
coherent Gaussian attack of the optical modes. Analysing the statistics of the
shared classical data and assuming that Eve controls the entire environmental
purification compatible with this data, the two parties may compute the
experimental secret-key rate (see details in the Supplemental Material). As we
can see from Fig.~\ref{ExpOUT}, the experimental points are slightly below the
theoretical curve associated with the correlated-additive environment,
reflecting the fact that the additional loss at the relay tends to degrade the
performance of the protocol. The experimental rate is able to beat the
single-repeater bound for additive-noise Gaussian links~\cite{RepBound} and
remains positive after the entanglement-breaking threshold, so that\ the
non-Markovian reactivation of QKD is experimentally confirmed.

\section*{Discussion}

We have theoretically and experimentally demonstrated that the most important
protocols operated by quantum relays can work in conditions of extreme
decoherence thanks to the presence of non-Markovian memory effects in the
environment. Assuming high Gaussian noise in the links, we have considered a
regime where any form of entanglement (bipartite, tripartite or quadripartite)
is broken under Markovian memoryless conditions. By allowing for a suitable
amount of correlations in the environment, we have proven that we can
reactivate the distribution of $1\times3$ quadripartite entanglement, and this
resource can successfully be localised into a bipartite form exploitable by
Alice and Bob. As a result, all the basic protocols for quantum and private
communication can be progressively reactivated by the action of the relay.

Surprisingly, this reactivation is possible without the need of any injection
of entanglement from the environment, but just because of the presence of
weaker classical correlations (described by a separable state for the
environment). In particular, we have shown that these correlations lead to the
violation of the single-repeater bound for quantum and private communications.

Our results might open new perspectives for all quantum systems where
correlated errors and memory effects are typical forms of decoherence. This
may involve both short-distance implementations (e.g., chip-based), and
long-distance ones, as is the case of relay-based QKD. Non-Markovian memory
effects should therefore be regarded as a potential physical resource to be
exploited in various settings of quantum communication.

\section*{Methods}

Theoretical and experimental methods are given in the Supplemental Material.
Theoretical methods contain details about the following points: (i)~Study of
the Gaussian environment with correlated thermal noise, including a full
analysis of its correlations. (ii)~Study of the various forms of entanglement
available before the Bell detection of the relay. (iii)~Study of the
entanglement swapping protocol, i.e., the computation of the CM $\mathbf{V}%
_{ab|\gamma}$ in Eq.~(\ref{CMmain}) and the derivation of the eigenvalue
$\varepsilon$ in Eq.~(\ref{epsFINITI}). (iv)~Generalization of the
teleportation protocol with details on Bob's quantum operation $\mathcal{Q}%
_{\gamma}$ and the analytical formula for the fidelity $F(\mu,\kappa
,\kappa^{\prime})$. (v)~Details of the distillation protocol with the
analytical formula of $I_{\mathcal{C}}(\mu,\kappa,\kappa^{\prime})$.
(vi)~Details of the ideal key-distillation protocol, discussion on
MDI-security, and proof of the lower-bound $K\geq I_{\mathcal{C}}$.
(vii)~Derivation of the general secret-key rate $R(\xi,\mu,\kappa
,\kappa^{\prime})$ of the practical QKD protocol, assuming arbitrary
reconciliation efficiency $\xi$ and modulation variance $\mu$. (viii)~Explicit
derivation of the optimal rate $R_{\text{opt}}$ and the proof of the tight
lower bound in Eq.~(\ref{rateTEXT}). (ix)~Derivation of the
correlated-additive environment as a limit of the correlated-thermal one.
(x)~Study of entanglement swapping and practical QKD in the
correlated-additive environment, providing the formula of the secret-key rate
$R(\xi,\mu,n,c,c^{\prime})$.

\section*{Acknowledgements}

This work has been funded by the EPSRC via the projects `qDATA' (EP/L011298/1)
and `Quantum Communications hub' (EP/M013472/1, EP/T001011/1), and by the
European Union via \textquotedblleft Continuous Variable Quantum
Communications\textquotedblright\ (CiViQ, grant agreement No 820466). S.P also
thanks the Leverhulme Trust (research fellowship `qBIO'). G.S. has been
sponsored by the EU via a\ Marie Sk\l odowska-Curie Global Fellowship (grant
No. 745727). T.G. acknowledges support from the H. C. {\O }rsted postdoc
programme. U. L. A. thanks the Danish Agency for Science, Technology and
Innovation (Sapere Aude project).

\newpage

\newcounter{S} \setcounter{section}{0} \setcounter{subsection}{0}
\renewcommand{\thefigure}{\arabic{figure}} 
\renewcommand{\figurename}{Figure}




\renewcommand{\thesection}{Sec. \arabic{section}} 

\bigskip

\bigskip

\begin{center}
{\huge Supplemental Material}
\end{center}


\section{Basics of Gaussian formalism\label{APP_subGAUSS}}

This section aims to help readers not familiar with continuous-variable (CV)
systems and Gaussian states. Those familiar with this formalism may skip this~\ref{APP_subGAUSS}.

\subsection{Gaussian states and operations}

A bosonic system of $n$ modes is described by a vector of $2n$ quadrature
operators%
\[
\mathbf{\hat{x}}^{T}:=(\hat{q}_{1},\hat{p}_{1},\ldots,\hat{q}_{n},\hat{p}%
_{n})~,
\]
satisfying $[\hat{x}_{i},\hat{x}_{j}]=2i\Omega_{ij}$, where $i,j=1,\ldots,2n$
and $\Omega_{ij}$ is the generic element of the symplectic form%
\begin{equation}
\mathbf{\Omega}^{(n)}:=\bigoplus\limits_{k=1}^{n}\left(
\begin{array}
[c]{cc}%
0 & 1\\
-1 & 0
\end{array}
\right)  ~. \label{Symplectic_Form}%
\end{equation}
A bosonic state $\rho$ is \textquotedblleft Gaussian\textquotedblright\ when
its Wigner phase-space representation is Gaussian~\cite{RMP}, so that it is
fully characterized by its first and second-order statistical moments.

The first-order moment is the mean value $\mathbf{\bar{x}}:=\langle
\mathbf{\hat{x}}\rangle$, where $\langle\hat{O}\rangle:=\mathrm{Tr}(\hat
{O}\rho)$ denotes the average of the arbitrary operator $\hat{O}$ on the state
$\rho$. The second-order moment is the covariance matrix (CM) $\mathbf{V}$,
with element%
\[
V_{ij}:=\frac{1}{2}\left\langle \{\Delta\hat{x}_{i},\Delta\hat{x}%
_{j}\}\right\rangle ~,
\]
where $\Delta\hat{x}_{i}:=\hat{x}_{i}-\bar{x}_{i}$ is the deviation and
$\{,\}$ is the anticommutator. The CM\ is a $2n\times2n$ real symmetric
matrix, which is positive-definite and must satisfy the uncertainty
principle~\cite{RMP}%
\begin{equation}
\mathbf{V}+i\mathbf{\Omega}^{(n)}\geq0~. \label{unc_PRINC}%
\end{equation}

The simplest Gaussian states are thermal states. A single-mode thermal state
has zero mean and CM$\ \mathbf{V}=(2\bar{n}+1)\mathbf{I}$, where $\mathbf{I}$
is the $2\times2$ identity matrix and $\bar{n}\geq0$ is the mean number of
thermal photons (vacuum state for $\bar{n}=0$). Multimode thermal states are
constructed by tensor product. Tensor product of states $\rho_{1}\otimes
\rho_{2}$\ corresponds to direct sum of CMs $\mathbf{V}_{1}\oplus
\mathbf{V}_{2}$. Conversely, the partial trace $\rho_{1}=\mathrm{Tr}_{2}%
(\rho_{12})$ corresponds to collapsing the total CM $\mathbf{V}_{12}$ into the
block $\mathbf{V}_{1}$ spanned by $(\hat{q}_{1},\hat{p}_{1})$.

By definition, a Gaussian channel transforms Gaussian states into Gaussian
states. Its action $\rho\rightarrow\mathcal{E}(\rho)$ corresponds to the
following transformation for the CM%
\[
\mathbf{V}\rightarrow\mathbf{KVK}^{T}+\mathbf{N}~,
\]
where $\mathbf{K}$ and $\mathbf{N}=\mathbf{N}^{T}$ are $2n\times2n$ real
matrices, satisfying suitable bona-fide conditions~\cite{RMP}.

A reversible Gaussian channel is a Gaussian unitary $\rho\rightarrow U\rho
U^{\dagger}$, whose action can be described by%
\[
\mathbf{\bar{x}}\rightarrow\mathbf{S\bar{x}}+\mathbf{d},~\mathbf{V}%
\rightarrow\mathbf{SVS}^{T},
\]
where $\mathbf{d}$ is a real displacement vector and $\mathbf{S}$ is a
symplectic matrix, i.e., a real matrix preserving the symplectic form
$\mathbf{S\Omega}^{(n)}\mathbf{S}^{T}=\mathbf{\Omega}^{(n)}$. In the
Heisenberg picture, a Gaussian unitary corresponds to the affine map
\[
\mathbf{\hat{x}}\rightarrow\mathbf{S\hat{x}}+\mathbf{d~.}%
\]

The basic example of Gaussian channel is the one-mode lossy channel, defined
by the matrices%
\[
\mathbf{K}=\sqrt{\tau}\mathbf{I,~N}=(1-\tau)(2\bar{n}+1)\mathbf{I~,}%
\]
where $0\leq\tau\leq1$ is the transmissivity of the channel and $\bar{n}\geq0$
its thermal number. This channel can be dilated into a two-mode Gaussian
unitary mixing the input state with an environmental thermal state with
$\bar{n}$ mean photons. This Gaussian unitary is the beam-splitter
transformation, characterized by the symplectic matrix%
\begin{equation}
\mathbf{S}(\tau)=\left(
\begin{array}
[c]{cc}%
\sqrt{\tau}\mathbf{I} & \sqrt{1-\tau}\mathbf{I}\\
-\sqrt{1-\tau}\mathbf{I} & \sqrt{\tau}\mathbf{I}%
\end{array}
\right)  ~. \label{BSsymplectic}%
\end{equation}

\subsection{Symplectic spectrum\label{APP_subSPECTRUM}}

According to Williamson's theorem~\cite{Williamson,RMP}, an arbitrary
CM\ $\mathbf{V}$ can be diagonalized by a symplectic matrix $\mathbf{S}$ as
\[
\mathbf{V=S~}\left[  \bigoplus\limits_{k=1}^{n}\nu_{k}\mathbf{\mathbf{I}%
}\right]  \mathbf{~S}^{T}\mathbf{,}%
\]
where $\{\nu_{1},\cdots,\nu_{n}\}$ are the $n$ symplectic eigenvalues. Using
the symplectic spectrum, we write the uncertainty principle in a simple form.
Assuming that $\mathbf{V}>0$ holds, then Eq.~(\ref{unc_PRINC}) is equivalent
to $\nu_{k}\geq1$.

Given the symplectic spectrum, we can compute the von Neumann entropy
$S(\rho):=-$Tr$(\rho\log\rho)$ of an arbitrary $n$-mode Gaussian state as
follows~\cite{RMP}%
\begin{equation}
S(\rho)=\sum_{k=1}^{n}h(\nu_{k})~, \label{VN_Entropy}%
\end{equation}
where%
\[
h(x):=\frac{x+1}{2}\log_{2}\frac{x+1}{2}-\frac{x-1}{2}\log_{2}\frac{x-1}{2}~.
\]
Whereas the symplectic eigenvalues are large, we can use the asymptotic
expansion~\cite{RMP}%
\begin{equation}
h(x)\simeq\log_{2}\frac{e}{2}x+O\left(  \frac{1}{x}\right)  ~. \label{EXPA}%
\end{equation}

\subsection{Two-mode Gaussian states\label{APP_sub2modes}}

Let us consider two modes only, say $A$ and $B$, in a zero-mean Gaussian state
$\rho_{AB}$ with CM in the blockform
\begin{equation}
\mathbf{V}=\left(
\begin{array}
[c]{cc}%
\mathbf{A} & \mathbf{C}\\
\mathbf{C}^{T} & \mathbf{B}%
\end{array}
\right)  ~, \label{BlockFORM}%
\end{equation}
where $\mathbf{A}$, $\mathbf{B}$ and $\mathbf{C}$ are $2\times2$ matrices.
Finding the symplectic spectrum $\{\nu_{-},\nu_{+}\}$ is straightforward,
since~\cite{Sera, RMP}
\begin{equation}
\nu_{\pm}=\sqrt{\frac{\Delta\pm\sqrt{\Delta^{2}-4\det\mathbf{V}}}{2}}~,
\label{symFORM}%
\end{equation}
where $\Delta:=\det\mathbf{A}+\det\mathbf{B}+2\det\mathbf{C}$. \

It is easy to study the separability properties of a two-mode Gaussian state.
Let us introduce the reflection matrix $\mathbf{Z}:=\mathrm{diag}(1,-1)$ and
define the partial transposition (PT) matrix
\begin{equation}
\boldsymbol{\Lambda}:=\mathbf{Z}\oplus\mathbf{I}=\mathrm{diag}(1,-1,1,1),
\label{lambda2modes}%
\end{equation}
so that we can compute the partial transpose $\mathbf{\tilde{V}}%
=\boldsymbol{\Lambda}\mathbf{V}\boldsymbol{\Lambda}$. Then, the state is
separable if and only if
\begin{equation}
\mathbf{\tilde{V}}+i\mathbf{\Omega}^{(2)}\geq0. \label{sepaCONDapp}%
\end{equation}
The latter condition is the positive partial transpose (PPT) criterion
expressed in terms of CMs.

Then, to quantify entanglement, we derive the smallest symplectic eigenvalue
of $\mathbf{\tilde{V}}$, also known as the smallest partially-transposed
symplectic (PTS) eigenvalue. This eigenvalue $\varepsilon$ can directly be
computed from the formula of $\nu_{-}$ in Eq.~(\ref{symFORM}) up to replacing
$\Delta$ with $\tilde{\Delta}=\det\mathbf{A}+\det\mathbf{B}-2\det\mathbf{C}$.
The Gaussian state is entangled if and only if $\varepsilon<1$, and its
log-negativity~\cite{logNEG} is equal to%
\begin{equation}
\mathcal{N}=\max\left\{  0,-\log_{2}\varepsilon\right\}  ~. \label{logNEGdef}%
\end{equation}

By means of local symplectic transformations $\mathbf{S}_{A}\oplus
\mathbf{S}_{B}$ (preserving the correlations of the state), we can always
transform an arbitrary CM into the normal form
\[
\mathbf{V}(a,b,c,c^{\prime})=\left(
\begin{array}
[c]{cccc}%
a & 0 & c & 0\\
0 & a & 0 & c^{\prime}\\
c & 0 & b & 0\\
0 & c^{\prime} & 0 & b
\end{array}
\right)  ~,
\]
where the four parameters are connected to the original CM by the relations
$a^{2}=\det\mathbf{A}$, $b^{2}=\det\mathbf{B}$, $cc^{\prime}=\det\mathbf{C}$
and $\det\mathbf{V}=(ab-c^{2})(ab-c^{\prime2})$. This normal-form CM
$\mathbf{V}(a,b,c,c^{\prime})$ is the starting point for computing the quantum
discord of the corresponding Gaussian state~\cite{OptimalDiscord}.

Quantum discord~\cite{RMPdis} is defined by the difference
\begin{equation}
D(A|B)=I(\rho_{AB})-C(A|B)~,\label{DISCORDeq}%
\end{equation}
where $I(\rho_{AB})=S(\rho_{A})+S(\rho_{B})-S(\rho_{AB})$ is the quantum
mutual information of the state, and $C(A|B)$ quantifies its (non-discordant)
purely-classical correlations. These are given by
\begin{equation}
C(A|B)=S(\rho_{A})-\inf_{M}H_{M}(A|B)~,\label{CDEVE}%
\end{equation}
where $M=\{M_{k}\}$ is a POVM acting on mode $B$ and
\begin{equation}
H_{M}(A|B):=\sum_{k}p_{k}S(\rho_{A|k})~,\label{accaM}%
\end{equation}
where $p_{k}$ is the probability of the outcome $k$, and $\rho_{A|k}$ is the
conditional state of mode $A$.

In the case of Gaussian states, quantum discord can be upper-bounded by
Gaussian discord~\cite{GerryD,ParisD}, which restricts the minimization above
to Gaussian POVMs. Ref.~\cite{OptimalDiscord} showed that quantum discord and
Gaussian discord actually coincide for a large family of Gaussian states. This
family includes all Gaussian states with CMs in the normal-form $\mathbf{V}%
(a,b,c,c^{\prime})$ with $\left\vert c\right\vert =\left\vert c^{\prime
}\right\vert $. This further includes the class of two-mode squeezed thermal
states for which $c^{\prime}=-c$ (those considered in Ref.~\cite{ParisD}).

\subsection{Three-mode Gaussian states}

Here we provide the basic criteria to study the separability
properties of three-mode Gaussian states. These states can display different
types of tripartite entanglement, whose classification is based on the
generalization of the PPT criterion to multimode Gaussian
states~\cite{quadripartite}. In general, let us consider one mode $\{0\}$\ for
Alice and $m$ modes $\{1,\cdots,m\}$ for Bob. Let $\mathbf{V}$ be the CM of a
Gaussian state $\rho$\ of such a $1\times m$ system and denote by
\[
\boldsymbol{\Lambda}_{A}:=\mathbf{Z}\oplus\underset{m}{\underbrace
{\mathbf{I}\oplus\cdots\oplus\mathbf{I}}}%
\]
the partial transposition with respect to Alice's mode. Then $\rho$ is
separable, with respect to the grouping $A=\{0\}$ and $B=\{1,...,m\}$, if and
only if~\cite{quadripartite}%
\begin{equation}
\boldsymbol{\Lambda}_{A}\mathbf{V}\boldsymbol{\Lambda}_{A}+i\mathbf{\Omega
}^{(1+m)}\geq0~. \label{PPT_1N}%
\end{equation}

\begin{figure}[ptbh]
\vspace{-1.1cm}
\par
\begin{center}
\includegraphics[width=0.5\textwidth] {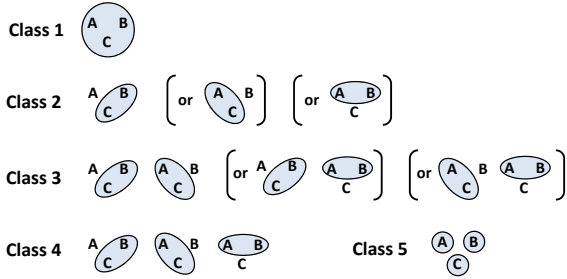}
\end{center}
\par
\vspace{-1.7cm}\caption{Classification of tripartite entanglement. See text
for more details and Ref.~\cite{tripartite} for the definitions of the
classes.}%
\label{TripartiteAPP}%
\end{figure}

In the case of three-mode Gaussian states, this criterion can be applied to
all possible groupings of modes. Let us consider three modes $A$, $B$ and $C$,
described by a Gaussian state with CM $\mathbf{V}$. We may define the three PT
matrices%
\[
\boldsymbol{\Lambda}_{A}:=\mathbf{Z}\oplus\mathbf{I}\oplus\mathbf{I,~}%
\boldsymbol{\Lambda}_{B}:=\mathbf{I}\oplus\mathbf{Z}\oplus\mathbf{I,~}%
\boldsymbol{\Lambda}_{C}:=\mathbf{I}\oplus\mathbf{I}\oplus\mathbf{Z,}%
\]
and we compute the partial transpose $\mathbf{\tilde{V}}_{k}%
:=\boldsymbol{\Lambda}_{k}\mathbf{V}\boldsymbol{\Lambda}_{k}$ for mode
$k=A,B,C$. Then, the state is~\cite{tripartite}:

\begin{description}
\item[Class 1] Fully entangled if $\mathbf{\tilde{V}}_{k}+i\mathbf{\Omega
}^{(3)}\ngeq0$ for all modes.

\item[Class 2] One-mode biseparable if $\mathbf{\tilde{V}}_{k}+i\mathbf{\Omega
}^{(3)}\geq0$ for one mode only, e.g., $k=A$ (or $k=B$ or $k=C$).

\item[Class 3] Two-mode biseparable if $\mathbf{\tilde{V}}_{k}+i\mathbf{\Omega
}^{(3)}\geq0$ for two modes only, e.g., $k=A$ and $k=B$(or the other two combinations).

\item[Class 4 or 5] Either three-mode biseparable (class 4) or fully separable
(class 5) if $\mathbf{\tilde{V}}_{k}+i\mathbf{\Omega}^{(3)}\geq0$ for all
modes. See Fig.~\ref{TripartiteAPP} for a schematic.
\end{description}

Note that the tripartite PPT condition
\begin{equation}
\mathbf{\tilde{V}}_{k}+i\mathbf{\Omega}^{(3)}\geq0~\text{for}~k=A,B,C
\label{3PPT}%
\end{equation}
is not able to distinguish the fully separable states from the three-mode
biseparable states (bound entangled). To distinguish between classes $4$ and
$5$, we need an additional criterion. Put the CM in the block-form%
\[
\mathbf{V}=\left(
\begin{array}
[c]{cc}%
\mathbf{A} & \mathbf{W}\\
\mathbf{W}^{T} & \mathbf{V}_{BC}%
\end{array}
\right)  ,
\]
where $\mathbf{A}$ is the reduced CM of mode $A$, $\mathbf{V}_{BC}$ is reduced
CM of modes $B$ and $C$, while $\mathbf{W}$ is a $2\times4$ block. Using the
pseudoinverse, we may construct the test-matrices%
\begin{align}
\mathbf{T}  &  :=\mathbf{A}-\mathbf{W}(\mathbf{V}_{BC}+i\mathbf{\Omega}%
^{(2)})^{-1}\mathbf{W}^{T},\label{test1}\\
\mathbf{\tilde{T}}  &  :=\mathbf{A}-\mathbf{W}(\mathbf{V}_{BC}%
+i\boldsymbol{\Lambda}\mathbf{\Omega}^{(2)}\boldsymbol{\Lambda})^{-1}%
\mathbf{W}^{T}, \label{test2}%
\end{align}
where $\boldsymbol{\Lambda}$ is the two-mode PT matrix of
Eq.~(\ref{lambda2modes}). Then, a CM satisfying Eq.~(\ref{3PPT}) is fully
separable if and only if there exists a single-mode pure-state CM
$\mathbb{\sigma}$ such that~\cite{tripartite}
\begin{equation}
\mathbf{T}\geq\mathbb{\sigma},~\mathbf{\tilde{T}}\geq\mathbb{\sigma}.
\label{boundENT}%
\end{equation}

\section{Theory: Correlated-Thermal Noise\label{TheoMETH}}

In this section, we consider the Gaussian environment with
correlated thermal noise. In such an environmnent, we derive the dynamics of
the bosonic modes involved in the basic protocol of entanglement swapping: We
study the multipartite separability properties of the state before Bell
detection and we compute the CM of the final swapped state analyzing its
entanglement. We then study the protocols of quantum teleportation,
entanglement distillation, key distillation and practical QKD. In more
details, we provide the following elements:

\bigskip

$\bullet~$\ref{EnvREVapp}: We briefly describe the model of Gaussian
environment with correlated-thermal noise, studying its correlations.

\bigskip

$\bullet~$\ref{APPevoCM}: Considering the basic swapping protocol, we
study the evolution of the bosonic modes, in particular, of the global CM.

\bigskip

$\bullet~$\ref{APPentFORMS}: We analyze the various forms of entanglement
(bipartite, tripartite, and quadripartite) in the output state before the Bell detection.

\bigskip

$\bullet~$\ref{APPcmCALCOLO}: We apply the Bell detection and compute the
CM $\mathbf{V}_{ab|\gamma}(\mu,\kappa,\kappa^{\prime})$ of the swapped state
$\rho_{ab|\gamma}$.

\bigskip

$\bullet~$\ref{swappedENT}: We derive the analytical formula for the
smallest PTS eigenvalue $\varepsilon(\mu,\kappa,\kappa^{\prime})$ associated
with the swapped CM. We can therefore quantify the swapped entanglement. In
particular, we discuss the reactivation condition $\kappa\kappa^{\prime}<1$
and its independence from the input entanglement $\mu$. Finally, we derive the
asymptotic optimum $\varepsilon_{\text{opt}}$ for large $\mu$.

\bigskip

$\bullet~$\ref{APPtele}: We discuss our generalized protocol for
teleporting coherent states, where Bob's conditional quantum operation
$\mathcal{Q}_{\gamma}$ is tailored to deal with the correlated-thermal
environment. We then provide the closed analytical formula for the average
teleportation fidelity $F(\mu,\kappa,\kappa^{\prime})$ and we derive its
asymptotic optimum $F_{\text{opt}}$ for large $\mu$, connecting $F_{\text{opt}%
}$ with $\varepsilon_{\text{opt}}$.

\bigskip

$\bullet~$\ref{DisPROTapp}: We study the distillation protocols. In
particular, in~\ref{DistillRATE}, we study the protocol of entanglement
distillation (with one-way classical communication) which is operated on top
of entanglement swapping. We compute the analytical formula for the coherent
information $I_{\mathcal{C}}(\mu,\kappa,\kappa^{\prime})$, and we derive its
optimal expression for large $\mu$, where it becomes a simple function of
$\varepsilon_{\text{opt}}$. Then, in~\ref{keyDISTapp}, we discuss the
ideal key-distillation protocol based on quantum memories, and we easily show
that its rate $K$ is lower-bounded by the coherent information.

\bigskip

$\bullet~$\ref{QKDapp}: We consider the practical QKD\ protocol where
coherent states are sent to a midway relay (which can be untrusted). For this
protocol, we derive a closed formula for the secret-key rate $R(\xi,\mu
,\kappa,\kappa^{\prime})$ for arbitrary reconciliation efficiency $\xi$ and
modulation variance $\mu$. We study the corresponding security threshold $R=0$
for achievable values $\mu\simeq50$, showing that there is a small difference
between ideal reconciliation ($\xi=1$) and realistic reconciliation efficiency
($\xi\simeq0.97$). Then, we derive the asymptotic optimal rate $R_{\text{opt}%
}$ considering $\xi=1$ and large $\mu$. The security threshold $R_{\text{opt}%
}=0$ is shown to be comparable with those achieved at $\mu\simeq50$. From
$R_{\text{opt}}$, we then derive a lower-bound $R_{\text{LB}}$, expressed in
terms of $\varepsilon_{\text{opt}}$ and $F_{\text{opt}}$. This bound is
sufficiently tight, with its security threshold $R_{\text{LB}}=0$\ being very
close to $R_{\text{opt}}=0$.

\subsection{Environment with correlated-thermal noise\label{EnvREVapp}}

\begin{figure}[ptbh]
\vspace{-1.5cm}
\par
\begin{center}
\includegraphics[width=0.6\textwidth] {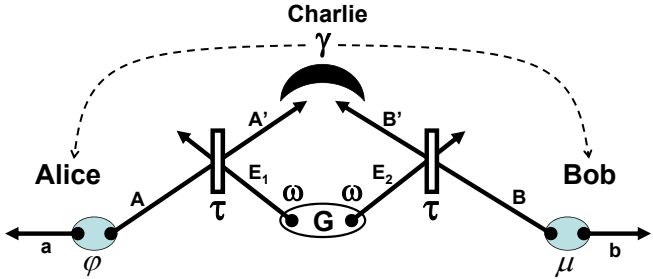}
\end{center}
\par
\vspace{-2.0cm}\caption{Swapping protocol in the presence of a correlated
Gaussian environment with transmissivity $\tau$, thermal noise $\omega$ and
noise-correlations $\mathbf{G}$. Bell detector is simplified.}%
\label{swapLOSS}%
\end{figure}

Let us first describe the Gaussian environment with correlated-thermal noise.
As discussed in the main text, this is modelled by two beam-splitters with
transmissivity $\tau$ which mix the input modes $A$ and $B$, with two
environmental modes $E_{1}$ and $E_{2}$, prepared in a correlated-noise
Gaussian state. This is taken to have zero-mean and CM in the symmetric normal
form%
\begin{equation}
\mathbf{V}_{E_{1}E_{2}}(\omega,g,g^{\prime})=\left(
\begin{array}
[c]{cc}%
\omega\mathbf{I} & \mathbf{G}\\
\mathbf{G} & \omega\mathbf{I}%
\end{array}
\right)  ~, \label{CMenvi}%
\end{equation}
where $\omega\geq1$ is the variance of thermal noise in each mode, while the
block $\mathbf{G}=\mathrm{diag}(g,g^{\prime})$ describes the correlations
between modes $E_{1}$ and $E_{2}$. See also Fig.~\ref{swapLOSS}, which shows
the swapping protocol performed in the presence of this
environment.

One can derive~\cite{NJPpirs} simple bona-fide conditions in terms of the
parameters $\omega\geq1$, $g$ and $g^{\prime}$. By imposing
Eq.~(\ref{unc_PRINC}) to the CM\ of Eq.~(\ref{CMenvi}), one finds the
conditions%
\begin{equation}
|g|<\omega,~~|g^{\prime}|<\omega,~~\omega\left\vert g+g^{\prime}\right\vert
\leq\omega^{2}+gg^{\prime}-1. \label{gCON1}%
\end{equation}
Then, by imposing the separability, i.e., Eq.~(\ref{sepaCONDapp}), one finds
the additional condition~\cite{NJPpirs}%
\begin{equation}
\omega\left\vert g-g^{\prime}\right\vert \leq\omega^{2}-gg^{\prime}-1~.
\label{sepaBONA}%
\end{equation}
For any fixed $\omega\geq1$, the state of the environment is one-to-one with a
point in the correlation plane $(g,g^{\prime})$. Previous conditions in
Eq.~(\ref{gCON1}) identify which part of this plane is physically accessible.
Then, the addition of Eq.~(\ref{sepaBONA}) further identifies the region
associated with separable environments.

Despite being void of entanglement, separable environments still possess
residual quantum correlations. The residual quantum correlations between the
two ancillas, $E_{1}$ and $E_{2}$, can be quantified by their quantum discord
$D$~\cite{RMPdis}, which is here symmetric $D(E_{1}|E_{2})=D(E_{2}|E_{1})$.
This environmental discord can be expressed in terms of correlation parameters
$D=D(g,g^{\prime})$ at any value of thermal noise $\omega$. Similarly, we can
compute the (non-discordant)\ purely-classical correlations $C=C(g,g^{\prime
})$ and, therefore, the total separable correlations $I(g,g^{\prime})=C+D$. In Fig.~\ref{ENCcorr}, we show the typical maps for $I(g,g^{\prime})$ and
$D(g,g^{\prime})$.\begin{figure}[ptbh]
\vspace{+0.1cm}
\par
\begin{center}
\includegraphics[width=0.49\textwidth]{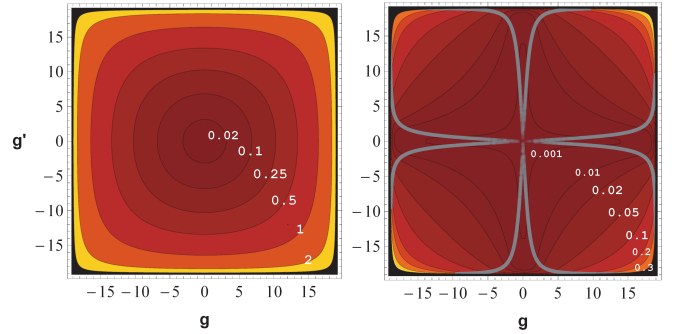}
\end{center}
\par
\vspace{-0.55cm}\caption{Map of the environmental correlations at fixed
thermal noise $\omega=19$. The external black region is excluded, as it
corresponds to entangled or unphysical environments. \textbf{Left panel}.~We
plot the total separable correlations $I(g,g^{\prime})$, corresponding to bits
of quantum mutual information. $I(g,g^{\prime})$ increases almost uniformly
away from the origin. Note that there is a corresponding plot in the main
paper where we consider $\omega=19.38$. \textbf{Right panel}. We plot the
Gaussian discord~\cite{GerryD,ParisD} (in bits) which is proven to be equal to
the (unrestricted) quantum discord $D(g,g^{\prime})$ inside the four delimited
lobes~\cite{OptimalDiscord}. Discord is non-uniform and rapidly increases
along the diagonals of the plane.}%
\label{ENCcorr}%
\end{figure}

\subsection{Evolution of the bosonic modes\label{APPevoCM}}

Here we consider the basic swapping protocol of Fig.~\ref{swapLOSS}, and we
study how Alice's modes ($a$ and $A$) and Bob's modes ($b$ and $B$) evolve
under the action of Gaussian environment with correlated-thermal noise. Since
we are interested in the dynamics of their correlations, we study the
evolution of their global CM $\mathbf{V}_{aAbB}\rightarrow\mathbf{V}%
_{aA^{\prime}bB^{\prime}}$. We start by considering a more general scenario
where Alice's and Bob's two mode squeezed vacuum (TMSV) states have different
variances $\varphi$\ and $\mu$. We then specialize our study to the case of
symmetric setting $\varphi=\mu$.

As depicted in Fig.~\ref{swapLOSS}, we have a total of six input modes:
Alice's modes $a$ and $A$, Bob's modes $b$ and $B$, and Eve's modes $E_{1}$
and $E_{2}$. The global input state is the tensor product%
\[
\rho_{aA}\otimes\rho_{bB}\otimes\rho_{E_{1}E_{2}},
\]
where $\rho_{aA}$ and $\rho_{bB}$ are two TMSV states, with CMs $\mathbf{V}%
(\varphi)$ and $\mathbf{V}(\mu)$, respectively. These are specified by%
\[
\mathbf{V}(\mu)=\left(
\begin{array}
[c]{cc}%
\mu\mathbf{I} & \sqrt{\mu^{2}-1}\mathbf{Z}\\
\sqrt{\mu^{2}-1}\mathbf{Z} & \mu\mathbf{I}%
\end{array}
\right)  ~,
\]
where $\mu\geq1$, $\mathbf{I}:=\mathrm{diag}(1,1)$ and $\mathbf{Z}%
:=\mathrm{diag}(1,-1)$. The environmental state $\rho_{E_{1}E_{2}}$ is
Gaussian with zero-mean and CM $\mathbf{V}_{E_{1}E_{2}}(\omega,g,g^{\prime})$
given in Eq.~(\ref{CMenvi}). The global input state is a zero-mean Gaussian
state with CM%
\[
\mathbf{V}_{aAbBE_{1}E_{2}}=\mathbf{V}(\varphi)\oplus\mathbf{V}(\mu
)\oplus\mathbf{V}_{E_{1}E_{2}}(\omega,g,g^{\prime})~.
\]

It is helpful to permute the modes so to have the ordering $abAE_{1}E_{2}B$,
where the upper-case modes are those transformed by the beam splitters. After
reordering, the input CM\ has the explicit form%
\[
\mathbf{V}_{abAE_{1}E_{2}B}=\left(
\begin{array}
[c]{cccccc}%
\varphi\mathbf{I} & \mathbf{0} & \tilde{\varphi}\mathbf{Z} & \mathbf{0} &
\mathbf{0} & \mathbf{0}\\
\mathbf{0} & \mu\mathbf{I} & \mathbf{0} & \mathbf{0} & \mathbf{0} & \tilde
{\mu}\mathbf{Z}\\
\tilde{\varphi}\mathbf{Z} & \mathbf{0} & \varphi\mathbf{I} & \mathbf{0} &
\mathbf{0} & \mathbf{0}\\
\mathbf{0} & \mathbf{0} & \mathbf{0} & \omega\mathbf{I} & \mathbf{G} &
\mathbf{0}\\
\mathbf{0} & \mathbf{0} & \mathbf{0} & \mathbf{G} & \omega\mathbf{I} &
\mathbf{0}\\
\mathbf{0} & \tilde{\mu}\mathbf{Z} & \mathbf{0} & \mathbf{0} & \mathbf{0} &
\mu\mathbf{I}%
\end{array}
\right)  ~,
\]
where $\mathbf{0}$ is the $2\times2$ zero matrix, and we use the notation%
\[
\tilde{\mu}:=\sqrt{\mu^{2}-1},~\tilde{\varphi}:=\sqrt{\varphi^{2}-1}.
\]
The global action of the two beam splitters can be represented by the
symplectic matrix%
\[
\mathbf{S=\mathbf{I}}\oplus\mathbf{\mathbf{I}}\oplus\mathbf{S\mathbf{(}}%
\tau\mathbf{\mathbf{)}\oplus S(}\tau\mathbf{)}^{T}~,
\]
where the identity matrices $\mathbf{\mathbf{I}}\oplus\mathbf{\mathbf{I}}$ act
on the remote modes, $a$ and $b$, the beam splitter matrix of
Eq.~(\ref{BSsymplectic}) acts on modes $A$ and $E_{1}$, and its transposed
$\mathbf{S(}\tau\mathbf{)}^{T}$ acts on modes $E_{2}$ and $B$. In the
following calculations we exclude the trivial and singular case of $\tau=0$.

The output state of modes $abA^{\prime}E_{1}^{\prime}E_{2}^{\prime}B^{\prime}$
after the action of the interferometer is a Gaussian state with zero mean and
CM equal to%
\[
\mathbf{V}_{abA^{\prime}E_{1}^{\prime}E_{2}^{\prime}B^{\prime}}=\mathbf{S~V}%
_{abAE_{1}E_{2}B}~\mathbf{S}^{T}~.
\]
Since we are interested in the CM\ of Alice and Bob, we trace out the two
environmental modes $E_{1}^{\prime}$ and $E_{2}^{\prime}$. As a result, we get
the following CM for modes $abA^{\prime}B^{\prime}$%
\begin{align}
&  \mathbf{V}_{abA^{\prime}B^{\prime}}(\varphi,\mu,\tau,\omega,g,g^{\prime
})\nonumber\\
&  =\left(
\begin{array}
[c]{cccc}%
\varphi\mathbf{I} & \mathbf{0} & \tilde{\varphi}\sqrt{\tau}\mathbf{Z} &
\mathbf{0}\\
\mathbf{0} & \mu\mathbf{I} & \mathbf{0} & \tilde{\mu}\sqrt{\tau}\mathbf{Z}\\
\tilde{\varphi}\sqrt{\tau}\mathbf{Z} & \mathbf{0} & y\mathbf{I} &
(1-\tau)\mathbf{G}\\
\mathbf{0} & \tilde{\mu}\sqrt{\tau}\mathbf{Z} & (1-\tau)\mathbf{G} &
x\mathbf{I}%
\end{array}
\right)  , \label{VabApBp}%
\end{align}
where%
\[
y:=\tau\varphi+(1-\tau)\omega,~x:=\tau\mu+(1-\tau)\omega~.
\]

\subsection{Output entanglement before Bell detection\label{APPentFORMS}}

It is important to study the evolution of quantum entanglement under the
Gaussian environment with correlated-thermal noise. For this analysis we
consider the symmetric case $\varphi=\mu$, so that
\begin{equation}
\mathbf{V}_{abA^{\prime}B^{\prime}}=\left(
\begin{array}
[c]{cccc}%
\mu\mathbf{I} & \mathbf{0} & \tilde{\mu}\sqrt{\tau}\mathbf{Z} & \mathbf{0}\\
\mathbf{0} & \mu\mathbf{I} & \mathbf{0} & \tilde{\mu}\sqrt{\tau}\mathbf{Z}\\
\tilde{\mu}\sqrt{\tau}\mathbf{Z} & \mathbf{0} & x\mathbf{I} & (1-\tau
)\mathbf{G}\\
\mathbf{0} & \tilde{\mu}\sqrt{\tau}\mathbf{Z} & (1-\tau)\mathbf{G} &
x\mathbf{I}%
\end{array}
\right)  . \label{VforENT}%
\end{equation}
From this CM, we can derive all the reduced CMs and analyze all the various
forms of output entanglement: Bipartite, tripartite and quadripartite. For
simplicity, we also consider the limit of large entanglement at the input,
i.e., $\mu\gg1$. This limit not only simplifies the analytical formulas but
also optimizes the scheme: If entanglement is broken in this limit, then it
must be broken for any finite value of $\mu$. Indeed, for $\mu\gg1$, a TMSV
state becomes an ideal EPR source, i.e., a CV maximally-entangled (asymptotic)
state, and the entanglement-breaking conditions for this state can be extended
to all the others. In the derivations of this section we also implicitly
assume that $\tau<1$.

\subsubsection{Bipartite entanglement}

Let us start from bipartite entanglement. By symmetry, it is sufficient to
study the pairings $aA^{\prime}$, $aB^{\prime}$, $ab$ and $A^{\prime}%
B^{\prime}$. The pairing $ab$ is trivial to consider since the two remote
modes are manifestly separable before Bell detection, as one can also check
from their reduced CM\ $\mathbf{V}_{ab}=\mu\mathbf{I}\oplus\mu\mathbf{I}$. It
is also trivial to check the pairing $aB^{\prime}$, for which we have
$\mathbf{V}_{aB^{\prime}}=\mu\mathbf{I}\oplus x\mathbf{I}$. Regarding the
pairing $A^{\prime}B^{\prime}$, we can check that $\rho_{A^{\prime}B^{\prime}%
}$ is separable for large $\mu$. In fact, from the reduced CM $\mathbf{V}%
_{A^{\prime}B^{\prime}}$ we compute the log-negativity%
\[
\mathcal{N}_{A^{\prime}B^{\prime}}=\max\left\{  0,-\log_{2}\left[  \tau
\mu+\frac{1-\tau}{2}(2\omega-|g-g^{\prime}|)\right]  \right\}  ,
\]
which is always zero for large $\mu$ and $\tau>0$. In the singular case
$\tau=0$, we have $\rho_{A^{\prime}B^{\prime}}=\rho_{E_{1}E_{2}}$ so that
separability directly comes from the (separable) environment.

The most interesting pairing is $aA^{\prime}$. From the reduced CM
$\mathbf{V}_{aA^{\prime}}$, we can compute the log-negativity $\mathcal{N}%
_{aA^{\prime}}$. For large $\mu$, we find%
\[
\mathcal{N}_{aA^{\prime}}=\max\left\{  0,\log_{2}\left[  \frac{1+\tau}%
{(1-\tau)\omega}\right]  \right\}  ,
\]
so that bipartite entanglement is lost ($\mathcal{N}_{aA^{\prime}}=0$) for
\[
\omega\geq\omega_{\text{EB}}(\tau):=\frac{1+\tau}{1-\tau},
\]
which is the known entanglement-breaking threshold of the lossy channel.

\begin{figure*}[t]
\vspace{-1.5cm}
\par
\begin{center}
\includegraphics[width=0.68\textwidth] {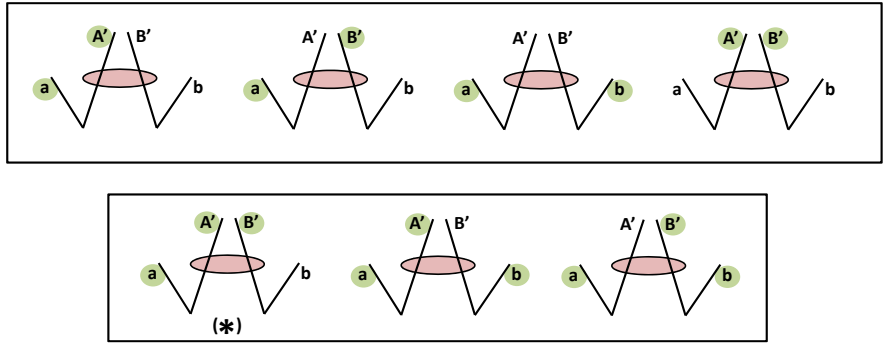}
\end{center}
\par
\vspace{-1.99cm}\caption{\textbf{Upper panel.} Study of the bipartite
entanglement. At the threshold condition $\omega=\omega_{\text{EB}}$, there is
no entanglement surviving between any two modes in the scheme. By symmetry it
is sufficient to consider the pairings $aA^{\prime}$, $aB^{\prime}$, $ab$, and
$A^{\prime}B^{\prime}$ (from left to right). \textbf{Lower Panel.} Study of
the tripartite entanglement. At the threshold condition $\omega=\omega
_{\text{EB}}$, there is no tripartite entanglement in the triplets
$aA^{\prime}b$ and $aB^{\prime}b$ (fully separable, i.e., class 5). Different
is the case for the triplet $aA^{\prime}B^{\prime}$, denoted by (*) in the
panel. Here the tripartite entanglement is broken if we assume the strict
inequality $\omega>\omega_{\text{EB}}$. At exactly $\omega=\omega_{\text{EB}}$
the tripartite state of $aA^{\prime}B^{\prime}$ is guaranteed to be fully
separable only if there are no correlations in the environment ($g=g^{\prime
}=0$), otherwise it is a one-mode biseparable state (class 2 entanglement).}%
\label{casi}%
\end{figure*}

Thus, the threshold condition $\omega=\omega_{\text{EB}}(\tau)$ guarantees
that bipartite entanglement is broken between any pairing of two modes, no
matter how strong are the correlations in the separable environment. See also Fig.~\ref{casi}.

\subsubsection{Tripartite entanglement}

Thanks to the symmetry of the configuration, it is sufficient to study the
triplets of modes $aA^{\prime}B^{\prime}$, $aA^{\prime}b$ and $aB^{\prime}b$,
shown in Fig.~\ref{casi}. Two of these cases are very easy to study. In fact,
from the CM of Eq.~(\ref{VforENT}) we see that%
\begin{align*}
\mathbf{V}_{aA^{\prime}b}  &  =\mathbf{V}_{aA^{\prime}}\oplus\mathbf{V}_{b},\\
\mathbf{V}_{aB^{\prime}b}  &  =\mathbf{V}_{a}\oplus\mathbf{V}_{B^{\prime}b},
\end{align*}
which means that $\rho_{aA^{\prime}b}=\rho_{aA^{\prime}}\otimes\rho_{b}$ and
$\rho_{aB^{\prime}b}=\rho_{a}\otimes\rho_{B^{\prime}b}$. Because of this
tensor product structure, the absence of bipartite entanglement (in
$aA^{\prime}$ and $B^{\prime}b$) implies the absence of tripartite
entanglement. Thus, the previous threshold condition $\omega=\omega
_{\text{EB}}(\tau)$ also breaks tripartite entanglement in $aA^{\prime}b$ and
$aB^{\prime}b$.

More involved is the situation for the triplet $aA^{\prime}B^{\prime}$. First
of all we study the positivity of the three matrices%
\[
\mathbf{W}_{k}:=\boldsymbol{\Lambda}_{k}\mathbf{V}_{aA^{\prime}B^{\prime}%
}\boldsymbol{\Lambda}_{k}+i\mathbf{\Omega}^{(3)},
\]
with $k=a$, $A^{\prime}$ and $B^{\prime}$ (here $\boldsymbol{\Lambda}_{k}$\ is
the usual PT matrix with block $\mathbf{Z}$ being applied to mode $k$). Since
the matrices $\mathbf{W}_{k}$ are Hermitian, the positive-semidefiniteness
$\mathbf{W}_{k}\geq0$ is equivalent to check the non-negativity of their
eigenvalues or, equivalently, their principal minors (more easily, the
positive-definiteness $\mathbf{W}_{k}>0$ is\ equivalent to check the strict
positivity of the eigenvalues or the \textit{leading} principal minors). If
all these matrices are $\mathbf{W}_{k}\geq0$, i.e., the tripartite state is
PPT, then we apply the criterion of Eq.~(\ref{boundENT}) to distinguish class
4 and 5. In particular, note that $\mathbf{T}-\mathbb{\sigma}$ and
$\mathbf{\tilde{T}}-\mathbb{\sigma}$ are Hermitian matrices.

\subparagraph{Markovian case}

In the absence of correlations ($g=g^{\prime}=0$), one can check that the
threshold condition $\omega=\omega_{\text{EB}}(\tau)$ is sufficient to destroy
tripartite entanglement in $aA^{\prime}B^{\prime}$. In fact, in these
conditions, the eigenvalues of the three matrices $\mathbf{W}_{k}$ are all
non-negative, which means that $\rho_{aA^{\prime}B^{\prime}}$ is a PPT state.
Then, we also find that the test matrices of Eqs.~(\ref{test1})
and~(\ref{test2}) satisfy $\mathbf{T}\geq\mathbf{I}$ and $\mathbf{\tilde{T}%
}\geq\mathbf{I}$\ (where the identity $\mathbf{I}$ is the CM of the vacuum
state). More precisely, the two Hermitian matrices $\mathbf{T}-\mathbf{I}$ and
$\mathbf{\tilde{T}}-\mathbf{I}$\ have the same non-negative spectrum of
eigenvalues $\left\{  0,2(\mu-1)[2+\tau(\mu+1)]^{-1}\right\}  $ for any
$\mu\geq1$. As a result, we find that $\rho_{aA^{\prime}B^{\prime}}$ is a
fully separable state (class 5).

\subparagraph{Non-Markovian case}

In the presence of correlations, i.e., for $(g,g^{\prime})\neq(0,0)$, the
threshold condition $\omega=\omega_{\text{EB}}(\tau)$ does not break
tripartite entanglement. In fact, in the limit of large $\mu$, we find that
both $\det\mathbf{W}_{1}$ and $\det\mathbf{W}_{2}$ are negative, so that
$\mathbf{W}_{1},\mathbf{W}_{2}\ngeq0$. Then, by studying its leading principal
minors, we find that $\mathbf{W}_{3}>0$. This means that the state
$\rho_{aA^{\prime}B^{\prime}}$ remains one-mode biseparable (entanglement
class 2).

However, the strict violation $\omega>\omega_{\text{EB}}(\tau)$ is sufficient
to break tripartite entanglement. In this case, for large $\mu$, we find that
the leading principal minors of $\mathbf{W}_{1}$, $\mathbf{W}_{2}$, and
$\mathbf{W}_{3}$ are all strictly positive. Then, we also find that
$\mathbf{T}-\mathbf{I}$ and $\mathbf{\tilde{T}}-\mathbf{I}$\ have the same
spectrum with strictly-positive eigenvalues. As a result, the state
$\rho_{aA^{\prime}B^{\prime}}$ becomes fully separable (class 5).

\subsubsection{Quadripartite entanglement}

From the previous discussion, we conclude that the condition $\omega
>\omega_{\text{EB}}(\tau)$ is able to break both bipartite and tripartite
entanglement, considering all possible combinations of bipartite and
tripartite states in our scheme. This is true both in the Markovian and
non-Markovian case (with separable correlations in the environment). But what
about the separability properties of the global quadri-partite state
$\rho_{abA^{\prime}B^{\prime}}$?

By exploiting the multipartite PPT criterion of Eq.~(\ref{PPT_1N}), we can
study the separability of the quadri-partite state with respect to the
$1\times3$ groupings of the modes $abA^{\prime}B^{\prime}$. Because of the
symmetry between Alice and Bob, it is sufficient to consider the two groupings%
\begin{equation}
\{a\}\{bA^{\prime}B^{\prime}\},~\{A^{\prime}\}\{abB^{\prime}\}, \label{groups}%
\end{equation}
which can be labelled by $k=a$ and $k=A^{\prime}$, respectively (with
corresponding PT matrices $\boldsymbol{\Lambda}_{a}$ and $\boldsymbol{\Lambda
}_{A^{\prime}}$). Thus, we compute the two matrices%
\[
\mathbf{M}_{k}:=\boldsymbol{\Lambda}_{k}\mathbf{V}_{abA^{\prime}B^{\prime}%
}\boldsymbol{\Lambda}_{k}+i\mathbf{\Omega}^{(4)}.
\]

Assuming $\omega>\omega_{\text{EB}}(\tau)$ and large $\mu$, we study the
positivity properties of the matrices $\mathbf{M}_{k}$, finding that they can
be expressed in terms of analytical functions. Set $\omega=r~\omega
_{\text{EB}}(\tau)$ with $r>1$. Then we define%
\[
\Sigma^{\prime}:=\min\{f,f^{\prime}\},~\Sigma^{\prime\prime}:=\min
\{f,f^{\prime\prime}\},
\]
where%
\begin{align*}
f  &  :=(1+\tau)^{2}(r^{2}-1)-g^{2}(1-\tau)^{2},\\
f^{\prime}  &  :=(1-\tau)^{2}[1+\tau-gg^{\prime}(1-\tau)]^{2}+\zeta,\\
f^{\prime\prime}  &  :=(1-\tau)^{2}[1+\tau+gg^{\prime}(1-\tau)]^{2}+\zeta,
\end{align*}
and%
\[
\zeta:=(1+\tau)^{4}r^{4}-(1+\tau)^{2}[2+g^{2}(1-\tau)^{2}+g^{\prime2}%
(1-\tau)^{2}+2\tau^{2}]r^{2}.
\]

Using these functions, we can write the implications
\begin{align*}
\Sigma^{\prime}  &  >0\Longleftrightarrow\mathbf{M}_{a}>0\Longrightarrow
\{a\}\{bA^{\prime}B^{\prime}\}\text{ separable,}\\
\Sigma^{\prime}  &  <0\Longrightarrow\mathbf{M}_{a}\ngeq0\Longleftrightarrow
\{a\}\{bA^{\prime}B^{\prime}\}\text{ entangled.}%
\end{align*}
Despite the fact that the border condition $\Sigma^{\prime}=0$ is
inconclusive, we find that it only occurs in a set of zero measure within the
correlation plane $(g,g^{\prime})$. As a result, $\Sigma^{\prime}=0$ clearly
distinguishes the region where the state is separable from that where it is
entangled, with respect to the grouping $\{a\}\{bA^{\prime}B^{\prime}\}$.

Similarly, we can write the following implications for the other grouping of
modes
\begin{align*}
\Sigma^{\prime\prime}  &  >0\Longleftrightarrow\mathbf{M}_{A^{\prime}%
}>0\Longrightarrow\{A^{\prime}\}\{abB^{\prime}\}\text{ separable,}\\
\Sigma^{\prime\prime}  &  <0\Longrightarrow\mathbf{M}_{A^{\prime}}%
\ngeq0\Longleftrightarrow\{A^{\prime}\}\{abB^{\prime}\}\text{ entangled,}%
\end{align*}
with the border condition $\Sigma^{\prime\prime}=0$ distinguishing between
regions of separability and entanglement.

Altogether, we can identify four regions for the quadripartite state
$\rho_{abA^{\prime}B^{\prime}}$:

\begin{description}
\item[(I)] $\Sigma^{\prime},\Sigma^{\prime\prime}>0$:\ Separable in all the
$1\times3$ groupings,

\item[(II)] $\Sigma^{\prime}>0$ and $\Sigma^{\prime\prime}<0$: Entangled in
$\{A^{\prime}\}\{abB^{\prime}\}$,

\item[(III)] $\Sigma^{\prime}<0$ and $\Sigma^{\prime\prime}>0$: Entangled in
$\{a\}\{bA^{\prime}B^{\prime}\}$,

\item[(IV)] $\Sigma^{\prime},\Sigma^{\prime\prime}<0$: Entangled in all the
$1\times3$ groupings.
\end{description}

\begin{figure}[t]
\vspace{0.1cm}
\par
\begin{center}
\includegraphics[width=0.48\textwidth] {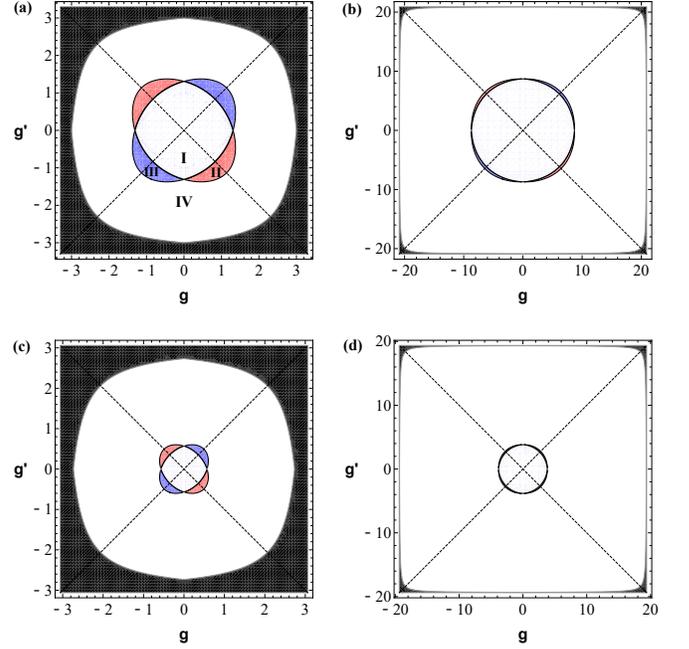}
\end{center}
\par
\vspace{-0.4cm}\caption{Study of the quadripartite entanglement of the state
$\rho_{abA^{\prime}B^{\prime}}$ on the correlation plane. In each panel, we
plot the border conditions $\Sigma^{\prime}=0$ and $\Sigma^{\prime\prime}=0$,
which result in two different intersecting curves. The most inner region (I)
corresponds to $1\times3$ separability. The red region (II) corresponds to
entanglement with respect to the grouping $\{A^{\prime}\}\{abB^{\prime}\}$.
The blue region (III) corresponds to entanglement with respect to the grouping
$\{a\}\{bA^{\prime}B^{\prime}\}$. Finally, the outer region (IV) corresponds
to entanglement in all the groupings. Here we consider: $r=1.1$ and $\tau=0.5$
in panel~(a), $r=1.1$ and $\tau=0.9$ in panel~(b), $r=1.02$ and $\tau=0.5$ in
panel~(c), and $r=1.02$ and $\tau=0.9$ in panel~(d).}%
\label{ent4}%
\end{figure}

These regions are numerically shown in Fig.~\ref{ent4}. As we can see from the
figure, $1\times3$ quadri-partite entanglement is reactivated after a certain
amount of separable correlations is injected by the environment. This critical
amount increases in the thermal noise $\omega>\omega_{\text{EB}}(\tau)$. For
instance, this is evident by comparing panel (c), where $r=1.02$, with panel
(a), where $r=1.1$. By increasing the thermal noise, region (I) widens while
region (IV) shrinks. Also note that, by increasing the transmissivity $\tau$
(while keeping $r$ fixed), the two regions (II) and (III) tend to coincide.
For instance, compare panel (a) with panel (b), and panel (c) with panel (d).
We have checked that this behavior is generic and also occurs at finite $\mu$,
where we have studied the positivity properties of the matrices $\mathbf{M}%
_{k}$ by (numerically)\ computing their spectra.\ 

In conclusion, we find that, despite the condition $\omega>\omega_{\text{EB}%
}(\tau)$ is able to break bipartite and tripartite entanglement, there could
be a survival of quadri-partite $1\times3$ entanglement, whose existence
depends on the amount of separable correlations injected by the environment.
In particular, this entanglement can be reactivated with respect to the
grouping $\{a\}\{bA^{\prime}B^{\prime}\}$. In this case, a quantum measurement
on modes $A^{\prime}$ and $B^{\prime}$ can localize this multi-partite
resource in the remaining modes $a$ and $b$ (thus generating bipartite
entanglement). A simple (but presumably sub-optimal) way to localize this
quadri-partite entanglement is the use of the Bell detection. If the procedure
is successful, then the protocol of entanglement swapping can be reactivated
from entanglement-breaking (see~\ref{APPcmCALCOLO}
and~\ref{swappedENT}).

Note that we have not analyzed quadripartite entanglement of the $2\times2$
type, associated with the groupings $\{ab\}\{A^{\prime}B^{\prime}\}$,
$\{aA^{\prime}\}\{bB^{\prime}\}$, or $\{aB^{\prime}\}\{bA^{\prime}\}$. In the
Markovian case ($g=g^{\prime}=0$), this is certainly absent for $\omega
>\omega_{\text{EB}}(\tau)$. In the non-Markovian case this type of
entanglement could potentially be reactivated by the separable correlations of
the environment. However, this analysis not only is involved but also
secondary, since the reactivation of $1\times3$ quadripartite entanglement
(always distillable~\cite{quadripartite}) is already sufficient to induce the
localization into a bipartite form.

\subsection{Covariance Matrix of the Swapped State\label{APPcmCALCOLO}}

To study the protocol of entanglement swapping in general non-Markovian
conditions, the first step is the computation of the CM $\mathbf{V}%
_{ab|\gamma}$ of the swapped state $\rho_{ab|\gamma}$, after Bell detection.
Here we start by considering an asymmetric scenario, where Alice and Bob may
have different EPR\ resources (TMSV states). We then specify the formula to
the case of identical resources.

Starting from the CM of Eq.~(\ref{VabApBp}), we compute the CM $\mathbf{V}%
_{ab|\gamma}$ of the conditional remote state $\rho_{ab|\gamma}$ by applying
the transformation rules for CMs under Bell-like measurements specified in
Ref.~\cite{BellFORMULA}. As a first step, we put $\mathbf{V}_{abA^{\prime
}B^{\prime}}$ in the blockform%
\[
\mathbf{V}_{abA^{\prime}B^{\prime}}=\left(
\begin{array}
[c]{ccc}%
\mathbf{V}_{ab} & \mathbf{C}_{1} & \mathbf{C}_{2}\\
\mathbf{C}_{1}^{T} & \mathbf{B}_{1} & \mathbf{D}\\
\mathbf{C}_{2}^{T} & \mathbf{D}^{T} & \mathbf{B}_{2}%
\end{array}
\right)  ~,
\]
where%
\[
\mathbf{B}_{1}=y\mathbf{I,~B}_{2}=x\mathbf{I,~D}=(1-\tau)\mathbf{G,}%
\]
and%
\[
\mathbf{C}_{1}=\left(
\begin{array}
[c]{c}%
\tilde{\varphi}\sqrt{\tau}\mathbf{Z}\\
\mathbf{0}%
\end{array}
\right)  ,~\mathbf{C}_{2}=\left(
\begin{array}
[c]{c}%
\mathbf{0}\\
\tilde{\mu}\sqrt{\tau}\mathbf{Z}%
\end{array}
\right)  .
\]

Then the conditional CM is given by the formula~\cite{BellFORMULA}%
\[
\mathbf{V}_{ab|\gamma}=\mathbf{V}_{ab}-\frac{1}{2\det\boldsymbol{\Theta}}%
\sum_{i,j=1}^{2}\mathbf{C}_{i}(\mathbf{X}_{i}^{T}\boldsymbol{\Theta}%
\mathbf{X}_{j})\mathbf{C}_{j}^{T}~,
\]
where%
\[
\mathbf{X}_{1}:=\left(
\begin{array}
[c]{cc}%
0 & 1\\
1 & 0
\end{array}
\right)  ,~\mathbf{X}_{2}:=\left(
\begin{array}
[c]{cc}%
0 & 1\\
-1 & 0
\end{array}
\right)  ~,
\]
and%
\[
\boldsymbol{\Theta}:=\frac{1}{2}\left(  \mathbf{ZB}_{1}\mathbf{Z}%
+\mathbf{B}_{2}-\mathbf{ZD}-\mathbf{D}^{T}\mathbf{Z}\right)  =\left(
\begin{array}
[c]{cc}%
\theta & 0\\
0 & \theta^{\prime}%
\end{array}
\right)
\]
with diagonal terms%
\[
\left\{
\begin{array}
[c]{c}%
\theta:=\frac{\tau}{2}(\varphi+\mu)+(1-\tau)(\omega-g),\\
\\
\theta^{\prime}:=\frac{\tau}{2}(\varphi+\mu)+(1-\tau)(\omega+g^{\prime}).
\end{array}
\right.
\]

After simple algebra, we derive the following expression for the conditional
CM
\begin{equation}
\mathbf{V}_{ab|\gamma}=\left(
\begin{array}
[c]{cc}%
\varphi\mathbf{I} & \mathbf{0}\\
\mathbf{0} & \mu\mathbf{I}%
\end{array}
\right)  -\frac{\tau}{2}\left(
\begin{array}
[c]{cccc}%
\frac{\tilde{\varphi}^{2}}{\theta} & 0 & -\frac{\tilde{\varphi}\tilde{\mu}%
}{\theta} & 0\\
0 & \frac{\tilde{\varphi}^{2}}{\theta^{\prime}} & 0 & \frac{\tilde{\varphi
}\tilde{\mu}}{\theta^{\prime}}\\
-\frac{\tilde{\varphi}\tilde{\mu}}{\theta} & 0 & \frac{\tilde{\mu}^{2}}%
{\theta} & 0\\
0 & \frac{\tilde{\varphi}\tilde{\mu}}{\theta^{\prime}} & 0 & \frac{\tilde{\mu
}^{2}}{\theta^{\prime}}%
\end{array}
\right)  . \label{CMgene}%
\end{equation}
In the symmetric case of identical EPR sources (TMSV states), i.e., for
$\varphi=\mu$, the conditional CM $\mathbf{V}_{ab|\gamma}$ of
Eq.~(\ref{CMgene}) takes the following simple form
\begin{equation}
\mathbf{V}_{ab|\gamma}=\left(
\begin{array}
[c]{cc}%
\mu\mathbf{I} & \mathbf{0}\\
\mathbf{0} & \mu\mathbf{I}%
\end{array}
\right)  -\frac{\mu^{2}-1}{2}\boldsymbol{\Psi}~, \label{CMmainAPP}%
\end{equation}
where%
\begin{equation}
\boldsymbol{\Psi}=\left(
\begin{array}
[c]{cccc}%
\frac{1}{\mu+\kappa} & 0 & \frac{-1}{\mu+\kappa} & 0\\
0 & \frac{1}{\mu+\kappa^{\prime}} & 0 & \frac{1}{\mu+\kappa^{\prime}}\\
\frac{-1}{\mu+\kappa} & 0 & \frac{1}{\mu+\kappa} & 0\\
0 & \frac{1}{\mu+\kappa^{\prime}} & 0 & \frac{1}{\mu+\kappa^{\prime}}%
\end{array}
\right)  , \label{bigPSI}%
\end{equation}
with
\begin{equation}
\left\{
\begin{array}
[c]{c}%
\kappa:=(\tau^{-1}-1)(\omega-g)\geq0~,\\
\\
\kappa^{\prime}:=(\tau^{-1}-1)(\omega+g^{\prime})\geq0~.
\end{array}
\right.  \label{kappas}%
\end{equation}

The CM of Eq.~(\ref{CMmainAPP}) can be put in the blockform%
\begin{equation}
\mathbf{V}_{ab|\gamma}=\left(
\begin{array}
[c]{cc}%
\mathbf{A} & \mathbf{C}\\
\mathbf{C}^{T} & \mathbf{B}%
\end{array}
\right)  , \label{CMgeneBLOCKS}%
\end{equation}
where%
\begin{align}
\mathbf{A}  &  =\mathbf{B}=\left(
\begin{array}
[c]{cc}%
\mu-\frac{\mu^{2}-1}{2(\mu+\kappa)} & 0\\
0 & \mu-\frac{\mu^{2}-1}{2(\mu+\kappa^{\prime})}%
\end{array}
\right)  ,\label{blockB}\\
\mathbf{C}  &  =\left(
\begin{array}
[c]{cc}%
\frac{\mu^{2}-1}{2(\mu+\kappa)} & 0\\
0 & -\frac{\mu^{2}-1}{2(\mu+\kappa^{\prime})}%
\end{array}
\right)  , \label{blockC}%
\end{align}
which is the expression in Eqs.~(1)-(3) of the main text.

It is clear that the CM\ $\mathbf{V}_{ab|\gamma}$ of the swapped state
$\rho_{ab|\gamma}$\ does not depend on the specific outcome $\gamma$\ of the
Bell detection, which only affects the first moments of the state (the CM is
only conditioned by the fact that the Bell detection has been performed and
the outcome communicated). Also note that the conditional CM\ $\mathbf{V}%
_{ab|\gamma}$ is symmetric under $a-b$ permutation.

For the next calculations it is helpful to derive the symplectic spectrum of
the CM\ $\mathbf{V}_{ab|\gamma}$. This is given by~\cite{RMP}%
\[
\nu_{\pm}=\sqrt{\frac{\Delta\pm\sqrt{\Delta^{2}-4\det\mathbf{V}_{ab|\gamma}}%
}{2}},
\]
where the symplectic invariant $\Delta:=\det\mathbf{A}+\det\mathbf{B}%
+2\det\mathbf{C}$ is computed from the blocks in Eqs.~(\ref{blockB})
and~(\ref{blockC}). After simple algebra we find%
\begin{equation}
\{\nu_{-},\nu_{+}\}=\left\{  \sqrt{\frac{\mu(1+\mu\kappa)}{\mu+\kappa}}%
,~\sqrt{\frac{\mu(1+\mu\kappa^{\prime})}{\mu+\kappa^{\prime}}}\right\}  .
\label{nimenopiu}%
\end{equation}
It is also useful to derive the symplectic eigenvalue of the reduced
CM\ $\mathbf{V}_{b|\gamma}=\mathbf{B}$, which describes Bob's reduced state
$\rho_{b|\gamma}$. This is just given by
\begin{equation}
\nu_{b}=\sqrt{\det\mathbf{B}}=\frac{1}{2}\sqrt{\frac{(1+2\mu\kappa+\mu
^{2})(1+2\mu\kappa^{\prime}+\mu^{2})}{(\mu+\kappa)(\mu+\kappa^{\prime})}}.
\label{niREDUCED}%
\end{equation}


\subsection{Quantification of the Swapped Entanglement\label{swappedENT}}

Let us consider the symmetric scenario $\varphi=\mu$. In order to quantify the
amount of entanglement which is present in the swapped state $\rho_{ab|\gamma
}$, we compute the smallest PTS eigenvalue of the CM\ $\mathbf{V}_{ab|\gamma}%
$. This is given by~\cite{RMP}%
\[
\varepsilon=\sqrt{\frac{\Sigma-\sqrt{\Sigma^{2}-4\det\mathbf{V}_{ab|\gamma}}%
}{2}},
\]
where $\Sigma:=\det\mathbf{A}+\det\mathbf{B}-2\det\mathbf{C}$ is computed from
the blocks in Eqs.~(\ref{blockB}) and~(\ref{blockC}). After simple algebra we
find%
\begin{equation}
\varepsilon=\sqrt{\frac{(1+\mu\kappa)(1+\mu\kappa^{\prime})}{(\mu+\kappa
)(\mu+\kappa^{\prime})}} \label{epsFIN}%
\end{equation}
with $\kappa$ and $\kappa^{\prime}$ specified by Eq.~(\ref{kappas}). Also note
that $\varepsilon=\nu_{-}\nu_{+}\mu^{-1}$, where the $\nu$'s are the
eigenvalues in Eq.~(\ref{nimenopiu}).

We can easily check that the presence of entanglement in the swapped state
($\varepsilon<1$) is equivalent to the condition $\kappa\kappa^{\prime}<1$ for
any $\mu>1$. In other words, as long as input entanglement is present (i.e.,
$\mu>1$), the \textit{success} of entanglement swapping corresponds to
$\kappa\kappa^{\prime}<1$, no matter how much entangled the input was (i.e.,
independently from the actual value of $\mu>1$). It is however true that the
\textit{amount} of the swapped entanglement, e.g., as quantified by the
log-negativity~\cite{logNEG,logNEG2} $\mathcal{N}=\max\{0,-\log_{2}%
\varepsilon\}$, depends on the value of $\mu$, as we can see from
Eq.~(\ref{epsFIN}). One can check that $\frac{d\mathcal{N}}{d\mu}>0$ for
$\kappa\kappa^{\prime}<1$, so that the amount of entanglement increases in
$\mu$. The maximal swapped entanglement is achieved in the limit of infinite
input entanglement $\mu\gg1$, so that the previous eigenvalue becomes%
\begin{equation}
\varepsilon\rightarrow\varepsilon_{\text{opt}}:=\sqrt{\kappa\kappa^{\prime}%
}=(\tau^{-1}-1)\sqrt{(\omega-g)(\omega+g^{\prime})}. \label{PTSeigAPP}%
\end{equation}

Note that, for antisymmetric correlations $g+g^{\prime}=0$ (i.e.,
$\kappa=\kappa^{\prime}$), we have
\[
\varepsilon=\frac{1+\mu\kappa}{\mu+\kappa}~,
\]
which is less than $1$ when $\kappa<1$. In the specific case of a Markovian
environment ($g=g^{\prime}=0$), we have $\kappa=\kappa^{\prime}=(\tau
^{-1}-1)\omega$ and the condition $\kappa<1$ corresponds to $\omega
<\tau(1-\tau)^{-1}$. Such condition is clearly not satisfied assuming
entanglement-breaking
\[
\omega>\omega_{\text{EB}}(\tau)=(1+\tau)(1-\tau)^{-1}~.
\]

\begin{figure}[ptbh]
\vspace{-0.0cm}
\par
\begin{center}
\includegraphics[width=0.48\textwidth] {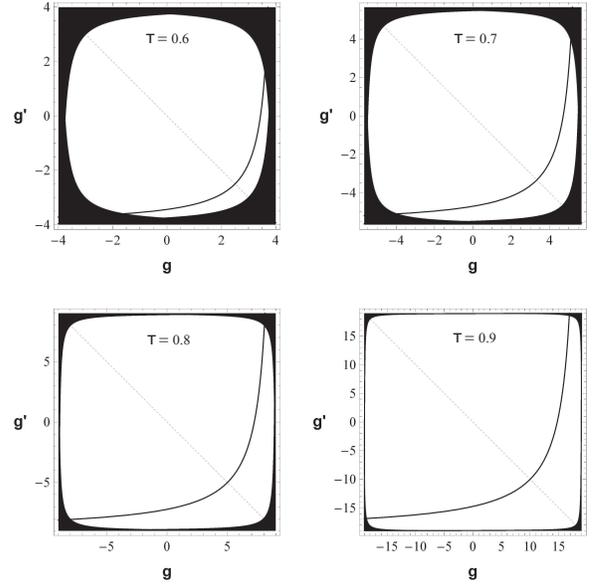}
\end{center}
\par
\vspace{-0.4cm}\caption{In each panel, the white regions correspond to
bona-fide separable Gaussian environments. We plot the threshold condition
$\kappa\kappa^{\prime}=1$ (solid curve). For any point $(g,g^{\prime})$ below
the threshold, we have $\kappa\kappa^{\prime}<1$\ which means that
entanglement swapping is reactivated. Here we consider $\tau=0.6$, $0.7$,
$0.8$, and $0.9$ from top-left to bottom-right, and corresponding
entanglement-breaking values of the thermal noise $\omega=r~\omega_{\text{EB}%
}(\tau)$, with $r=1+10^{-4}$. The value of $\mu$ is arbitrary as long as
$\mu>1$ (i.e., input entanglement is present).}%
\label{fourT}%
\end{figure}

As we discuss in the main text, the situation is different when the Gaussian
environment is non-Markovian with separable correlations. In this case, we can
swap entanglement ($\varepsilon<1$) even if $\omega>\omega_{\text{EB}}(\tau)$.
Imposing $\kappa\kappa^{\prime}=1$, we can write a threshold\ condition for
the correlation parameters $g$ and $g^{\prime}$ at any transmissivity $\tau$,
which can be expressed as%
\[
g=\frac{1+2\tau+g^{\prime}(1-\tau^{2})}{1-\tau^{2}+g^{\prime}(1-\tau)^{2}}~.
\]
For any $g$ exceeding such a threshold, entanglement swapping is reactivated.
In general, the reactivation condition $\kappa\kappa^{\prime}<1$ corresponds
to a region of the correlation plane $(g,g^{\prime})$ as shown in Fig.~\ref{fourT}.

\subsection{Quantum Teleportation of Coherent States\label{APPtele}}

\begin{figure}[ptbh]
\vspace{-0.0cm}
\par
\begin{center}
\includegraphics[width=0.51\textwidth] {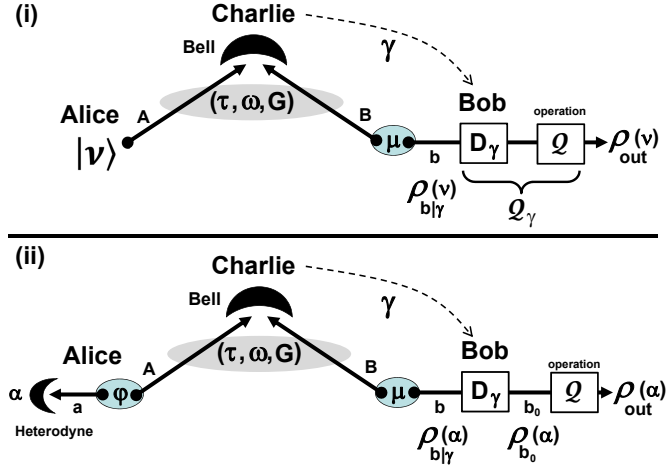}
\end{center}
\par
\vspace{-0.6cm}\caption{(i)~Teleportation and (ii)~Measurement-based scheme
for teleportation. See text for more explanations.}%
\label{equi}%
\end{figure}

Here we analytically compute the average fidelity for the teleportation
protocol in the presence of the non-Markovian Gaussian environment with
correlated thermal noise. Consider a symmetric protocol of teleportation,
where one party, Alice, aims to teleport an unknown coherent state $\left\vert
\nu\right\rangle $ to the another party, Bob, using a middle station as
teleporter, Charlie. As depicted in Fig.~\ref{equi}(i), Alice sends her
coherent state $\left\vert \nu\right\rangle $ to Charlie, who also receives
part $B$ of a TMSV\ state $\rho_{Bb}$ from Bob, with variance $\mu$. These two
transmissions are affected by the correlated-noise Gaussian environment with
links' transmissivity $\tau$, thermal noise variance $\omega$, and
correlations $\mathbf{G}=\mathrm{diag}(g,g^{\prime})$. Then, Charlie performs
a Bell detection and communicates the outcome $\gamma$ to Bob, therefore
projecting his mode $b$ onto a conditional state $\rho_{b|\gamma}(\nu)$.

On this state, Bob applies a conditional quantum operation $\mathcal{Q}%
_{\gamma}$ which provides the teleported output state $\rho_{\text{out}}%
(\nu)\approx\left\vert \nu\right\rangle \left\langle \nu\right\vert $. Bob's
conditional operation $\mathcal{Q}_{\gamma}$ can be broken down in two
subsequent operations, first a conditional displacement $D_{b|\gamma}$
(erasing the shift coming from the measurement), and then a suitably-optimized
quantum operation $\mathcal{Q}$ which aims to correct the perturbation of the
noisy environment (as we will see afterwards, this operation $\mathcal{Q}$ is
in turn broken down into a squeezing unitary followed by a quantum
amplifier~\cite{RMP}).

As shown in Fig.~\ref{equi}(ii), this protocol can equivalently be described
as a \textit{measurement-based teleportation}. Here Alice has another TMSV
state $\rho_{aA}$ (with variance $\varphi$) whose mode $a$ is subject to a
heterodyne detection with complex outcome $\alpha=(q+ip)/2$, equivalently
denoted by the real vector $\mathbf{a}=(q,p)^{T}$. This prepares a coherent
state on mode $A$, with randomly-modulated amplitude $\nu(\alpha
)=\tilde{\varphi}(\varphi+1)^{-1}\alpha^{\ast}$, where $\alpha$ follows a
complex Gaussian distribution $p(\alpha)$ with zero mean and variance
$\varphi-1$. Equivalently, the coherent state has mean-value $\mathbf{\bar{x}%
}_{A|\alpha}=\tilde{\varphi}(\varphi+1)^{-1}\mathbf{Za}$, which is modulated
by a bivariate Gaussian distribution with zero mean and variance $\varphi-1$.
This can be proven by using the formulas for the heterodyne detection which
can be found in Ref.~\cite{HET} (see also Supplementary Material of
Ref.~\cite{OptimalDiscord} for all details on the remote preparation of
one-mode Gaussian states by using local Gaussian measurements on two-mode
Gaussian states). In the limit $\varphi\gg1$, the coherent state has amplitude
$\alpha^{\ast}$ and mean-value $\mathbf{Za}$, uniformly picked from the entire phase-space.

From the point of view of Bob, the measurements of Alice and Charlie permute,
so that we can equivalently assume that the Bell detection occurs before the
heterodyne detection. As a result, Bob's conditional state $\rho_{b|\gamma
}(\alpha)$ can be derived by applying the heterodyne POVM $\{\Pi_{a}%
(\alpha)\}$ to the $a$ mode of the swapped state $\rho_{ab|\gamma}$, i.e., we
have%
\[
\rho_{b|\gamma}(\alpha)=p(\alpha|\gamma)^{-1}\mathrm{Tr}_{a}[\Pi_{a}%
(\alpha)\rho_{ab|\gamma}].
\]
The teleported state is therefore given by
\[
\rho_{\text{out}}(\alpha)=\mathcal{Q}[D_{b|\gamma}~\rho_{b|\gamma}%
(\alpha)~D_{b|\gamma}^{\dagger}],
\]
with outcome-dependent fidelity
\[
F(\alpha):=\left\langle \nu(\alpha)\right\vert \rho_{\text{out}}%
(\alpha)\left\vert \nu(\alpha)\right\rangle .
\]
The average teleportation fidelity is finally given by%
\begin{equation}
F=\int d^{2}\alpha~p(\alpha)F(\alpha). \label{meanF}%
\end{equation}

Note that we may alternatively write
\[
\rho_{\text{out}}(\alpha)=\mathcal{Q}[\rho_{b_{0}}(\alpha)]
\]
where%
\begin{align*}
\rho_{b_{0}}(\alpha)  &  =p(\alpha|\gamma)^{-1}\mathrm{Tr}_{a}[\Pi_{a}%
(\alpha)\rho_{ab|\gamma}(0)]~,\\
\rho_{ab|\gamma}(0)  &  :=D_{b|\gamma}\rho_{ab|\gamma}D_{b|\gamma}^{\dagger}~.
\end{align*}
In other words, we may consider the swapped state $\rho_{ab|\gamma}(0)$, after
its mean value has been erased by the conditional displacement $D_{b|\gamma}$.
Then, we heterodyne its mode $a$ to get the conditional state $\rho_{b_{0}%
}(\alpha)$, which is finally transformed into the output state $\rho
_{\text{out}}(\alpha)$.

To derive the teleportation fidelity, we start by computing the statistical
moments, $\mathbf{\bar{x}}_{b_{0}|\alpha}$ and $\mathbf{V}_{b_{0}|\alpha}$, of
the conditional Gaussian state $\rho_{b_{0}}(\alpha)$. These are derived by
applying the formulas for the heterodyne detection~\cite{HET} to the Gaussian
state $\rho_{ab|\gamma}(0)$ with zero mean-value and conditional CM
$\mathbf{V}_{ab|\gamma}$ specified by Eqs.~(\ref{CMgeneBLOCKS})-(\ref{blockC}%
). Thus, we find the CM%
\begin{equation}
\mathbf{V}_{b_{0}|\alpha}=\mathbf{V}_{b|\gamma\alpha}=\mu\mathbf{I}-\frac
{\tau(\mu^{2}-1)}{2}\left(
\begin{array}
[c]{cc}%
\frac{1}{\theta_{1}} & 0\\
0 & \frac{1}{\theta_{1}^{\prime}}%
\end{array}
\right)  , \label{Vb0}%
\end{equation}
where%
\begin{align}
\theta_{1}  &  :=\tau\left(  \frac{\mu+1}{2}+\kappa\right)  ,\label{teta1}\\
\theta_{1}^{\prime}  &  :=\tau\left(  \frac{\mu+1}{2}+\kappa^{\prime}\right)
, \label{teta2}%
\end{align}
and the mean value%
\begin{align}
\mathbf{\bar{x}}_{b_{0}|\alpha}  &  =\frac{\tau\tilde{\varphi}\tilde{\mu}%
}{2(\varphi+1)}\left(
\begin{array}
[c]{cc}%
\frac{1}{\theta_{1}} & 0\\
0 & -\frac{1}{\theta_{1}^{\prime}}%
\end{array}
\right)  \mathbf{a}\nonumber\\
&  \rightarrow\frac{\tau\tilde{\mu}}{2}\left(
\begin{array}
[c]{cc}%
\frac{1}{\theta_{1}} & 0\\
0 & -\frac{1}{\theta_{1}^{\prime}}%
\end{array}
\right)  \mathbf{a}~, \label{limit2}%
\end{align}
where Eq.~(\ref{limit2}) corresponds to the limit $\varphi\gg1$ (i.e., for a
completely unknown coherent state at the input).

\begin{figure}[ptbh]
\vspace{-2.2cm}
\par
\begin{center}
\includegraphics[width=0.60\textwidth] {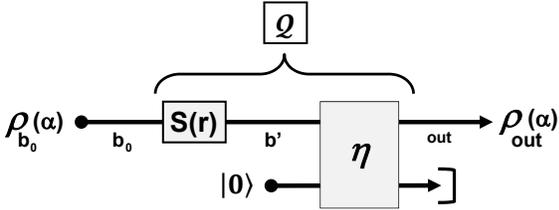}
\end{center}
\par
\vspace{-2.8cm}\caption{Decomposition of Bob's quantum operation $\mathcal{Q}%
$\ into a single-mode squeezer $S(r)$ followed by an amplifying channel
$\mathcal{A}(\eta)$, which can be dilated into a two-mode squeezer involving a
vacuum input.}%
\label{deco}%
\end{figure}

In the limit $\varphi\gg1$, Alice's input mode $A$ is projected onto a
coherent state $\left\vert \alpha^{\ast}\right\rangle $ with mean-value
$\mathbf{\bar{x}}_{A|\alpha}=\mathbf{Za}$, which is uniformly modulated in the
phase space. Correspondingly, the mean-value of the remote state $\rho_{b_{0}%
}(\alpha)$ is given by%
\begin{equation}
\mathbf{\bar{x}}_{b_{0}|\alpha}=\frac{\tau\tilde{\mu}}{2}\left(
\begin{array}
[c]{cc}%
\frac{1}{\theta_{1}} & 0\\
0 & \frac{1}{\theta_{1}^{\prime}}%
\end{array}
\right)  \mathbf{\bar{x}}_{A|\alpha}~. \label{xb0alpha}%
\end{equation}
Now Bob applies a quantum operation $\mathcal{Q}$ to $\rho_{b_{0}}(\alpha)$ in
such a way that the output state $\rho_{\text{out}}(\alpha)$ has the same mean
value of the input coherent state, i.e., $\mathbf{\bar{x}}_{\text{out}|\alpha
}=\mathbf{\bar{x}}_{A|\alpha}$. This operation can be decomposed into a
squeezing unitary $S(r)$, with real squeezing parameter $r$, followed by an
amplifying channel $\mathcal{A}(\eta)$ with gain parameter $\eta\geq1$, as
shown in Fig.~\ref{deco}. 

The action of the squeezer is to balance the diagonal terms in
Eq.~(\ref{xb0alpha}), so that the position and momentum components are equal.
This corresponds to apply a symplectic squeezing matrix%
\begin{equation}
\mathbf{S}(r)=\left(
\begin{array}
[c]{cc}%
r & 0\\
0 & \frac{1}{r}%
\end{array}
\right)  ~\text{with}~r=\sqrt{\frac{\theta_{1}}{\theta_{1}^{\prime}}},
\label{SqueezeAPP}%
\end{equation}
so that we have
\[
\mathbf{\bar{x}}_{b^{\prime}|\alpha}:=\mathbf{S}(r)~\mathbf{\bar{x}}%
_{b_{0}|\alpha}=\frac{\tau\tilde{\mu}}{2\sqrt{\theta_{1}\theta_{1}^{\prime}}%
}~\mathbf{\bar{x}}_{A|\alpha}.
\]
Now, we apply a phase-insensitive quantum-limited amplifier, i.e., a two-mode
squeezer combining the state with a vacuum state. This device realizes an
amplifying channel transforming the quadrature operators $\mathbf{\hat{x}%
}=(\hat{q},\hat{p})^{T}$ as%
\[
\mathbf{\hat{x}}_{b^{\prime}|\alpha}\rightarrow\mathbf{\hat{x}}_{\text{out}%
|\alpha}=\sqrt{\eta}\mathbf{\hat{x}}_{b^{\prime}|\alpha}+\sqrt{\eta
-1}\mathbf{Z\hat{x}}_{\text{vac}},
\]
where $\mathbf{\hat{x}}_{\text{vac}}$ are the quadrature operators of the
vacuum mode. Choosing the gain to be%
\begin{equation}
\eta=\frac{4\theta_{1}\theta_{1}^{\prime}}{\tau^{2}(\mu^{2}-1)}\geq1,
\label{etaAPP}%
\end{equation}
we have that the output state has the desired mean value
\[
\mathbf{\bar{x}}_{\text{out}|\alpha}=\sqrt{\eta}\mathbf{\bar{x}}_{b^{\prime
}|\alpha}=\mathbf{\bar{x}}_{A|\alpha}.
\]

It is clear that such quantum processing by $\mathcal{Q}$ does not come for
free. Recovering the mean-value of the input coherent state is achieved at the
cost of increasing the noise in the teleported state. In fact, the CM of the
output state $\rho_{\text{out}}(\alpha)$ is given by
\begin{equation}
\mathbf{V}_{\text{out}|\alpha}=\eta\left[  \mathbf{S}(r)\mathbf{V}%
_{b_{0}|\alpha}\mathbf{S}(r)^{T}\right]  +(\eta-1)\mathbf{I~.}
\label{voutalpha}%
\end{equation}
Using Eqs.~(\ref{Vb0}), (\ref{SqueezeAPP}) and (\ref{etaAPP}) in Eq.
(\ref{voutalpha}), we get%
\[
\mathbf{V}_{\text{out}|\alpha}=\frac{4\mu}{\tau^{2}(\mu^{2}-1)}\left(
\begin{array}
[c]{cc}%
\theta_{1}^{2} & 0\\
0 & \theta_{1}^{\prime2}%
\end{array}
\right)  -\frac{2}{\tau}\left(
\begin{array}
[c]{cc}%
\theta_{1} & 0\\
0 & \theta_{1}^{\prime}%
\end{array}
\right)  +(\eta-1)\mathbf{I}.
\]

Once we have derived its the first and second-order statistical moments, the
teleported Gaussian state $\rho_{\text{out}}(\alpha)$ is fully determined. For
a given outcome $\alpha$, the fidelity of teleportation
\[
F(\alpha)=\mathrm{Tr}\left[  \left\vert \nu(\alpha)\right\rangle \left\langle
\nu(\alpha)\right\vert \rho_{\text{out}}(\alpha)\right]
\]
can be computed using the trace-rule for Gaussian states. In general, for two
arbitrary single-mode Gaussian states, $\rho$ and $\rho^{\prime}$, with
statistical moments $\{\mathbf{\bar{x}},\mathbf{V}\}$ and $\{\mathbf{\bar{x}%
}^{\prime},\mathbf{V}^{\prime}\}$, we can write%
\[
\mathrm{Tr}\left(  \rho\rho^{\prime}\right)  =\frac{2\exp\left[  -\frac{1}%
{2}(\mathbf{\bar{x}-\bar{x}}^{\prime})^{T}(\mathbf{V}+\mathbf{V}^{\prime
})^{-1}(\mathbf{\bar{x}-\bar{x}}^{\prime})\right]  }{\sqrt{\det(\mathbf{V}%
+\mathbf{V}^{\prime})}}.
\]
Applying this formula to our specific case, we obtain $F(\alpha)=2N^{-1}$,
where%
\begin{align*}
N  &  :=\sqrt{\det(\mathbf{V}_{\text{out}|\alpha}+\mathbf{I})}\\
&  =\frac{2\sqrt{\theta_{1}\theta_{1}^{\prime}}}{\tau}\sqrt{\frac{2(\mu
\theta_{1}+\theta_{1}^{\prime})}{\tau(\mu^{2}-1)}-1}\sqrt{\frac{2(\mu
\theta_{1}^{\prime}+\theta_{1})}{\tau(\mu^{2}-1)}-1}\\
&  =\frac{2}{\mu^{2}-1}\sqrt{(1+\mu+2\kappa)(1+\mu+2\kappa^{\prime})}\times\\
&  \sqrt{1+\kappa^{\prime}+\mu(1+\kappa)}\sqrt{1+\kappa+\mu(1+\kappa^{\prime
})}~.
\end{align*}
Since input and output states have the same mean-value, $F(\alpha)$ is
constant in $\alpha$, so that it coincides with the average teleportation
fidelity of Eq.~(\ref{meanF}), i.e., we find
\begin{equation}
F=\frac{2}{N}:=F(\mu,\kappa,\kappa^{\prime}). \label{avF}%
\end{equation}

Besides $\mu$, this is clearly a function of the environmental parameters
($\tau$, $\omega$, $g$ and $g^{\prime}$) via $\kappa$ and $\kappa^{\prime}$.
We can then fix an experimentally achievable value for $\mu$ (in particular,
$\mu\simeq6.5$, corresponding to about 11dB of two-mode
squeezing~\cite{dB,TopENT,TobiasSqueezing}), and consider an environment with
transmissivity $\tau$ and entanglement-breaking thermal noise $\omega
>\omega_{\text{EB}}(\tau)$. We can therefore explore the points in the
correlation plane $(g,g^{\prime})$ where the protocol is quantum, i.e.,
$F>1/2$. This is done in Fig.~\ref{total} of the main text.

One can easily check that the average teleportation fidelity\ of
Eq.~(\ref{avF}) is an increasing function in the parameter $\mu$, as clearly
expected since this parameter quantifies the amount of entanglement in Bob's
TMSV state. The average teleportation fidelity is therefore maximum in the
limit $\mu\gg1$. At the leading order in $\mu$, we derive the asymptotic
expression%
\begin{equation}
F=F_{\text{opt}}+O(\mu^{-1}),~~F_{\text{opt}}=\frac{1}{\sqrt{(1+\kappa
)(1+\kappa^{\prime})}}, \label{FidOmu}%
\end{equation}
where $\kappa$ and $\kappa^{\prime}$ are given in Eq.~(\ref{kappas}). It is
easy to check that the fidelity in Eq.~(\ref{FidOmu}) may be written as%
\[
F_{\text{opt}}=\frac{1}{\sqrt{1+\varepsilon_{\text{opt}}^{2}+\Omega}},
\]
where%
\begin{equation}
\Omega:=\kappa+\kappa^{\prime}=(\tau^{-1}-1)(2\omega+g^{\prime}-g),
\label{OmegaAPPexp}%
\end{equation}
and $\varepsilon_{\text{opt}}=\sqrt{\kappa\kappa^{\prime}}$ is the asymptotic
PTS eigenvalue of Eq.~(\ref{PTSeigAPP}). This eigenvalue quantifies the amount
of entanglement which would be shared by Alice and Bob if we replaced the
teleportation protocol with an entanglement swapping protocol, where\ Alice
has the same TMSV state as Bob ($\varphi=\mu\gg1$) and receives the classical
communication from Charlie.

Note that, since $\Omega\geq2\varepsilon_{\text{opt}}$, we have the
upper-bound
\begin{equation}
F_{\text{opt}}\leq\frac{1}{1+\varepsilon_{\text{opt}}}, \label{derAPP}%
\end{equation}
with the equality $F=(1+\varepsilon_{\text{opt}})^{-1}$ holding for
environments with antisymmetric correlations $g+g^{\prime}=0$ (in fact this
implies $\kappa=\kappa^{\prime}$ and therefore $\Omega=2\varepsilon
_{\text{opt}}$). Thus, in these antisymmetric environments, we have a full
equivalence between the asymptotic protocols of quantum teleportation and
entanglement swapping: Teleportation of coherent states is quantum
($F_{\text{opt}}>1/2$) if and only if CV entanglement is swapped
($\varepsilon_{\text{opt}}<1$).

\subsection{Distillation\label{DisPROTapp}}

\subsubsection{Entanglement Distillation\label{DistillRATE}}

\begin{figure}[ptbh]
\vspace{-0.5cm}
\par
\begin{center}
\includegraphics[width=0.6\textwidth] {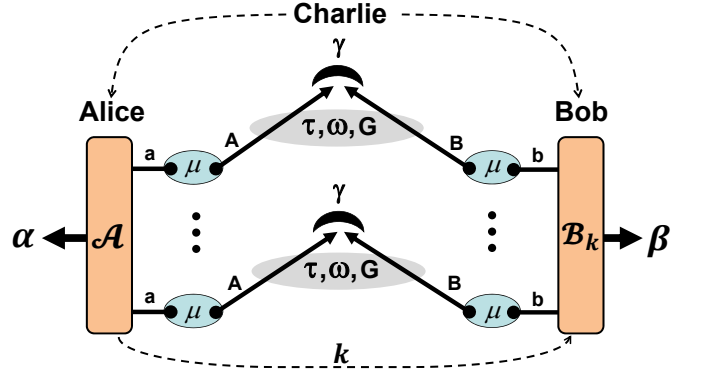}
\end{center}
\par
\vspace{-1.8cm}\caption{Entanglement distillation protocol based on one-way
classical communication. This is operated on top of entanglement swapping to
realize a non-Gaussian quantum repeater. See text for details.}%
\label{DISTapp}%
\end{figure}

Entanglement distillation can be operated on top of entanglement swapping.
After the parties have run the swapping protocol many times and stored their
remote modes in quantum memories, they can perform a one-way entanglement
distillation protocol on the whole set of swapped states. This consists of
Alice locally applying an optimal quantum instrument~\cite{QinstNOTE}
$\mathcal{A}$ on her modes $a$, whose quantum outcome $\boldsymbol{\alpha}%
$\ is a distilled system while the classical outcome $k$ is communicated. Upon
receipt of $k$, Bob performs a conditional quantum operation $\mathcal{B}_{k}%
$\ transforming his modes $b$ into a distilled system $\boldsymbol{\beta}$
(see Fig.~\ref{DISTapp} for a schematic). 

The process can be designed to be highly non-Gaussian so that the distilled
systems have discrete variables and are collapsed into a number of
entanglement bits (Bell state pairs). According to the hashing
inequality~\cite{Qinstrument}, the distillation rate achievable by one-way
distillation protocols is lower bounded by the coherent
information~\cite{CohINFO,CohINFO2}. In general, the coherent information of a
bipartite state $\rho_{ab}$ is defined as%
\[
I_{\mathcal{C}}(\rho_{ab}):=S(\rho_{b})-S(\rho_{ab})~,
\]
where $\rho_{b}=\mathrm{Tr}_{a}(\rho_{ab})$ and $S(\rho):=-$Tr$(\rho\log
_{2}\rho)$ is the von Neumann entropy. This is also denoted by $I(a\rangle
b)$. In particular, the coherent information of a two-mode Gaussian state
$\rho_{ab}$ depends only on its CM%
\[
\mathbf{V}=\left(
\begin{array}
[c]{cc}%
\mathbf{A} & \mathbf{C}\\
\mathbf{C}^{T} & \mathbf{B}%
\end{array}
\right)  .
\]
In fact, it can be written as
\begin{equation}
I_{\mathcal{C}}(\rho_{ab})=h(\nu_{b})-h(\nu_{-})-h(\nu_{+})~, \label{COHana}%
\end{equation}
where the entropic function~\cite{RMP}
\begin{equation}
h(x):=\frac{x+1}{2}\log_{2}\left(  \frac{x+1}{2}\right)  -\frac{x-1}{2}%
\log_{2}\left(  \frac{x-1}{2}\right)  \label{hENTROPIC}%
\end{equation}
is applied to $\nu_{b}=\sqrt{\det\mathbf{B}}$ (symplectic eigenvalue of
$\mathbf{B}$)\ and $\{\nu_{-},\nu_{+}\}$, which is the symplectic spectrum of
$\mathbf{V}$. In the specific case where all symplectic eigenvalues are large,
we can use the expansion%
\begin{equation}
h(x)=\log_{2}\left(  \frac{ex}{2}\right)  +O\left(  x^{-1}\right)  ~~(x\gg1)~,
\label{hEXPANSION}%
\end{equation}
which leads to the following asymptotic formula%
\begin{equation}
I_{\mathcal{C}}(\rho_{ab})\simeq\log_{2}\left(  \frac{2}{e}\sqrt{\frac
{\det\mathbf{B}}{\det\mathbf{V}}}\right)  ~. \label{I_div2}%
\end{equation}

In our analysis, the coherent information $I_{\mathcal{C}}$ has to be computed
on the swapped state $\rho_{ab|\gamma}$ whose CM\ is given in
Eqs.~(\ref{CMgeneBLOCKS}), (\ref{blockB}) and~(\ref{blockC}). It is clear that
$I_{\mathcal{C}}(\rho_{ab|\gamma})$ does not depend on the specific value of
the outcome $\gamma$ (for Gaussian states, the von Neumann entropy and the
coherent information do not depend on the first statistical moments, which are
those encoding the specific value $\gamma$ of the Bell detection). By
replacing the symplectic eigenvalues in Eqs. (\ref{nimenopiu})
and~(\ref{niREDUCED}) into Eq.~(\ref{COHana}), we derive a closed analytical
expression for $I_{\mathcal{C}}(\rho_{ab|\gamma})$ as function of the main
parameters of the problem. Explicitly, we have%
\begin{align}
I_{\mathcal{C}}(\rho_{ab|\gamma})  &  =h\left[  \frac{1}{2}\sqrt{\frac
{(1+2\mu\kappa+\mu^{2})(1+2\mu\kappa^{\prime}+\mu^{2})}{(\mu+\kappa
)(\mu+\kappa^{\prime})}}\right] \nonumber\\
&  -h\left[  \sqrt{\frac{\mu(1+\mu\kappa)}{\mu+\kappa}}\right]  -h\left[
\sqrt{\frac{\mu(1+\mu\kappa^{\prime})}{\mu+\kappa^{\prime}}}\right]
\nonumber\\
&  :=I_{\mathcal{C}}(\mu,\tau,\omega,g,g^{\prime}). \label{cohEXPLICIT}%
\end{align}

This quantity can numerically be studied considering low values of the
parameter $\mu$, i.e., for experimentally achievable values of the input
entanglement (in particular, $\mu\simeq6.5$). As we show in Fig.~\ref{total} of the main
text, entanglement distillation is possible in the presence of
entanglement-breaking channels as long as sufficient amount of separable
correlations is present in the non-Markovian Gaussian environment.

As one intuitively expects and can easily verify via the computation of the
derivatives, the coherent information of Eq.~(\ref{cohEXPLICIT}) is increasing
for increasing $\mu$, reaching its optimal value for large input entanglement
($\mu\gg1$). The spectra of the two Gaussian states $\rho_{ab|\gamma}$ and
$\rho_{b|\gamma}$\ are both diverging in $\mu$, as one can check directly from
the CM $\mathbf{V}_{ab|\gamma}$ and the reduced CM $\mathbf{V}_{b|\gamma}$,
which is just the block $\mathbf{B}$ in Eq.~(\ref{blockB}). For large $\mu$,
we find%
\[
\det\mathbf{B}\simeq\mu^{2}/4,~\det\mathbf{V}_{ab|\gamma}\simeq(\varepsilon
_{\text{opt}}\mu)^{2}~,
\]
where $\varepsilon_{\text{opt}}$ is the asymptotic expression of the smallest
PTS eigenvalue given in Eq.~(\ref{PTSeigAPP}). Thus, using Eq.~(\ref{I_div2}),
we find the following asymptotic expression for the coherent information%
\[
I_{\mathcal{C}}(\rho_{ab|\gamma})\simeq I_{\mathcal{C}\text{,opt}}:=-\log
_{2}\left(  e\varepsilon_{\text{opt}}\right)  ~,
\]
which is the one given in the main text. Asymptotically in the input
resources, we have that entanglement can be distilled ($I_{\mathcal{C}%
\text{,opt}}>0$) for $\varepsilon_{\text{opt}}<e^{-1}\simeq0.367$. Such
condition is more demanding to be satisfied with respect to that of simple
entanglement swapping ($\varepsilon_{\text{opt}}<1$), so that it requires the
presence of more separable correlations in the environment, as shown by Fig.~\ref{total} in the main text.

Finally, we remark that what we computed is the rate achievable by one-way
coherent protocols operated on top of entanglement swapping, i.e., after a
large amount of swapped states are available in Alice's and Bob's quantum
memories. It is interesting to note that this approach seems to be more robust
than the other one where the two procedures are inverted (so that sessions of
entanglement distillation are performed with the relay, followed by
entanglement swapping on the distilled states). In a correlated Gaussian
environment with entanglement-breaking noise, we have that `swapping plus
distillation' can work, while `distillation plus swapping' tends to fail if
the environmental correlations are washed out during the distillation stage.

\subsubsection{Secret-Key Distillation\label{keyDISTapp}}

The previous one-way entanglement distillation protocol of Fig.~\ref{DISTapp}
can be modified into a one-way key distillation protocol, where Charlie is a
generally-untrusted relay distributing secret correlations to Alice and Bob.
Despite the fact that Charlie could be played by an eavesdropper (Eve), the
action of the Bell detection does not give Eve any information. Furthermore,
if Eve tries to tamper with the working mechanism of the relay, Alice and Bob
can always undo this action on the relay and absorb its effects in the
environment. This is a key point of measurement-device-independent (MDI)
QKD~\cite{mdiQKD,Untrusted}.

The environment must be interpreted as the effect of a coherent attack of the
eavesdropper. This can be reduced to a two-mode coherent attack within each
single use of the relay (by adopting quantum de Finetti arguments) and, in
particular, to a two-mode Gaussian attack, by using the extremality of
Gaussian states~\cite{Untrusted}. In such a Gaussian attack, Eve's output
modes $\mathbf{E}$ (not shown in the figure) are stored in a quantum memory
and finally detected. Alice's quantum instrument $\mathcal{A}$ is here a
quantum measurement with classical outputs $\boldsymbol{\alpha}$ (the secret
key) and $k$ (assisting data for Bob). Bob's operation $\mathcal{B}_{k}$ is a
coherent measurement conditioned on $k$, which provides the classical output
$\boldsymbol{\beta}$ (key estimate).

This is an ideal key distribution protocol~\cite{KeyCAP} whose rate $K$ is
lower-bounded by the Devetak-Winter rate $R_{\text{DW}}$~\cite{DW,SciREP}. In
fact, let us restrict Alice to individual measurements, each one applied to
one mode $a$ with outcome $\alpha$. Then, we can write%
\begin{equation}
K\geq R_{\text{DW}}=\chi_{ab|\gamma}-\chi_{a\mathbf{E}|\gamma}~,
\label{RateDW}%
\end{equation}
where $\chi_{ab(\mathbf{E})|\gamma}$ is the conditional Holevo information
between mode $a$ and mode $b$ (modes $\mathbf{E}$). Explicitly,
\begin{align}
\chi_{ab|\gamma}  &  =S(\rho_{b|\gamma})-S(\rho_{b|\gamma\alpha}%
)~,\label{chi1}\\
\chi_{a\mathbf{E}|\gamma}  &  =S(\rho_{\mathbf{E}|\gamma})-S(\rho
_{\mathbf{E}|\gamma\alpha})~, \label{chi2}%
\end{align}
where $\rho_{b|\gamma}$ ($\rho_{\mathbf{E}|\gamma}$) is the conditional state
of Bob (Eve) after each use of the relay, and $\rho_{b|\gamma\alpha}$
($\rho_{\mathbf{E}|\gamma\alpha}$) is the corresponding projected state for
Bob (Eve) after Alice's further measurement, with outcome $\alpha$.

Note that the Bell detection is a rank-1 measurement, therefore projecting
pure states into pure states. For this reason we have that the global state
$\rho_{ab\mathbf{E}|\gamma}$ is pure and therefore
\begin{equation}
S(\rho_{\mathbf{E}|\gamma})=S(\rho_{ab|\gamma})~. \label{S1}%
\end{equation}
If we now restrict Alice's individual measurements to be rank-1, then we also
have that $\rho_{b\mathbf{E}|\gamma\alpha}$ is pure, so that
\begin{equation}
S(\rho_{\mathbf{E}|\gamma\alpha})=S(\rho_{b|\gamma\alpha})~. \label{S2}%
\end{equation}
Thus, using Eqs.~(\ref{chi1})-(\ref{S2}), we may write%
\begin{equation}
R_{\text{DW}}\geq R_{\text{DW}}^{\text{rank-1}}=S(\rho_{b|\gamma}%
)-S(\rho_{ab|\gamma})=I_{\mathcal{C}}(\rho_{ab|\gamma}). \label{DWrank1}%
\end{equation}
Combining Eqs.~(\ref{RateDW}) and~(\ref{DWrank1}), we finally achieve%
\[
K\geq I_{\mathcal{C}}(\rho_{ab|\gamma}),
\]
which is the result stated in the main text.

\subsection{Relay-based practical QKD\label{QKDapp}}

\subsubsection{Description and security analysis\label{QKDappGEN}}

\begin{figure}[t]
\vspace{-0.3cm}
\par
\begin{center}
\includegraphics[width=0.57\textwidth] {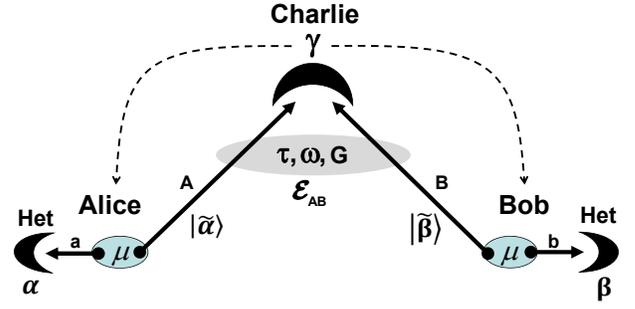}
\end{center}
\par
\vspace{-3.3cm}\caption{Entanglement-based representation of a practical QKD
protocol where coherent states are prepared and sent to Charlie for Bell
detection. Charlie is an untrusted relay, meaning that he could be Eve (relay
is an MDI-QKD node).}%
\label{QKDPICapp}%
\end{figure}

The previous ideal key-distillation\ protocol can be simplified by removing
quantum memories and using one-mode measurements for the parties, in
particular, heterodyne detections. This becomes the entanglement-based
representation of an equivalent `prepare and measure' protocol where
amplitude-modulated coherent states ($\left\vert \tilde{\alpha}\right\rangle $
on Alice's mode $A$, and $|\tilde{\beta}\rangle$ on Bob's mode $B$) are
prepared and sent to the relay (see Fig.~\ref{QKDPICapp}). In particular, the
values of the amplitudes are simply connected with the outcomes of the
heterodyne detectors in the equivalent entanglement-based representation. The
amplitudes satisfy the relations
\[
\left(
\begin{array}
[c]{c}%
\tilde{\alpha}\\
\tilde{\beta}%
\end{array}
\right)  =\frac{\sqrt{\mu^{2}-1}}{\mu+1}\left(
\begin{array}
[c]{c}%
\alpha^{\ast}\\
\beta^{\ast}%
\end{array}
\right)  ,
\]
where $\alpha$ and $\beta$ are the outcomes of the virtual heterodyne
detectors and $\mu$ is the variance of the virtual TMSV states at Alice's and
Bob's stations. As a result the amplitudes $\tilde{\alpha}$ and $\tilde{\beta
}$ of the coherent states are Gaussianly modulated with a variance $\mu-1$
(see Ref.~\cite{HET} and also the Supplementary Material of
Ref.~\cite{OptimalDiscord}).

At the relay the transmitted states are subject to Bell detection and the
result $\gamma$ is communicated back to the parties. Since $\gamma\simeq
\alpha-\beta^{\ast}$, we have that classical correlations are remotely created
between Alice's and Bob's complex variables. As mentioned before, the
knowledge of $\gamma$ alone does not help Eve as long as the variance of the
modulation $\mu-1$ is sufficiently high. Experimentally, values of modulation
$\mu\gtrsim50$ are easily achievable and already well approximate the
performance of the asymptotic scenario $\mu\gg1$ (in terms of secret key rate
assuming ideal reconciliation performances).

The non-Markovian environment is the result of Eve's attack. Eve stores all
her output ancillas $\mathbf{E}$ (not shown) in a quantum memory which is
subject to a final optimized coherent measurement. As previously discussed,
this environment may also absorb the effects of an attack directed at the
working mechanism of the middle relay. Furthermore, suitable classes of
side-channel attacks which directly affect the optical preparations inside
Alice's and Bob's private spaces can also be treated as part of the external environment.

Despite the fact that the protocol is performed as a prepare and measure
protocol, its security is more easily studied considering its
entanglement-based representation. In this equivalent representation, Alice
and Bob can estimate the post-relay conditional quantum state $\rho
_{ab|\gamma}$ by comparing a subset of their data and analyzing the joint
classical statistics $p(\tilde{\alpha},\tilde{\beta},\gamma)$. Then, Alice and
Bob purify $\rho_{ab|\gamma}$ into an environment $\mathbf{E}$\ which is fully
controlled by Eve. In these general conditions, the two parties are able to
compute the secret key rate directly from the CM $\mathbf{V}_{ab|\gamma}$ of
$\rho_{ab|\gamma}$. Because of the extremality properties of Gaussian states,
Alice and Bob can always assume that $\rho_{ab|\gamma}$ is
Gaussian~\cite{Untrusted}.

Let us discuss in detail how the rate of the protocol can be computed from the
second-order statistical moments $\mathbf{V}_{ab|\gamma}$. After the action of
the relay, Alice and Bob's mutual information
\[
I_{AB|\gamma}:=I(\tilde{\alpha}:\tilde{\beta}|\gamma)=I(\alpha:\beta|\gamma)
\]
is given by~\cite{Untrusted}
\begin{equation}
I_{AB|\gamma}=\frac{1}{2}\log_{2}\Sigma,~~\Sigma:=\frac{1+\det\mathbf{V}%
_{b|\gamma}+\mathrm{Tr}\mathbf{V}_{b|\gamma}}{1+\det\mathbf{V}_{b|\gamma
\alpha}+\mathrm{Tr}\mathbf{V}_{b|\gamma\alpha}}. \label{mutuaABapp}%
\end{equation}
Here $\mathbf{V}_{b|\gamma}$ is the CM of Bob's reduced state $\rho_{b|\gamma
}$ and $\mathbf{V}_{b|\gamma\alpha}$ is the CM of Bob's state $\rho
_{b|\gamma\alpha}$ after the detections of both the relay and Alice. The
latter CM\ can easily be computed from the CM $\mathbf{V}_{ab|\gamma}$ using
the formulas for the heterodyne detection~\cite{HET}.

To bound Eve's stolen information on Alice's variable $\alpha$, we use the
conditional Holevo information $\chi_{a\mathbf{E}|\gamma}$ between Alice's
(detected) mode $a$ and Eve's output ancillas $\mathbf{E}$ for the single use
of the relay. Since the Bell detection at the relay is rank-1, the conditional
global state $\rho_{ab\mathbf{E}|\gamma}$ of Alice, Bob and Eve is pure.
Furthermore, Alice's heterodyne detection is also rank-1, so that the
double-conditional state $\rho_{b\mathbf{E}|\gamma\alpha}$ is also pure. This
means that we can exploit the entropic equalities $S(\rho_{\mathbf{E}|\gamma
})=S(\rho_{ab|\gamma})$ and $S(\rho_{\mathbf{E}|\gamma\alpha})=S(\rho
_{b|\gamma\alpha})$. Thus, we may write%
\begin{equation}
\chi_{a\mathbf{E}|\gamma}=S(\rho_{ab|\gamma})-S(\rho_{b|\gamma\alpha}).
\label{EveHOLEapp}%
\end{equation}
This quantity can be computed from the symplectic spectra of the CMs
$\mathbf{V}_{ab|\gamma}$ and $\mathbf{V}_{b|\gamma\alpha}$. We have%
\begin{equation}
\chi_{a\mathbf{E}|\gamma}=h(\nu_{-})+h(\nu_{+})-h(\nu_{\text{c}}),
\label{ChiEVEcalcolo}%
\end{equation}
where the function $h(x)$ of Eq.~(\ref{hENTROPIC}) is applied to the
symplectic spectrum $\{\nu_{-},\nu_{+}\}$ of $\mathbf{V}_{ab|\gamma}$ and
$\nu_{\text{c}}=\sqrt{\det\mathbf{V}_{b|\gamma\alpha}}$.

The secret key rate is finally given by the difference
\begin{equation}
R=\xi I_{AB|\gamma}-\chi_{a\mathbf{E}|\gamma} \label{rateAPPn}%
\end{equation}
where $\xi\leq1$ is the reconciliation efficiency (due to the finite
efficiency of realistic codes for error correction and privacy amplification).
Thus, the rate can be computed from the second-order moments, in particular,
from $\mathbf{V}_{ab|\gamma}$. This procedure is very general: In the next
section it is used to derive the analytical expression of the key rate from
the main parameters of a two-mode Gaussian attack against the two links;
afterwards, in~\ref{ExpMETH}, it is used to derive the experimental key
rate from the statistics of the shared classical data.

\subsubsection{Analytical expression of the key rate}

Let us consider a two-mode Gaussian attack of the links which results into a
non-Markovian Gaussian environment with correlated-thermal noise, with
parameters $\tau$, $\omega$, $g$ and $g^{\prime}$, as described in~\ref{EnvREVapp}. Then, the conditional CM\ $\mathbf{V}_{ab|\gamma}$\ is
specified by Eqs.~(\ref{CMgeneBLOCKS})-(\ref{blockC}). Bob's reduced
CM\ $\mathbf{V}_{b|\gamma}$ is the block $\mathbf{B}$ given in
Eq.~(\ref{blockB}). The expression of $\mathbf{V}_{b|\gamma\alpha}$ has been
already obtained in Eqs.~(\ref{Vb0})-(\ref{teta2}). Thus, for Alice and Bob's
mutual information $I_{AB|\gamma}$, we find%
\[
\Sigma=\frac{\left(  1+\mu+2\kappa\right)  ^{2}\left(  1+\mu+2\kappa^{\prime
}\right)  ^{2}}{16(1+\kappa)(1+\kappa^{\prime})(\mu+\kappa)(\mu+\kappa
^{\prime})}~.
\]

For the computation of Eve's Holevo information $\chi_{a\mathbf{E}|\gamma}$,
we see that the symplectic spectrum $\{\nu_{-},\nu_{+}\}$ of $\mathbf{V}%
_{ab|\gamma}$ is given in Eq.~(\ref{nimenopiu}), and we compute%
\[
\nu_{\text{c}}=\sqrt{\frac{\left(  1+\mu+2\mu\kappa\right)  \left(  1+\mu
+2\mu\kappa^{\prime}\right)  }{\left(  1+\mu+2\kappa\right)  \left(
1+\mu+2\kappa^{\prime}\right)  }}~.
\]

The rate is an analytical but cumbersome function of the relevant parameters
of the problem, i.e., the finite reconciliation efficiency $\xi$, the signal
modulation variance $\mu$ of the coherent states, and the $\kappa$-parameters
of the environment. Explicitly, we have
\begin{align}
R  &  =\frac{\xi}{2}\log_{2}\frac{\left(  1+\mu+2\kappa\right)  ^{2}\left(
1+\mu+2\kappa^{\prime}\right)  ^{2}}{16(1+\kappa)(1+\kappa^{\prime}%
)(\mu+\kappa)(\mu+\kappa^{\prime})}\nonumber\\
&  -h\left[  \sqrt{\frac{\mu(1+\mu\kappa)}{\mu+\kappa}}\right]  -h\left[
\sqrt{\frac{\mu(1+\mu\kappa^{\prime})}{\mu+\kappa^{\prime}}}\right]
\nonumber\\
&  +h\left[  \sqrt{\frac{\left(  1+\mu+2\mu\kappa\right)  \left(  1+\mu
+2\mu\kappa^{\prime}\right)  }{\left(  1+\mu+2\kappa\right)  \left(
1+\mu+2\kappa^{\prime}\right)  }}\right] \nonumber\\
&  :=R(\xi,\mu,\kappa,\kappa^{\prime}) \label{rateKAPPAapp}%
\end{align}

\begin{figure}[ptbh]
\vspace{-0.3cm}
\par
\begin{center}
\includegraphics[width=0.33\textwidth] {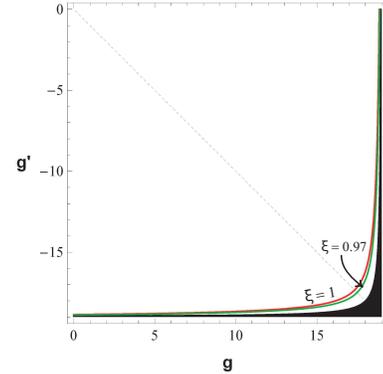}
\end{center}
\par
\vspace{-0.7cm}\caption{Security thresholds $R=0$ in the correlation plane,
considering $\mu\simeq50$ and $\xi=1$ (red line) and $\xi\simeq0.97$ (green
line). Below the thresholds the corresponding rates are positive. Other
parameters are $\tau=0.9$ and $\omega\simeq19>\omega_{\text{EB}}.$}%
\label{EfficiencyAPP}%
\end{figure}

Using the expressions for the $\kappa$-parameters in Eq.~(\ref{kappas}), we
can write the rate as $R=R(\xi,\mu,\tau,\omega,g,g^{\prime})$, i.e., directly
in terms of the parameters of the two-mode Gaussian attack, i.e., the
transmissivity $\tau$, the variance of the thermal noise $\omega$\ and the
correlation parameters $g$ and $g^{\prime}$. Fixing the reconciliation
efficiency (e.g., to be ideal $\xi=1$ or achievable $\xi\simeq0.97~$%
\cite{Jouguetapp}) and the modulation $\mu\simeq50$, we can study the rate $R$
in an entanglement-breaking Gaussian attack with $\tau=0.9$ and $\omega
=(1+10^{-4})\omega_{\text{EB}}\simeq19$ (about 9 thermal photons). The
security threshold $R=0$ can be plotted in the correlation plane
$(g,g^{\prime})$ as shown in Fig~\ref{EfficiencyAPP}.

As we can see from Fig.~\ref{EfficiencyAPP}, the secret-key rate can be
positive for sufficiently high separable correlations in the Gaussian attack.
Furthermore, there is no much difference between the cases with ideal ($\xi
=1$) or achievable ($\xi\simeq0.97$) reconciliation efficiency. For this
reason, in the theoretical discussions of the main text we have only
considered the simpler case of ideal reconciliation, i.e., the rate
$R=R(1,\mu,\kappa,\kappa^{\prime})=R(1,\mu,\tau,\omega,g,g^{\prime})$.

In order to show the optimal performance of the QKD\ protocol, we consider
ideal reconciliation ($\xi=1$) and we perform the limit of large modulation
$\mu\gg1$ (in fact, by assuming ideal reconciliation, the rate is increasing
in $\mu$). Let us derive the asymptotic optimal rate
\[
R(1,\mu\gg1,\kappa,\kappa^{\prime})\simeq R_{\text{opt}}(\kappa,\kappa
^{\prime})=R_{\text{opt}}(\tau,\omega,g,g^{\prime}).
\]
At the leading order in $\mu$, we find
\begin{align}
\Sigma &  =\frac{\mu^{2}}{16(1+\kappa)(1+\kappa^{\prime})}+O(\mu
),\label{eq1}\\
\nu_{\text{c}}  &  =\sqrt{(1+2\kappa)(1+2\kappa^{\prime})}+O(\mu
^{-1}),\label{eq2}\\
\nu_{-}  &  =\sqrt{\kappa\mu}+O(\mu^{-1/2}),\label{eq3}\\
\nu_{+}  &  =\sqrt{\kappa^{\prime}\mu}+O(\mu^{-1/2}), \label{eq4}%
\end{align}
where $\kappa$ and $\kappa^{\prime}$ are given in Eq.~(\ref{kappas}). Using
the previous Eqs.~(\ref{eq1})-(\ref{eq4}) and the expansion in
Eq.~(\ref{hEXPANSION}) we find the simple formula%
\begin{align*}
R_{\text{opt}}  &  =\log_{2}\left[  \frac{1}{e^{2}\sqrt{(1+\kappa
)(1+\kappa^{\prime})\kappa\kappa^{\prime}}}\right] \\
&  +h\left[  \sqrt{(1+2\kappa)(1+2\kappa^{\prime})}\right]  .
\end{align*}

\begin{figure}[ptbh]
\vspace{-0.3cm}
\par
\begin{center}
\includegraphics[width=0.35\textwidth] {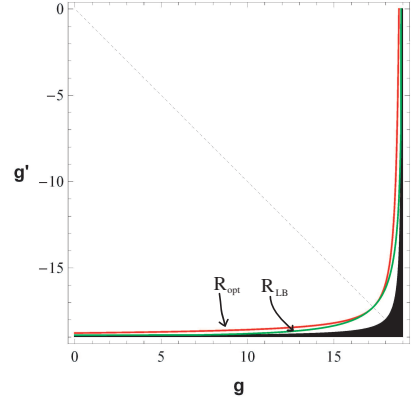}
\end{center}
\par
\vspace{-0.7cm}\caption{Security thresholds $R_{\text{opt}}=0$ (red line) and
$R_{\text{LB}}=0$ (green line). Other parameters are $\tau=0.9$ and
$\omega\simeq19>\omega_{\text{EB}}$.}%
\label{LBapp}%
\end{figure}

We can easily connect this asymptotic key rate with the asymptotic
PTS\ eigenvalue $\varepsilon_{\text{opt}}$ of Eq.~(\ref{PTSeigAPP}) and the
asymptotic fidelity $F_{\text{opt}}$ of Eq.~(\ref{FidOmu}). In fact, we may
write%
\begin{equation}
R_{\text{opt}}=\log_{2}\left(  \frac{F_{\text{opt}}}{e^{2}\varepsilon
_{\text{opt}}}\right)  +h\left[  \sqrt{1+(2\varepsilon_{\text{opt}}%
)^{2}+2\Omega}\right]
\end{equation}
where we have also used $\Omega:=\kappa+\kappa^{\prime}\geq2\varepsilon
_{\text{opt}}$. Using the latter inequality, we may write the lower bound%
\begin{equation}
R_{\text{opt}}\geq R_{\text{LB}}:=\log_{2}\left(  \frac{F_{\text{opt}}}%
{e^{2}\varepsilon_{\text{opt}}}\right)  +h(1+2\varepsilon_{\text{opt}}).
\label{LBappendix}%
\end{equation}
As we can see from Fig.~\ref{LBapp}, this bound is sufficiently tight for the
most interesting range of parameters, i.e., for environments with
antisymmetric correlations $g+g^{\prime}\simeq0$ (around the diagonal of the
correlation plane). In particular, for environments with exactly $g+g^{\prime
}=0$, we have the equality
\begin{equation}
R_{\text{opt}}=R_{\text{LB}}=\log_{2}\left[  \frac{1}{e^{2}\varepsilon
_{\text{opt}}(1+\varepsilon_{\text{opt}})}\right]  +h(1+2\varepsilon
_{\text{opt}}), \label{simAPP}%
\end{equation}
since we have $\kappa=\kappa^{\prime}$ which implies both $\Omega
=2\varepsilon_{\text{opt}}$ and $F_{\text{opt}}=(1+\varepsilon_{\text{opt}%
})^{-1}$.

Note that the left hand sides of Eqs.~(\ref{LBappendix}) and~(\ref{simAPP})
can be positive only for $\varepsilon_{\text{opt}}\lesssim0.192$.
Asymptotically, the practical QKD protocol appears to be the most difficult to
reactivate: Its reactivation implies that of entanglement/key distillation
($\varepsilon_{\text{opt}}<0.367$) and that of entanglement swapping
($\varepsilon_{\text{opt}}<1$). \begin{figure}[ptbh]
\vspace{-0.2cm}
\par
\begin{center}
\includegraphics[width=0.35\textwidth] {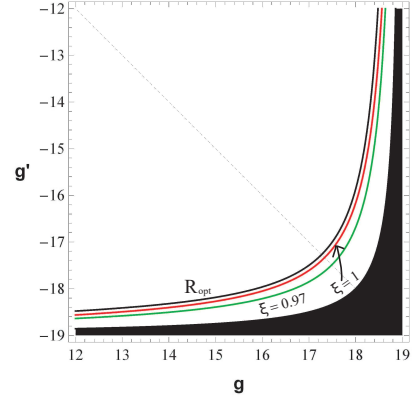}
\end{center}
\par
\vspace{-0.7cm}\caption{Comparison between $R_{\text{opt}}=0$ (black line),
$R=0$ with $\mu\simeq50$ and $\xi=1$ (red line), and $R=0$ with $\mu\simeq50$
and $\xi=0.97$ (green line). Other parameters are $\tau=0.9$ and $\omega
\simeq19>\omega_{\text{EB}}$. }%
\label{zoomAPP}%
\end{figure}

In conclusion, we also show that the asymptotic key rate $R_{\text{opt}}$ is
well approximated by the finite modulation rate $R$\ with $\mu\simeq50$
(assuming both the case of ideal reconciliation $\xi=1$ and realistic
reconciliation efficiency $\xi\simeq0.97$). This is evident from Fig.~\ref{zoomAPP}, which is a zoom on the most interesting part of the
correlation plane. We see that considering realistic finite modulations and
reconciliation efficiencies does not sensibly degrade the reactivation process
(determined by the performances of the various security thresholds).

\section{Theory: Correlated-Additive Noise\label{AdditiveAPP}}

Here we extend our previous analysis to another kind
of non-Markovian environment, which is the Gaussian environment with
correlated-additive noise. Here we provide full details on the following
theoretical elements:

\bigskip

$\bullet~$S\ref{LimitenvAPP}: We show how the correlated-additive
environment can be obtained as a suitable limit of the previous
correlated-thermal environment.

\bigskip

$\bullet~$\ref{swapLIMapp}: We derive the CM of the swapped state in this
environment. We discuss the condition for swapping reactivation, and we
identify optimal and suboptimal points for reactivation.

\bigskip

$\bullet~$\ref{practlimAPP}: We study the security of the practical
relay-based QKD\ protocol, providing the general formula for its secret-key
rate $R(\xi,\mu,n,c,c^{\prime})$. We then discuss the reactivation of this
practical protocol from entanglement-breaking.

\subsection{Additive-Noise Limit\label{LimitenvAPP}}

The correlated-additive Gaussian environment can be obtained from the previous
correlated-thermal Gaussian environment by taking a suitable continuous limit
(where the continuity is guaranteed by the Gaussian nature of all the
process). In particular, we consider the limit for $\tau\rightarrow1$ and
$\omega\rightarrow+\infty$, while keeping constant
\begin{equation}
n:=(1-\tau)\omega,~c:=\frac{g}{\omega-1},~c^{\prime}:=\frac{g^{\prime}}%
{\omega-1}. \label{limitAPP}%
\end{equation}

The effect of this limit can be understood studying the CM\ of the modes
$A^{\prime}$ and $B^{\prime}$\ after the action of the channel and before the
Bell detection. For the correlated-thermal Gaussian environment, we have the
following CM%
\begin{equation}
\mathbf{V}_{A^{\prime}B^{\prime}}=\left(
\begin{array}
[c]{cc}%
\lbrack\tau\mu+(1-\tau)\omega]\mathbf{I} & (1-\tau)\mathbf{G}\\
(1-\tau)\mathbf{G} & [\tau\mu+(1-\tau)\omega]\mathbf{I}%
\end{array}
\right)  ,
\end{equation}
which can easily be derived from Eq.~(\ref{VabApBp}) by setting $\varphi=\mu$
and deleting the entries of modes $a$ and $b$. Now taking the previous limit,
it leads to%
\begin{equation}
\mathbf{V}_{A^{\prime}B^{\prime}}\rightarrow\mathbf{V}_{A^{\prime}B^{\prime}%
}^{\text{add}}=\left(
\begin{array}
[c]{cc}%
(\mu+n)\mathbf{I} & n\mathbf{C}\\
n\mathbf{C} & (\mu+n)\mathbf{I}%
\end{array}
\right)  , \label{ClassCM}%
\end{equation}
where%
\[
\mathbf{C}:=\left(
\begin{array}
[c]{cc}%
c & 0\\
0 & c^{\prime}%
\end{array}
\right)  ,
\]
and we have used $(1-\tau)\mathbf{G}=(1-\tau)(\omega-1)\mathbf{C}\rightarrow
n\mathbf{C}$.

As we can easily check, the CM in Eq.~(\ref{ClassCM}) can be decomposed as
follows%
\[
\mathbf{V}_{A^{\prime}B^{\prime}}^{\text{add}}=\mathbf{V}_{AB}+n\left(
\begin{array}
[c]{cc}%
\mathbf{I} & \mathbf{C}\\
\mathbf{C} & \mathbf{I}%
\end{array}
\right)  ,
\]
so that the environment adds classical Gaussian noise to the input CM
$\mathbf{V}_{AB}=\mu(\mathbf{I}\oplus\mathbf{I})$ with variance $n\geq0$ in
each quadrature, and noise-correlations described by the off-diagonal block
$n\mathbf{C}$, with parameters $-1\leq c,c^{\prime}\leq1$.

In terms of input-output quadrature transformations, the action of the
asymptotic environment is therefore described by%
\begin{equation}
\left\{
\begin{array}
[c]{c}%
\hat{q}_{A^{\prime}}=\hat{q}_{A}+\xi_{1},\\
\hat{p}_{A^{\prime}}=\hat{p}_{A}+\xi_{2},\\
\hat{q}_{B^{\prime}}=\hat{q}_{B}+\xi_{3},\\
\hat{p}_{B^{\prime}}=\hat{p}_{B}+\xi_{4},
\end{array}
\right.  \label{tracsi}%
\end{equation}
where the $\xi_{i}$'s are zero-mean Gaussian variables\ whose covariances
$\left\langle \xi_{i}\xi_{j}\right\rangle $ are given by the classical CM
\begin{equation}
\mathbf{V}\left(  n,c,c^{\prime}\right)  =n\left(
\begin{array}
[c]{cc}%
\mathbf{I} & \mathbf{C}\\
\mathbf{C} & \mathbf{I}%
\end{array}
\right)  . \label{CMeveAPPE}%
\end{equation}

It is straightforward to extend the previous calculation to include the remote
modes $a$ and $b$, so that we find the following CM after the action of the
correlated-additive environment
\begin{equation}
\mathbf{V}_{abA^{\prime}B^{\prime}}^{\text{add}}=\left(
\begin{array}
[c]{cccc}%
\mu\mathbf{I} & \mathbf{0} & \tilde{\mu}\mathbf{Z} & \mathbf{0}\\
\mathbf{0} & \mu\mathbf{I} & \mathbf{0} & \tilde{\mu}\mathbf{Z}\\
\tilde{\mu}\mathbf{Z} & \mathbf{0} & (\mu+n)\mathbf{I} & n\mathbf{C}\\
\mathbf{0} & \tilde{\mu}\mathbf{Z} & n\mathbf{C} & (\mu+n)\mathbf{I}%
\end{array}
\right)  . \label{startAPP}%
\end{equation}
From Eq.~(\ref{startAPP}), we see that the CM of Alice's modes $a$ and
$A^{\prime}$, and that of Bob's modes $b$ and $B^{\prime}$ are equal to%
\[
\mathbf{V}_{aA^{\prime}}^{\text{add}}=\mathbf{V}_{bB^{\prime}}^{\text{add}%
}=\left(
\begin{array}
[c]{cc}%
\mu\mathbf{I} & \tilde{\mu}\mathbf{Z}\\
\tilde{\mu}\mathbf{Z} & (\mu+n)\mathbf{I}%
\end{array}
\right)  ,
\]
whose smallest PTS\ eigenvalue is $\geq1$ (i.e., bipartite entanglement is
lost) when $n\geq2$. This entanglement-breaking condition can be strengthened
into the strict inequality $n>2$ in order to exclude also the possible
presence of tripartite entanglement.

Note that $n>2$ can equivalently be\ obtained by the taking the limit of the
previous entanglement-breaking condition $\omega>\omega_{\text{EB}}%
=(1+\tau)(1-\tau)^{-1}$. In fact, using the latter inequality and taking the
limit, we find%
\begin{align*}
\left\langle \hat{q}_{A^{\prime}}^{2}\right\rangle  &  =\tau\mu+(1-\tau
)\omega>\tau\mu+(1-\tau)\omega_{\text{EB}}\\
&  =\tau\mu+1+\tau\rightarrow\mu+2=\left\langle \hat{q}_{A}^{2}\right\rangle
+2.
\end{align*}
Comparing the latter equation with $\left\langle \hat{q}_{A^{\prime}}%
^{2}\right\rangle =\left\langle \hat{q}_{A}^{2}\right\rangle +n$, we see that
the entanglement breaking condition is asymptotically mapped into $n>2$.

\subsection{Entanglement Swapping\label{swapLIMapp}}

In order to compute the CM of the conditional state $\rho_{ab|\gamma}$ after
Bell detection, we can equivalently start from the CM in Eq.~(\ref{startAPP})
and repeat the derivation of~\ref{APPcmCALCOLO}, or just taking the limit
in the CM\ $\mathbf{V}_{ab|\gamma}$ given in Eqs.~(\ref{CMmainAPP})
and~(\ref{bigPSI}). The final result is achieved by taking the limit in the
$\kappa$-parameters of Eq.~(\ref{kappas}), which become%
\begin{equation}
\left\{
\begin{array}
[c]{c}%
\kappa\rightarrow(1-c)n,\\
\\
\kappa^{\prime}\rightarrow(1+c^{\prime})n.
\end{array}
\right.  \label{reDEFkappa}%
\end{equation}
Thus, the CM of the swapped state is given by%
\begin{align}
\mathbf{V}_{ab|\gamma}^{\text{add}}  &  =\left(
\begin{array}
[c]{cc}%
\mu\mathbf{I} & \mathbf{0}\\
\mathbf{0} & \mu\mathbf{I}%
\end{array}
\right)  -\frac{\mu^{2}-1}{2}\left[  \boldsymbol{\Psi}\right]  _{\kappa
:=(1-c)n,\kappa^{\prime}:=(1+c^{\prime})n}\nonumber\\
&  =\left(
\begin{array}
[c]{cc}%
\mu\mathbf{I} & \mathbf{0}\\
\mathbf{0} & \mu\mathbf{I}%
\end{array}
\right)  -\frac{\mu^{2}-1}{2}\times\nonumber\\
&  \times\left(
\begin{array}
[c]{cccc}%
\frac{1}{\mu+(1-c)n} & 0 & \frac{-1}{\mu+(1-c)n} & 0\\
0 & \frac{1}{\mu+(1+c^{\prime})n} & 0 & \frac{1}{\mu+(1+c^{\prime})n}\\
\frac{-1}{\mu+(1-c)n} & 0 & \frac{1}{\mu+(1-c)n} & 0\\
0 & \frac{1}{\mu+(1+c^{\prime})n} & 0 & \frac{1}{\mu+(1+c^{\prime})n}%
\end{array}
\right)  . \label{bigCMapp}%
\end{align}

\begin{figure}[ptbh]
\vspace{+0.2cm}
\par
\begin{center}
\includegraphics[width=0.28\textwidth] {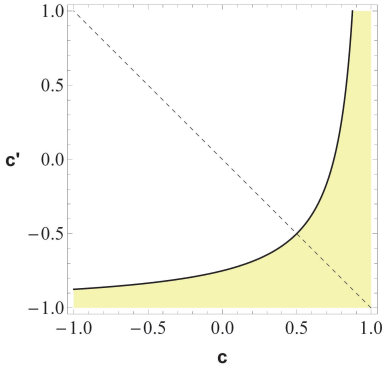}
\end{center}
\par
\vspace{-0.4cm}\caption{Assuming entanglement-breaking in each link
($n=2+10^{-4}\gtrsim2$) and arbitrary entanglement at the input ($\mu>1$), we
characterize the region of the classical correlation plane $(c,c^{\prime})$,
depicted in yellow, where entanglement swapping is reactivated, i.e., the
condition of Eq.~(\ref{reactSWAPadd}) is satisfied.}%
\label{swapadd}%
\end{figure}

It is clear that all the quantities previously derived for the
correlated-thermal environment can be extended via the continuous limit to the
correlated-additive environment, by just re-defining the $\kappa$-parameters
according to Eq.~(\ref{reDEFkappa}). Thus, the condition for swapping
entanglement $\kappa\kappa^{\prime}<1$ here becomes
\begin{equation}
(1-c)(1+c^{\prime})<n^{-2}~. \label{reactSWAPadd}%
\end{equation}
We can see that this condition does not depend on the amount of input
entanglement ($\mu>1$) and can always be satisfied for $c\rightarrow1$ or
$c^{\prime}\rightarrow-1$, no matter what the value of $n$ is, even
entanglement-breaking ($n>2$). As shown in Fig.~\ref{swapadd}, there is a wide
region in the classical correlation plane $(c,c^{\prime})$ where entanglement
swapping can be reactivated.

In the correlation plane, we can identify an optimal point $(c,c^{\prime
})=(1,-1)$. This extremal point describes an environment whose classical
correlations are able to completely cancel the noise from the output modes. In
fact, by replacing in $\mathbf{V}_{ab|\gamma}^{\text{add}}(\mu,n,c,c^{\prime
})$ of Eq.~(\ref{bigCMapp}), we retrieve%
\[
\mathbf{V}_{ab|\gamma}^{\text{add}}(\mu,n,1,-1)=\frac{1}{2\mu}\left(
\begin{array}
[c]{cc}%
(\mu^{2}+1)\mathbf{I} & (\mu^{2}-1)\mathbf{Z}\\
(\mu^{2}-1)\mathbf{Z} & (\mu^{2}+1)\mathbf{I}%
\end{array}
\right)  ,
\]
which is the CM\ of a swapped state in absence of loss and noise (as one can
also double-check by applying the formulas in Refs.~\cite{GaussSWAP}). Such an
environment has such a unique property since it corresponds to the quadrature
transformations of Eq.~(\ref{tracsi}) with $\xi_{2}=\xi_{1}$ and $\xi_{4}%
=-\xi_{3}$, so that, after the beam splitter of the relay, we have%
\begin{align*}
\hat{q}_{-}  &  =\frac{\hat{q}_{A^{\prime}}-\hat{q}_{B^{\prime}}}{\sqrt{2}%
}=\frac{\hat{q}_{A}-\hat{q}_{B}}{\sqrt{2}},\\
\hat{p}_{+}  &  =\frac{\hat{p}_{A^{\prime}}+\hat{p}_{B^{\prime}}}{\sqrt{2}%
}=\frac{\hat{p}_{A}+\hat{p}_{B}}{\sqrt{2}}.
\end{align*}
Thanks to the global cancelation effect induced by its correlations, the
optimal environment $(1,-1)$ not only reactivates entanglement swapping but
any other quantum protocol.

Besides the optimal point $(1,-1)$\ and the Markovian point $(0,0)$ (which is
unable to reactivate), there are infinite other points in the plane
$(c,c^{\prime})$ with intermediate performances. An interesting environment
corresponds to the sub-optimal point $(1,1)$ for which we have the following
CM for the swapped state%
\begin{align}
\mathbf{V}_{ab|\gamma}^{\text{add}}(\mu,n,1,1)  &  =\left(
\begin{array}
[c]{cc}%
\mu\mathbf{I} & \mathbf{0}\\
\mathbf{0} & \mu\mathbf{I}%
\end{array}
\right)  -\frac{\mu^{2}-1}{2}\times\nonumber\\
&  \times\left(
\begin{array}
[c]{cccc}%
\frac{1}{\mu} & 0 & \frac{-1}{\mu} & 0\\
0 & \frac{1}{\mu+2n} & 0 & \frac{1}{\mu+2n}\\
\frac{-1}{\mu} & 0 & \frac{1}{\mu} & 0\\
0 & \frac{1}{\mu+2n} & 0 & \frac{1}{\mu+2n}%
\end{array}
\right)  .
\end{align}
This environment cancels the noise in only one quadrature and corresponds to
the transformations of Eq.~(\ref{tracsi}) with $\xi_{2}=\xi_{1}$ and $\xi
_{4}=\xi_{3}$, so that, after the beam splitter of the relay, we have%
\[
\hat{q}_{-}=\frac{\hat{q}_{A}-\hat{q}_{B}}{\sqrt{2}},~\hat{p}_{+}=\frac
{\hat{p}_{A}+\hat{p}_{B}+2\xi_{3}}{\sqrt{2}}.
\]
Despite the fact that this sub-optimal environment reactivates entanglement
swapping, its effects on the other quantum protocols, in particular, the
practical QKD\ protocol, need to be investigated (see below).

\subsection{Relay-based practical QKD\label{practlimAPP}}

In order to study the reactivation properties of the correlated-additive
environment, we consider the quantum protocol most difficult to reactivate,
i.e., the practical QKD protocol (such property is inherited via the
continuous limit from the previous correlated-thermal environment). The analysis of~\ref{QKDapp}, based on the use of the entropic
equalities $S(\rho_{\mathbf{E}|\gamma})=S(\rho_{ab|\gamma})$ and
$S(\rho_{\mathbf{E}|\gamma\alpha})=S(\rho_{b|\gamma\alpha})$, is valid for any
physical value of the parameters $\tau$, $\omega$, $g$ and $g^{\prime}$ of the
correlated-thermal environment. Therefore, it continues to be valid in the
considered limit for $\tau\rightarrow1$ and $\omega\rightarrow+\infty$ with
the constraints specified by Eq.~(\ref{limitAPP}).

\begin{figure}[ptbh]
\vspace{+0.1cm}
\par
\begin{center}
\includegraphics[width=0.28\textwidth] {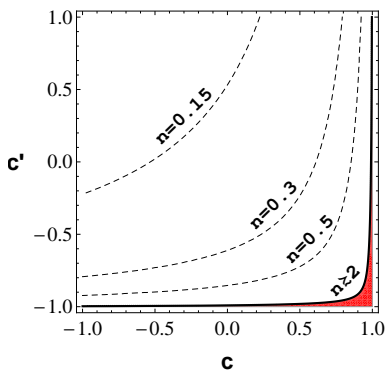}
\end{center}
\par
\vspace{-0.4cm}\caption{Security thresholds $R=0$ plotted on the correlation
plane $(c,c^{\prime})$ for increasing values of the additive noise $n$ (signal
modulation $\mu\simeq52$ and reconciliation efficiency $\xi=1$). The solid
curve is the security threshold for entanglement-breaking links ($n=2+10^{-4}%
\gtrsim2$): The points in the red region are environments whose classical
correlations are strong enough to reactivate the QKD protocol.}%
\label{CLASSapp}%
\end{figure}

The net effect of this limit is the re-definition~(\ref{reDEFkappa}) of the
$\kappa$-parameters in the secret-key rate $R(\xi,\mu,\kappa,\kappa^{\prime})$
of Eq.~(\ref{rateKAPPAapp}). Thus, the analytical expression of the key rate
in the correlated-additive environment is given by
\[
R_{\text{add}}(\xi,\mu,n,c,c^{\prime})=R[\xi,\mu,(1-c)n,(1+c^{\prime})n].
\]
Let us fix a value for the reconciliation efficiency (e.g., $\xi=1$) and a
finite value for the modulation variance (e.g., $\mu=52$). Then, for any value
of the additive noise $n$, we can plot the security threshold $R_{\text{add}%
}=0$ on the classical correlation plane $(c,c^{\prime})$, as done in Fig.~\ref{CLASSapp}.

\begin{figure}[t]
\vspace{-4.8cm}
\par
\begin{center}
\includegraphics[width=0.61\textwidth] {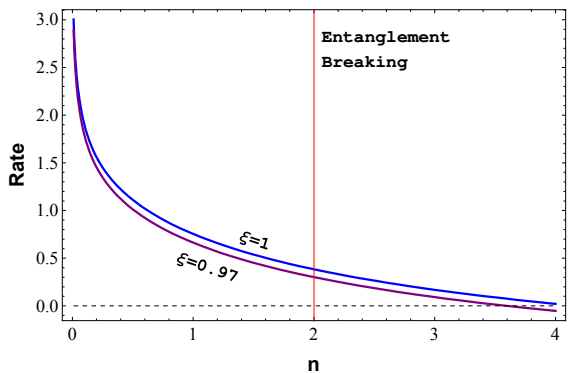}
\end{center}
\par
\vspace{-5.1cm}\caption{Rate $R_{\text{add}}(\xi,52,n,1,1)$ in bits/use as a
function of the additive noise $n$, for $\xi=1$ and $\xi\simeq0.97$.}%
\label{TwoRates}%
\end{figure}

As we can see, a positive key rate can be extracted in the presence of
entanglement-breaking channels as long as the classical correlations of the
environment are sufficiently high (see the red `reactivating region' in the
figure). The best reactivating environment is clearly the extremal
bottom-right point $(c,c^{\prime})=(1,-1)$. At some specified point, e.g., the
suboptimal point $(1,1)$, we can plot the rate $R_{\text{add}}$ as a function
of $n$. This is done in Fig.~\ref{TwoRates}, where the rate is shown to be
positive in the entanglement-breaking range $2<n\leq4$. In the following
experimental implementation, we show that this behavior is robust to the
presence of loss.

\section{Experimental Methods\label{ExpMETH}}

\subsection{Description of the setup}

Our theoretical results are confirmed by the proof-of-principle experiment,
whose setup is schematically depicted in Fig.~\ref{SetupFIGmain} of the main text. 
Here Alice and Bob receive 1064 nm light from the
same laser source (common local oscillator), divided amongst them by a
balanced beamsplitter. At both stations, the incoming beams are Gaussianly
modulated in phase and amplitude using independent electro-optical modulators
driven by uncorrelated signal generators. Unwanted correlations between
different quadratures are remedied by purifying the polarization of the light
entering the modulators with a combination of waveplates and polarizing
beamsplitters. In this way the two parties are able to generate random
coherent states with independent Gaussian modulations in the two quadratures
on top of the common local oscillator.

Simultaneously, the phase and amplitude modulators are subject to a
side-channel attack~\cite{Gisin,mdiQKD}: Additional electrical inputs are
introduced by Eve, whose effect is to generate additional and unknown
phase-space displacements. In particular, Eve's electrical inputs are
perfectly correlated so that the resulting optical displacements introduce a
correlated-additive Gaussian environment with CM~(\ref{CMeveAPPE})
$\mathbf{V}\left(  n,c\simeq1,c^{\prime}\simeq1\right)  $, i.e., the
suboptimal point $(1,1)$ previously discussed. The magnitudes of the
correlated noise modulations are incrementally increased from $n=0$ to $n=4$
($\simeq4.8$dB) via $0.2$dB-steps, and kept symmetric between the quadratures.
Simultaneously, the signal modulations are kept constant at the same level in
both quadratures for both Alice and Bob, realizing the constant modulation
variance of $\mu\simeq52$ shot noise units.

The optical modes then reach the midway Charlie, i.e., the relay. Here the two
modes interfere at a balanced beam splitter with very high visibility
($>99\%$) and their relative phase is controlled by using a piezo mounted
mirror in such a way to produce equally intense beams at the output. The
output beams are then focused onto two balanced and highly-efficient
photodetectors. Their photocurrents are subtracted and added to produce both
the difference of the amplitude quadratures and the sum of the phase
quadratures, respectively. The overall quantum efficiency of the relay is
around $98\%$.

Even if small, the additional loss associated with the experimental
imperfections must be ascribed to Eve. This means that, besides the
side-channel attack of Alice's and Bob's private spaces, Eve is also assumed
to actively attack the two external links with the relay. Globally, we then
assume that Eve performs a coherent two-mode Gaussian attack affecting the
modes both inside and outside the private spaces. As explained before, we can
handle this worst-case scenario because we can derive the key rate directly
from Alice and Bob's shared data, assuming that Eve possesses the whole
environmental purification compatible with this data (see~\ref{QKDappGEN}). The presence of additional loss clearly worsens the performance of the
protocol but also makes the implementation more interesting since it proves
that the reactivation phenomenon may indeed occur in a realistic lossy
environment.


In our experiment, it is worth noticing that the detection method is enabled
by the brightness of the carrier and represents a simple alternative to the
standard eight-port measurement setup which is typically needed for
implementing the CV Bell detection~\cite{UlrikPRL}. Furthermore, as the
subtraction/addition processes are performed in a software program, an
imbalanced hardware-system can be compensated during the post-processing.

All measurements are done at a sideband frequency of $10.5$MHz. This is done
in order to avoid the low frequency noise close to the carrier frequency and
in turn provides a quantum noise limited signal. The power of the individual
beams before the detectors was about $1.4$mW. The received signal is mixed
down to dc from the measurement frequency of $10.5$MHz. The dc signal is low
pass filtered at $100$kHz to set the detection bandwidth and is digitized with
$500$kHz sampling rate and $14$bit resolution. Our data blocks are $10^6$ and thus long enough for our secret-key rate to converge to its asymptotic value, which is achieved after $10^6$ data points.


\subsection{Experimental secret key rate}

Eve's electric signals sent to the modulators have the effect to create random
displacements on the optical modes, in such a way to generate an optical
Gaussian environment with correlated-additive noise. Besides this, we also
have loss at the untrusted relay which must be ascribed to Eve in the
worst-case scenario. From the point of view of Alice and Bob, Eve's actions
are globally perceived as a coherent Gaussian attack of the two optical modes.
All the environmental ancillas used by Eve are stored in a quantum memory,
which is coherently detected at the end of the protocol.

As typical in QKD, Alice and Bob publicly disclose a subset of their data.
Thus, they are able to reconstruct the joint Gaussian statistics of the three
main variables of the protocol, i.e., their encodings and the outcome of the
relay $\gamma$. From the second-order statistical moments they can compute the
experimental CM $\mathbf{V}_{ab|\gamma}^{\text{exp}}$ associated with the
entanglement-based representation of the protocol.

\begin{figure}[t]
\vspace{-5.4cm}
\par
\begin{center}
\includegraphics[width=0.70\textwidth]{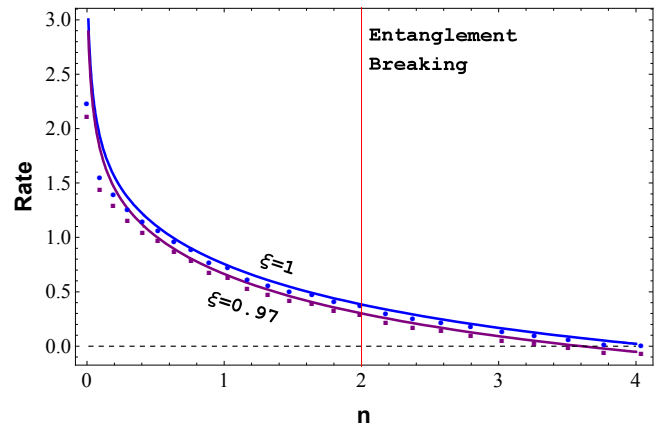}
\end{center}
\par
\vspace{-5.4cm}\caption{Experimental key rate (bits/use) assuming ideal
reconciliation ($\xi=1$, blue circles) and realistic reconciliation efficiency
($\xi\simeq0.97$, purple squares). Modulation variance is $\mu\simeq52$. Due
to loss, the experimental rates are slightly below the theoretical curves
associated with the side-channel attack, corresponding to a
correlated-additive environment with $(c,c^{\prime})=(1,1)$. Additive noise
$n$ is increased beyond the entanglement-breaking threshold ($n>2$). We can
see that the experimental key rate is positive in the region $2<n\leq4$. The
reactivation of QKD from entanglement-breaking is experimentally confirmed.}%
\label{finAPP}%
\end{figure}

Once this CM is known, they can derive the experimental secret-key rate,
following the steps of~\ref{QKDappGEN}. From $\mathbf{V}_{ab|\gamma
}^{\text{exp}}$ they can compute its symplectic spectrum and the matrices
$\mathbf{V}_{b|\gamma}^{\text{exp}}$ and $\mathbf{V}_{b|\gamma\alpha
}^{\text{exp}}$. Thus, they can compute the mutual information via
Eq.~(\ref{mutuaABapp}) and Eve's Holevo information via
Eq.~(\ref{ChiEVEcalcolo}). By replacing these quantities in
Eq.~(\ref{rateAPPn}), they then derive the experimental value of the secret
key rate $R_{\text{exp}}(\xi)$. This is the rate that Alice and Bob would
achieve by post-processing their data via classical codes with reconciliation
efficiency $\xi$.

As we can see from Fig.~\ref{finAPP} (and the corresponding Fig.~\ref{ExpOUT} in the main
text), the experimental key rate is slightly below the theoretical prediction
which is computed for the correlated-additive environment induced by the
side-channel attack. This discrepancy comes from the presence of additional
(small) loss at the relay station which clearly degrades the performance of
the realistic protocol. From Fig.~\ref{finAPP} we see that the experimental
rate remains positive after the entanglement-breaking threshold. (Note that
the entanglement-breaking condition $n>2$, derived for the additive-noise
environment, continues to hold, approximately, when small loss is present.)

\section{Further Discussion}

We clarify some points which may help the readers to better
understand the impact of our results. In the first subsection we discuss a
simple example of reactivation with qubits. This example is rather artificial
and is only provided to clarify the conditions where the phenomenon of
reactivation becomes non-trivial. In the second subsection we explain the
relations with previous literature.

\subsection{Reactivation with discrete variables}

Consider the protocol of entanglement swapping in a lossless environment known
as $U\otimes U^{\ast}$-twirling. This is realized by a classical mixture of
operators of the type $U\otimes U^{\ast}$, with $U$ being a unitary. Suppose
that Alice and Bob possess two Bell pairs, $\rho_{aA}$ and $\rho_{bB}$,
respectively. For instance, each pair may be a singlet state $\left(
\left\vert 0,1\right\rangle -\left\vert 1,0\right\rangle \right)  /\sqrt{2}$.
Qubits $a$ and $b$ are retained, while traveling qubits $A$ and $B$ are
subject to twirling, so that their reduced state $\rho_{AB}$ is transformed
as
\begin{equation}
\rho_{A^{\prime}B^{\prime}}=\int dU~(U\otimes U^{\ast})~\rho_{AB}~(U\otimes
U^{\ast})^{\dagger}~, \label{twirl}%
\end{equation}
where the integral is over the entire unitary group $\mathcal{U}(2)$ acting on
the bi-dimensional Hilbert space and $dU$ is the Haar measure.

On the one hand, this environment is locally entanglement breaking. In fact,
by taking the partial traces of Eq.~(\ref{twirl}), we can see that the two
channels $\rho_{A}\rightarrow\rho_{A^{\prime}}$ and $\rho_{B}\rightarrow
\rho_{B^{\prime}}$ are completely depolarizing. On the other, the application
of a Bell detection on the output qubits $A^{\prime}$ and $B^{\prime}$ has the
effect to completely cancel the environmental noise. In fact, one can easily
check, the output state of the remote qubits $a$ and $b$ will be projected
onto a singlet state up to a Pauli operator, which can be \textquotedblleft
undone\textquotedblright\ thanks to the communication of the Bell outcome.

This example is artificial because it heavily relies on the fact that
environment is lossless (no qubit is lost) and the action of the unitaries is
very specific, i.e., they are perfectly correlated and of the twirling type
$U\otimes U^{\ast}$. In the presence of loss, this perfect noise-cancelation
rapidly tends to disappear. This is why the study of the reactivation
phenomenon becomes non-trivial in realistic lossy environments. From this
point of view, it is known that the quantum systems which are more fragile to
loss are CV systems, which is why the study of reactivation is absolutely
non-trivial for bosonic modes in lossy Gaussian environments.

\subsection{Main results and relations with other literature}

To our knowledge ours is the first work where:

\begin{enumerate}
\item The basic CV relay-based protocols of entanglement swapping, quantum
teleportation, entanglement/key distillation are studied and extended to
non-Markovian conditions.

\item Weak non-Markovian effects (modelled by separable correlations) are
shown to reactivate these relay-based protocols back from standard (Markovian)
conditions of entanglement breaking.

\item The survival of a multi-partite form of entanglement provides a physical
resource which can be localized and then exploited by the previous protocols
(directly or indirectly).

\item The reactivation of a quantum relay is experimentally demonstrated.

\item As explained in the main text, we experimentally show that the
single-repeater bound~\cite{RepBound} can be overcome by the presence of
classical (separable) correlations in the bosonic environment.
\end{enumerate}

These main achievements have a limited overlap with previous results in the
literature. Ref.~\cite{NJPpirs} is a theoretical-only study which considered a
different configuration, that where Charlie (in the middle) has an entangled
source to be distributed to Alice and Bob. Despite this configuration of
direct entanglement distribution can be seen (by some authors) as a reverse
formulation of entanglement swapping, the two schemes are inequivalent and
very well distinguished by the community.

The distinction between direct entanglement distribution and entanglement
swapping is really important and is at the basis of different branches of
quantum information protocols. For instance, direct entanglement distribution
plays an important role in device-independent QKD, testing of non-locality
etc. By contrast, entanglement swapping is the core technique for quantum
repeaters, MDI-QKD (i.e., semi device independent QKD), etc. It is clear that
showing a new effect or property for one of the two configurations does not
automatically extend to the other.

Let us further discuss the basic differences between direct entanglement
distribution and entanglement swapping:

\begin{itemize}
\item In CVs, entanglement swapping is much more fragile than direct
entanglement distribution. In fact the two protocols have well-known different
performances in the presence of loss. For example, in the non-Markovian
environment considered in our paper and considering the limit of large $\mu$,
the remote entanglement distributed to Alice and Bob is quantified by smallest
(and asymptotical) PTS eigenvalue%
\[
\varepsilon_{\text{opt}}=(1-\tau)\sqrt{(\omega-g)(\omega+g^{\prime})},
\]
for the case of direct distribution, and by%
\[
\varepsilon_{\text{opt}}=\frac{1-\tau}{\tau}\sqrt{(\omega-g)(\omega+g^{\prime
})},
\]
for the case of entanglement swapping. The analytical simplification induced
by the limit $\mu\rightarrow+\infty$ clearly shows the extra factor $\tau
^{-1}$, which makes the performances of the two configurations completely
inequivalent in the presence of loss.

\item Entanglement swapping is conceptually more interesting for its
connections with network implementations and the end-to-end principle.
Contrarily to the case of direct entanglement distribution, where the central
node (Charlie) must prepare quantum resources to be distributed, in the case
of entanglement swapping Charlie needs only to perform a very cheap and
efficient detection on the incoming systems. Removing quantum resources from
intermediate nodes is a key step for the scalability of quantum protocols to
large quantum networks.

\item Specifically about the phenomenon of reactivation:\ this is possible in
both configurations, but this is based on two inequivalent dynamics of the
quantum correlations. In the direct distribution of entanglement, the
injection of correlations from the environment can reactivate the transmission
of \textbf{bipartite} entanglement from Charlie to Alice and Bob. In the
entanglement swapping configuration, no bipartite entanglement (or even
tripartite) can be transmitted even with the injection of separable
correlations from the environment. The key resource is here
\textbf{quadripartite} entanglement which is not directly exploitable by the
parties but must be localized by the action of the relay. Thus the phenomenon
relies on the survival of a multipartite form of quantum entanglement.

\item Finally, we stress that the non-Markovian study of all the other
relay-based protocols, i.e., quantum teleportation, quantum repeater (swapping
plus distillation), key distillation and practical QKD, were not treated
before (and no experimental implementation was done).
\end{itemize}


For completeness, we also discuss the relations between our work and previous
literature on QKD, specifically Ref.~\cite{Untrusted}, where MDI-QKD\ with CV
systems has been introduced. Together with Ref.~\cite{Untrusted}, the present
work shares the basic structure of the QKD\ protocol and the necessity to
perform a security analysis in the presence of environmental correlations. As
discussed in Ref~\cite{Untrusted}, random permutations and quantum de Finetti
arguments do not allow to reduce the most general coherent attack into simple
one-mode Gaussian attacks of the links. In other words, the unconditional
security must be tested against a two-mode Gaussian attack of the links which
therefore involves the presence of correlations and non-Markovian effects.

Apart from this common ground, the novelties of the present work with respect
to previous Ref.~\cite{Untrusted} are several and non-trivial. These include
the following points:

\begin{itemize}
\item In the present work we study MDI-QKD in the presence of entanglement
breaking. This is an extremely insecure scenario which has never been
considered by previous literature.

\item For the first time we show that non-Markovian effects (in the form of a
suitable coherent Gaussian attack) may actually be beneficial for QKD. The
fact that some coherent attacks may actually help the key distribution is an
interesting new feature whose potentialities should further be explored.

\item We experimentally realize the first two-mode side-channel attack of a CV
QKD protocol.
\end{itemize}


\begin{thebibliography}{99}                                                                                               %


\bibitem {CoverThomas}Cover, T. M. \& Thomas, J. A. Elements of Information
Theory (2nd edition, John Wiley \& Sons, Inc., Hoboken, New Jersey 2006).

\bibitem {book1}Nielsen, M. A. \& Chuang, I. L. \textit{Quantum Computation
and Quantum Information} (Cambridge University Press, Cambridge, 2000).

\bibitem {book2}Bouwmeester, D. \textit{The Physics of Quantum Information:
Quantum Cryptography, Quantum Teleportation, Quantum Computation}
(Springer-Verlag, Berlin, 2000).

\bibitem {book3}Vedral, V. \textit{Introduction to Quantum Information
Science} (Oxford University Press, 2006).

\bibitem {book4}Bengtsson I. \& \.{Z}yczkowski, K. \textit{Geometry of quantum
states: An Introduction to Quantum Entanglement} (Cambridge University Press,
Cambridge 2006).

\bibitem {book5}Barnett, S. \textit{Quantum Information} (Oxford University
Press, 2009)

\bibitem {book6}Schumacher, B. \& Westmoreland, M. \textit{Quantum Processes
Systems, and Information} (Cambridge University Press, Cambridge, 2010).

\bibitem {book7}Holevo, A. \textit{Quantum Systems, Channels, Information: A
Mathematical Introduction} (De Gruyter, Berlin-Boston, 2012).

\bibitem {book8}Watrous, J. \textit{The theory of quantum information}
(Cambridge University Press, Cambridge, 2018).

\bibitem {RMP}Weedbrook, C., Pirandola, S., Garcia-Patron, R., Cerf, N. J.,
Ralph, T. C., Shapiro, J. H. \& Lloyd, S. Gaussian quantum information.
\textit{Rev. Mod. Phys.} \textbf{84}, 621 (2012).

\bibitem {RMP2}Braunstein, S. L. \& van Loock, P. Quantum information with
continuous variables. \textit{Rev. Mod. Phys.} \textbf{77}, 513 (2005).

\bibitem {hybrid1}Andersen, U. L., Neergaard-Nielsen, J. S., van Loock, P. \&
Furusawa, A. Hybrid quantum information processing. \textit{Nat. Phys.}
\textbf{11}, 713--719 (2015)

\bibitem {hybrid2}Kurizki, G., Bertet, P., Kubo, Y., M\o lmer, K., Petrosyan,
D., Rabl, P., \& Schmiedmayer, J. Quantum technologies with hybrid systems.
\textit{Proc. Natl. Acad. Sci. USA} \textbf{112}, 3866-73 (2015).

\bibitem {Zukowski}Zukowski, M., Zeilinger, A., Horne, M. A. \& Ekert, A.
\textquotedblleft Event ready detectors\textquotedblright\ Bell experiment via
entanglement swapping. \textit{Phys. Rev. Lett.} \textbf{71}, 4287 (1993).

\bibitem {EntSwap}van Loock, P. \& Braunstein, S. L. Unconditional
teleportation of continuous-variable entanglement.\ \textit{Phys. Rev. A}
\textbf{61}, 010302(R) (1999).

\bibitem {EntSwap2}Polkinghorne, R.E.S. \& Ralph, T. C. Continuous Variable
Entanglement Swapping. \textit{Phys. Rev. Lett.} \textbf{83}, 2095 (1999).

\bibitem {GaussSWAP}Pirandola, S., Vitali, D., Tombesi, P. \& Lloyd, S.
Macroscopic Entanglement by Entanglement Swapping. \textit{Phys. Rev. Lett.}
\textbf{97}, 150403 (2006).

\bibitem {Briegel}Briegel, H.-J., D\"{u}r, W., Cirac, J. I. \& Zoller, P.
Quantum Repeaters: The Role of Imperfect Local Operations in Quantum
Communication. \textit{Phys. Rev. Lett.} \textbf{81}, 5932 (1998)

\bibitem {Tele}Bennett, C. H. \textit{et al}. Teleporting an unknown quantum
state via dual classical and Einstein-Podolsky-Rosen channels.\ \textit{Phys.
Rev. Lett.} \textbf{70}, 1895 (1993).

\bibitem {Tele2}Furusawa, A., et al. Unconditional quantum teleportation.
\textit{Science} \textbf{282}, 706 (1998).

\bibitem {telereview}Pirandola, S., Eisert, J., Weedbrook, C., Furusawa, A. \&
Braunstein, S. L. Advances in Quantum Teleportation. \textit{Nature Photon.}
\textbf{9}, 641-652 (2015).



\bibitem {mdiQKD}Braunstein, S. L. \& Pirandola, S., Side-Channel-Free Quantum
Key Distribution. \textit{Phys. Rev. Lett.} \textbf{108}, 130502 (2012).


\bibitem {Lo}Lo, H.-K., Curty, M. \& Qi, B. Measurement-Device-Independent
Quantum Key Distribution. \textit{Phys. Rev. Lett.} \textbf{108}, 130503 (2012).

\bibitem {Untrusted}Pirandola, S. \textit{et al}. High-Rate
Measurement-Device-Independent Quantum Cryptography. \textit{Nature Photon.}
\textbf{9}, 397-402 (2015).

\bibitem {TFQKD}Lucamarini, M., Yuan, Z. L., Dynes, J. F. \& Shields, A. J.
Overcoming the rate-distance limit of quantum key distribution without quantum
repeaters. \textit{Nature} \textbf{557}, 400 (2018).


\bibitem {SNSwang}Wang, X.-B., Yu, Z.-W. \& Hu, X.-L. Twin-field quantum key
distribution with large misalignment error. \textit{Phys. Rev. A} \textbf{98},
062323 (2018).



\bibitem {QKDrev}Pirandola, S., Andersen, U. L., Banchi, L., Berta, M.,
Bunandar, D., Colbeck, R., Englund, D., Gehring, T., Lupo, C., Ottaviani, C.,
Pereira, J., Razavi, M., Shaari, J. S., Tomamichel, M., Usenko, V. C.,
Vallone, G., Villoresi, P. \& Wallden, P. Advances in quantum cryptography.
\textit{Preprint arXiv}:1906.01645 (2019).



\bibitem {photonic1}Metcalf, B. J., et al. Quantum teleportation on a photonic
chip. \textit{Nature Photon.} \textbf{8}, 770--774 (2014).

\bibitem {photonic2}Masada, G., et al. Continuous-variable entanglement on a
chip. \textit{Nature Photon.} \textbf{9}, 316--319 (2015)

\bibitem {chip3}L. Steffen, L., et al. Deterministic quantum teleportation
with feed-forward in a solid state system. \textit{Nature} \textbf{500}, 319 (2013).

\bibitem {Petruccione}Breuer, H.-P. \& Petruccione, F. \textit{The Theory of
Open Quantum Systems} (Oxford University Press, Oxford, 2002).

\bibitem {UlrikCORR}Lassen, M., Berni, A., Madsen, L .S., Filip, R. \&
Andersen U. L. Gaussian Error Correction of Quantum States in a Correlated
Noisy Channel. \textit{Phys. Rev. Lett.} \textbf{111}, 180502 (2013).

\bibitem {Tyler09}Tyler, G. A. \& Boyd, R. W. Influence of atmospheric
turbulence on the propagation of quantum states of light carrying orbital
angular momentum. \textit{Opt. Lett.} \textbf{34}, 142 (2009).

\bibitem {Semenov09}Semenov, A. A. \& Vogel, W. Quantum light in the turbulent
atmosphere. \textit{Phys. Rev. A} \textbf{80}, 021802(R) (2009).

\bibitem {Boyd11}Boyd, R. W., Rodenburg, B., Mirhosseini, M. \& Barnett, S. M.
Influence of atmospheric turbulence on the propagation of quantum states of
light using plane-wave encoding. \textit{Opt. Express} \textbf{19}, 18310 (2011).


\bibitem {Renner1}Renner, R. Symmetry of large physical systems implies
independence of subsystems. \textit{Nature Phys.} \textbf{3}, 645-649 (2007).

\bibitem {Renner2}Renner, R. \& Cirac, J. I. de Finetti representation theorem
for infinite-dimensional quantum systems and applications to quantum
cryptography. \textit{Phys. Rev. Lett.} \textbf{102}, 110504 (2009).


\bibitem {EBchannels}Horodecki, M., Shor, P. W. \& Ruskai, M. B. General
Entanglement Breaking Channels. \textit{Rev. Math. Phys.} \textbf{15}, 629 (2003).

\bibitem {HolevoEB}Holevo, A. S. Entanglement-breaking channels in infinite
dimensions. \textit{Problems of Information\ Transmission} \textbf{44}, 3 (2008).

\bibitem {RepBound}Pirandola, S. End-to-end capacities of a quantum
communication network. \textit{Commun. Phys.} \textbf{2}, 51 (2019). See also
\textit{preprint arXiv}:1601.00966 (2016).

\bibitem {BellFORMULA}Spedalieri, G., Ottaviani, C. \& Pirandola, S.
Covariance matrices under Bell-like detections. \textit{Open Syst. Inf. Dyn.}
\textbf{20}, 1350011 (2013).



\bibitem {TwomodePRA}Pirandola, S., Serafini, A. \& Lloyd, S. Correlation
matrices of two-mode bosonic systems. \textit{Phys. Rev. A \textbf{79}},
052327 (2009).

\bibitem {NJPpirs}Pirandola, S. Entanglement reactivation in separable
environments. \textit{New J. Phys.} \textbf{15}, 113046 (2013).


\bibitem {tripartite}Giedke, G., Kraus, B., Lewenstein, M. \& Cirac, J. I.
Separability Properties of Three-mode Gaussian States. \textit{Phys. Rev. A}
\textbf{64}, 052303 (2001).

\bibitem {quadripartite}Werner, R. F. \& Wolf, M. M. Bound entangled Gaussian
states. \textit{Phys. Rev. Lett.} \textbf{86}, 3658 (2001).

\bibitem {logNEG}Vidal, G. \& Werner, R. F. Computable measure of
entanglement. \textit{Phys. Rev. A} \textbf{65}, 032314 (2002).

\bibitem {logNEG1}Eisert, J. Entanglement in quantum information theory. PhD
thesis (Potsdam, February 2001).

\bibitem {logNEG2}Plenio, M. B. The logarithmic negativity: A full
entanglement monotone that is not convex. \textit{Phys. Rev. Lett}.
\textbf{95}, 090503 (2005).

\bibitem {Qinstrument}Devetak, I. \& Winter, A. Distillation of secret key and
entanglement from quantum states. \textit{Proc. R. Soc. Lond. A} \textbf{461},
207 (2005).

\bibitem {CohINFO}Schumacher, B. \& Nielsen, M. A. Quantum data processing and
error correction. \textit{Phys. Rev. A} \textbf{54}, 2629 (1996).

\bibitem {CohINFO2}Lloyd, S. Capacity of the noisy quantum
channel.\ \textit{Phys. Rev. A}\textbf{\ 55}, 1613 (1997).

\bibitem {KeyCAP}Pirandola, S., Garc\'{\i}a-Patr\'{o}n, R. Braunstein, S.~L.
\& Lloyd, S. Direct and reverse secret-key capacities of a quantum channel.
\textit{Phys. Rev. Lett.} \textbf{102}, 050503 (2009).


\bibitem {TopENT}Eckstein, A., Christ, A., Mosley, P. J. \& Silberhorn, C.
Highly Efficient Single-Pass Source of Pulsed Single-Mode Twin Beams of Light.
\textit{Phys. Rev. Lett.} \textbf{106}, 013603 (2011).

\bibitem {TobiasSqueezing}Eberle, T., H\"{a}ndchen, V. \& Schnabel R. Stable
control of 10 dB two-mode squeezed vacuum states of light. \textit{Opt.
Express} \textbf{21}, 11546 (2013).

\bibitem {PLOB}Pirandola, S., Laurenza, R., Ottaviani, C. \& Banchi, L.
Fundamental Limits of Repeaterless Quantum Communications. \textit{Nature
Commun.} \textbf{8}, 15043 (2017).






\bibitem {Williamson}J. Williamson, Am. J. Math. \textbf{58}, 141 (1936).

\bibitem {Sera}A. Serafini, F. Illuminati, and S. De Siena, J. Phys. B
\textbf{37}, L21 (2004).

%

\bibitem {HET}Let us consider an arbitrary two-mode Gaussian state $\rho_{ab}$
with the following mean value and CM
\[
\mathbf{\bar{x}}=\left(
\begin{array}
[c]{c}%
\mathbf{\bar{x}}_{a}\\
\mathbf{\bar{x}}_{b}%
\end{array}
\right)  \in\mathbb{R}^{4},~\mathbf{V}=\left(
\begin{array}
[c]{cc}%
\mathbf{A} & \mathbf{C}\\
\mathbf{C}^{T} & \mathbf{B}%
\end{array}
\right)  ,
\]
where $\mathbf{A=A}^{T}$, $\mathbf{B=B}^{T}$ and $\mathbf{C}$ are $2\times2$
real blocks. Let heterodyne mode $a$ with complex outcome $\alpha=(q+ip)/2$.
The corresponding real outcome $\mathbf{a}=(q,p)^{T}$ is achieved with
probability
\[
p(\mathbf{a})=\frac{\exp\left[  -\frac{1}{2}\mathbf{d}^{T}(\mathbf{A}%
+\mathbf{I})^{-1}\mathbf{d}\right]  }{2\pi\sqrt{\det(\mathbf{A}+\mathbf{I})}%
},~\mathbf{d}:=\mathbf{\bar{x}}_{a}-\mathbf{a},
\]
which is Gaussian with classical CM $\mathbf{A}+\mathbf{I}$. Correspondingly,
mode $b$ is projected on a conditional Gaussian state $\rho_{b|\alpha}$ with
mean value $\mathbf{\bar{x}}_{b|\alpha}=\mathbf{\bar{x}}_{b}-\mathbf{C}%
^{T}(\mathbf{A}+\mathbf{I})^{-1}\mathbf{d}$ and CM$\ \mathbf{V}_{b|\alpha
}=\mathbf{B}-\mathbf{C}^{T}(\mathbf{A}+\mathbf{I})^{-1}\mathbf{C}$.

\bibitem {OptimalDiscord}Pirandola, S., Spedalieri, G., Braunstein, S. L.,
Cerf, N. J. \& Lloyd, S. Optimality of Gaussian Discord. \textit{Phys. Rev.
Lett}. \textbf{113}, 140405 (2014).

\bibitem {RMPdis}K. Modi, A. Brodutch, H. Cable, T. Paterek, and V. Vedral,
Rev. Mod. Phys. \textbf{84}, 1655-1707 (2012).

\bibitem {ParisD}P. Giorda and M. G. A. Paris, Phys. Rev. Lett. \textbf{105},
020503 (2010).

\bibitem {GerryD}G. Adesso and A. Datta, Phys. Rev. Lett. \textbf{105}, 030501 (2010).



\bibitem {dB}Note that for a TMSV\ state with variance $\mu$, the amount of
two-mode squeezing in dB is given by the formula $-10\log_{10}\left(
\mu-\sqrt{\mu^{2}-1}\right)  $.



\bibitem {QinstNOTE}A quantum instrument is a quantum operation which can have
both classical and quantum outputs. For each classical outcome, there is a
corresponding completely positive map applied to the quantum
systems~\cite{Qinstrument}. For instance, a quantum instrument may describe
the global effect of a partial quantum measurement (i.e., applied on a subset
of the initial systems). A quantum measurement applied to all quantum systems
can be seen as quantum instrument with classical output only.


%


\bibitem {DW}Devetak, I. \&Winter, A. Relating Quantum Privacy and Quantum
Coherence: An Operational Approach. \textit{Phys. Rev. Lett}. \textbf{93},
080501 (2004).

\bibitem {SciREP}Pirandola, S. Quantum discord as a resource for quantum
cryptography. \textit{Sci. Rep.} \textbf{4}, 6956 (2014).



\bibitem {Jouguetapp}Jouguet, P., Kunz-Jacques, S., \& Leverrier, A.
Long-distance continuous-variable quantum key distribution with a Gaussian
modulation. \textit{Phys. Rev. A} \textbf{84}, 062317 (2011).


\bibitem {Gisin}Gisin, N., Ribordy, G., Tittel, W. \& Zbinden, H. Quantum
Cryptography. \textit{Rev. Mod. Phys.} \textbf{74}, 145 (2002).

\bibitem {UlrikPRL}Niset, J., Ac\'{\i}n, A., Andersen, U. L., Cerf, N. J.,
Garc\'{\i}a-Patr\'{o}n, R., Navascu\'{e}s, M., \& Sabuncu, M. Superiority of
entangled measurements over all local strategies for the estimation of product
coherent states. \textit{Phys. Rev. Lett.} \textbf{98}, 260404 (2007).




\end{thebibliography}
\end{document}